# Distortion Under Public-Spirited Voting

BAILEY FLANIGAN, Carnegie Mellon , USA

ARIEL D. PROCACCIA, Harvard University, USA

SVEN WANG, Massachusetts Institute of Technology, USA

A key promise of democratic voting is that, by accounting for all constituents' preferences, it produces decisions that benefit the constituency overall. It is alarming, then, that all deterministic voting rules have unbounded *distortion*: all such rules — even under reasonable conditions — will sometimes select outcomes that yield essentially no value for constituents. In this paper, we show that this problem is mitigated by voters being *public-spirited*: that is, when deciding how to rank alternatives, voters weigh the common good in addition to their own interests. We first generalize the standard voting model to capture this public-spirited voting behavior. In this model, we show that public-spirited voting can substantially — and in some senses, monotonically — reduce the distortion of several voting rules. Notably, these results include the finding that if voters are *at all* public-spirited, some voting rules have *constant distortion* in the number of alternatives. Further, we demonstrate that these benefits are robust to adversarial conditions likely to exist in practice. Taken together, our results suggest an implementable approach to improving the welfare outcomes of elections: *democratic deliberation*, an already-mainstream practice that is believed to increase voters' public spirit.

# 1 INTRODUCTION

Consider an election with two alternatives, $a$ and $b$; of the 100 voters, 50 prefer $a$ to $b$ and 50 prefer $b$ to $a$. Since the preference profile is symmetric, let us assume that $a$ is elected. Although their rankings are symmetric, voters may have highly asymmetric underlying *intensities* of preferences, perhaps capturing that they are affected to differing degrees by the outcome of the election. We capture these preference intensities with *utilities*, which can be interpreted as measuring the value a voter gains from a given alternative. In this case, suppose the supporters of $a$ are affected similarly by the alternatives, having utility 1 for $a$ and 0 for $b$, whereas the admirers of $b$ are, by comparison, are affected much more disparately by the alternatives, having utilities 0 for $a$ and 100 for $b$.

From a societal benefit standpoint, $b$ would have been the better choice, as it would yield substantially more utility to voters overall. This intuition is captured by the *utilitarian social welfare*, defined as the sum of voters' utilities for a given alternative: $a$ (the winner) is severely suboptimal in terms of this measure, its social welfare being 100 times lower than that of $b$ (the alternative with optimal social welfare). This ratio can be made arbitrarily large by, say, making the supporters of $a$ arbitrarily unaffected by the decision.

The simple example above implies an alarming conclusion: that *any* deterministic rankings-based voting procedure will, in some instances, choose an alternative that yields arbitrarily suboptimal value for the population. From a technical standpoint, this welfare loss arises due to information lost between cardinal utilities and ordinal preferences; this was first observed by [Procaccia and Rosenschein, 2006], who quantified this loss with the notion of *distortion*. Assuming voters report rankings that are consistent with their underlying utilities, the distortion of a voting rule is the worst-case (over latent utilities) ratio between the utilitarian social welfare of the highest-welfare alternative and that of the elected alternative. By the example above, then, *all deterministic voting rules* must have unbounded distortion.

A natural question, then, is: under what assumptions *is* the distortion bounded? The rich literature on distortion — overviewed in an excellent recent survey by Anshelevich et al. [2021] — has largely taken one of two approaches to achieving bounded distortion. One line of research, originating from the work of Procaccia and Rosenschein [2006], assumes that each voter's utilities sum to 1, thereby eliminating the possibility of voters being affected by widely differing degrees by the decision. Another line of research, originating from the work of Anshelevich et al. [2018], assumes that voters' preferences are induced by distances in an underlying metric space.

Both of these lines of work rely on assumptions that restrict voters' possible latent utilities (or analogously in some models, costs). However, it is not clear whether we can rely on such assumptions to hold in practice. This is perhaps most directly illustrated by the fact that the core problem in our example above cannot be ruled out as a potential feature of real-world elections: the utilities are such that there is a minority group that is much more affected by the issue than a majority group with decisive voting power. Moreover, it seems unlikely that we can *promote* such conditions on the utilities, because voters' utilities — how much they fundamentally gain from a given election outcome — would likely arise from features that are difficult to change with simple interventions.

In this paper, we take a different approach to attaining bounded distortion. This approach begins from the realization that while underlying utilities like those in our example might unfortunately be realistic, the *behavioral model by which voters translate utilities into rankings* might be too pessimistic. The standard behavioral assumption made in the literature is that voters rank alternatives according to only the order of their own utilities. However, as many social scientists have observed, this model is unrealistic in a way that can potentially help us: voters can be *public-spirited* — that is, when they vote, they weigh not only how they themselves are impacted by each alternative, but also each alternative impacts their society as a whole.[1] This behavior of balancing self and societal interest can be captured in a natural generalization of the standard behavioral model of voters: instead of ranking alternatives according to only their own utilities, a $\gamma$-*public spirited* voter ranks alternatives according to values that place weight $1 - \gamma$ on their own utilities, and weight $\gamma$ on each alternative's utilitarian social welfare. It is then intuitive why public spirited voting could help decrease the distortion: it will cause voters to more highly rank higher-welfare alternatives, potentially increasing the social welfare of the election winner.

While existing work suggests that voters are willing and able to be public-spirited, we need not assume that these conditions are satisfied by default; instead, we can intentionally cultivate them within the democratic process. One promising innovation on this front that is currently gaining momentum globally[2] is *democratic deliberation*, summarized by Mendelberg [2002] as dialogues in which "people rely on reasons that speak to the needs or principles of everyone affected by the matter at hand." This description of deliberation already alludes to some of its key potential benefits, which roughly correspond to promoting our conditions. For instance, deliberation is theorized to lead to "citizens [being] more enlightened about their own and others' needs and experiences" [Mendelberg, 2002] — akin to promoting more accurate estimates of alternatives' welfares, and to "an increased willingness to recognize community values and to compromise in the interest of the common good" [Karpowitz and Mendelberg, 2011] — akin to promoting voters' levels of public spirit. These theorized benefits are supported by empirical evidence showing, for example, that deliberation can increase public-spiritedness [Wang et al., 2020], lead to more egalitarian values [Gastil et al., 2010], and increase empathy for members of social outgroups [Grönlund et al., 2017].

This evidence suggests that public-spirited voting behavior can be cultivated (or may already exist) among voters. This motivates our research question, which, if answered affirmatively, would lead to an actionable approach to decreasing deterministic voting rules' otherwise unbounded distortion:

> *To what extent is public-spirited voting guaranteed to decrease the distortion, and for which voting rules?*[3]

We aim to formally answer this question with the tools of social choice theory, as outlined in the results and contributions below. In our analysis, we focus on deterministic voting rules, owing to the several political hurdles to implementing randomized rules. We leave the study of randomized rules in our model to future work.

## 1.1 Results and Contributions

Throughout the rest of the paper, we will often use *PS* to refer to the concept of public spirit.

---

[1] Public-spirited behavior among voters has been demonstrated empirically [Kendall and Matsusaka, 2021, Zettler et al., 2011] and has long featured in economic theories of how people make decisions [Becker, 1976, Kangas, 1997].

[2] Democratic deliberation is commonly implemented through *deliberative polls* or *citizens' assemblies*, of which hundreds have been run in the past few years [Participedia, 2022]. Such processes have played a key role in major political decisions: for example, citizens' assemblies commissioned by Ireland's national legislature recently led to amending the Irish constitution on the issues of same-sex marriage and abortion [Irish Citizen's Assembly Project, 2019].

[3] A natural question here is, if *constituents* can learn alternatives' social welfares via, e.g., deliberation, why can't the *election designer* learn these values and directly select the highest-welfare alternative? One reason is that the election designer imposing such "complete" public spirit could be perceived as undemocratic and illegitimate. Underlying this point is the premise that in a democracy, it is voters' prerogative to decide *how strongly* to account for the social good, an interpretation which views deliberation as a process of *clarifying for voters* how much public spirit their values dictate they should have.

**Section 2: A model of public-spirited voters.** A precursor to answering the question above is formally modeling public-spirited voting behavior. Our model is a simple generalization of the standard model: voter $i$ has public spirit level $\gamma_i \in [0, 1]$, where higher $\gamma_i$ corresponds to more public spirit. Then, voter $i$ ranks alternatives in order of their *PS-value* for each alternative $a$, called $v_i(a, \boldsymbol{\gamma}, U)$. This value is a convex combination of their utility $u_i(a)$ and $a$'s social welfare $\text{sw}(a, U)$ — the sum of all voters' utilities for $a$, summarized in the utility matrix $U$:

$$v_i(a, \boldsymbol{\gamma}, U) = (1 - \gamma_i)u_i(a) + \gamma_i \cdot \text{sw}(a, U)/n. \tag{1}$$

The standard behavioral model of voters is then the special case of our model where $\gamma_i = 0$ for all $i$.

**Section 3: Distortion bounds for voting rules.** We begin by proving our key lemma, which upper bounds the extent to which the social welfare of an alternative $a$ can exceed that of another alternative $b$ — a bound which is decreasing in the fraction of voters who rank $b$ ahead of $a$ in the election, along with the minimum level of public spirit among voters, $\gamma_{min} := \min_i \gamma_i$. We then use this result, plus other techniques, to give tight bounds on the distortion of several popular voting rules. For consistency with the distortion literature, we consider these bounds asymptotic in $m$, the number of alternatives in the election. The main takeaway from these bounds is that when voters have *any* public spirit (i.e., if $\gamma_{min} > 0$), several voting rules' distortion drops from unbounded to linear (for the rules Borda, Plurality, Maximin) or even *constant* (for the rules Copeland and Slater). We emphasize that our bounds asymptotically — and for some settings of $\gamma_{min}$, non-asymptotically — either match or beat those possible in both aforementioned models, and moreover do so without any assumptions on voters' underlying utilities.

**Section 4: PS-Monotonicity.** The upper and lower bounds we give in Section 3 are decreasing in $\gamma_{min}$, hinting at a weak form of *PS-monotonicity* — i.e., that the distortion decreases as voters' public spirit increases. Although it seems intuitive that this property should hold, we show that, while some notions of PS-monotonicity are guaranteed, other natural notions do not hold. Working from weaker to stronger notions, we show first that if public spirit increases *uniformly* among voters, then the worst-case distortion of *all voting rules* decreases monotonically. Given that in reality voters' $\gamma_i$ levels are unlikely to be uniform, we then show that for Copeland and Plurality, the worst-case distortion decreases even if voters' public spirit is increased heterogeneously. This implies that cultivating greater public spirit among any voters to any extent is guaranteed to decrease the worst-case distortion over possible utility profiles — already a useful guarantee, since we cannot observe voters' initial levels of public spirit. Given that utilities are also unobservable, one might hope that PS-monotonicity holds for all fixed utility matrices *and* initial levels of public spirit. We soundly resolve this question by showing this is too much to hope for: applying classic axiomatic impossibilities by Muller and Satterthwaite, we prove that *no weakly unanimous, non-dictatorial voting rule* exhibits PS-monotonicity on an instance-by-instance basis.

**Section 5: Robustness of distortion bounds.** There are two key weaknesses, from a practical perspective, of our upper bounds in the Section 3. First, they are vacuous if $\gamma_{min} = 0$, and second, they *a priori* rely on voters using accurate and internally-consistent values of the inputs to Equation (1), $\gamma_i$, $u_i(a)$, and $\text{sw}(a, U)$. We provide robustness results that address both of these gaps. First, we show that our upper bounds degrade by only a constant factor if up to some fraction of voters has $\gamma_i = 0$; for Copeland this fraction is quite large — up to $1/2$ of voters. Second, we generalize our model to allow voters to deviate arbitrarily from correct and/or internally-consistent values of *any model input* $\gamma_i$, $u_i(a)$, and $\text{sw}(a, U)$. We then extend our distortion upper bounds to this generalized model, showing that our original bounds are robust to *all such deviations*: that is, our upper bounds degrade smoothly, by constant factors, in the magnitude of these deviations.

## 1.2 Related Work

**Distortion under existing models.** As discussed in the introduction, the main body of work achieving bounded distortion does so by assuming regularity conditions on voters' utilities. Under the assumption that voters' preferences can be embedded in a metric space, the well-known rule COPELAND has distortion of 5 and there are deterministic voting rules that achieve the best possible distortion of 3 [Gkatzelis et al., 2020, Kizilkaya and Kempe, 2022]. Under the assumption that each voter's utilities sum to 1, all deterministic rules have distortion at least $\Omega(m^2)$, where $m$ is the number of alternatives; the popular rule PLURALITY achieves a matching upper bound [Caragiannis and Procaccia, 2011]. More distantly, there is some work that achieves bounded distortion by assuming additional access to some cardinal information about voters' utilities (see Sec. 5 of Anshelevich et al. [2021] for an overview). In contrast to these lines of existing work, our distortion bounds require neither regularity conditions on voters' utilities, nor any information from voters beyond their rankings. Nonetheless, we can match or improve upon the metric model's upper bound of 5 on COPELAND's distortion when $\gamma_{min} \geq (\sqrt{5} - 1)/2 \approx 0.61$, and we can show that PLURALITY, along with several other deterministic rules, have linear or sub-linear distortion, improving upon the distortion achievable in the unit sum model by at least a factor of $m$ (Table 1).

**Related behavioral models.** Our model of public-spirited voting is a direct analog of a model used in the study of congestion games by Chen et al. [2014], who in turn attribute the idea to Ledyard [1997, p. 154]. Additionally, similar ideas appear in literature exploring altruistic behavior by agents in decision-making systems: for instance, Lindbeck and Weibull [1988] model agents as giving some linear weight $\alpha > 0$ to the interests of another entity as a form of altruism. We remark, however, that *altruism* in this work is distinct from public spirit, because it may involve accounting for only the interests of population subgroups or specific agents for strategic reasons, rather than arising from the motive of benefiting society at large. Other related models include Fehr and Schmidt's model of how economic agents incorporate social inequality into their utilities [Fehr and Schmidt, 1999], and political economic models of *sociotropic* voters, who weigh the economic interests of their country over their own [Kinder and Kiewiet, 1981]. For instance, Bechtel and Liesch [2020] aim to estimate from data *how* sociotropic voters are, corresponding to estimating $\gamma_i$ parameters in our model.

## 2 MODEL

### 2.1 Public-Spirited Voting Behavior

There are $n$ voters and $m$ alternatives. We refer to the set of voters as $[n]$ and alternatives as $[m]$. By default, individual voters are denoted $i \in [n]$ and individual alternatives are denoted $a \in [m]$.

**Public spirit (PS)** We represent voters' levels of public spirit with the *PS-vector* $\boldsymbol{\gamma} \in [0, 1]^n$, whose $i$-th entry $\gamma_i$ is voter $i$'s level of public spirit (higher $\gamma_i$ means more public spirit). Our upper bounds will be in terms of the minimum level of public spirit possessed by any voter, $\gamma_{min} := \min_{i \in [n]} \gamma_i$. We will also sometimes restrict our consideration to *uniform* PS-vectors $\boldsymbol{\gamma} = \gamma \mathbf{1}$, in which all voters have the same public spirit level $\gamma \in [0, 1]$.

**Utilities.** We define a *utility matrix* $U \in [0, 1]^{n \times m}$ to be a matrix whose $(i, a)$-th entry is $i$'s utility for alternative $a$, called $u_i(a)$. Let $\mathrm{sw}(a, U)$ denote the utilitarian *social welfare* of $a$ based on $U$, i.e.,

$$\mathrm{sw}(a, U) := \sum_{i \in [n]} u_i(a).$$

When $U$ is clear, we may denote the highest-welfare alternative in $U$ as $a^* := \arg\max_{a \in [m]} \mathrm{sw}(a, U)$.

**PS-values.** Together, a pair $\boldsymbol{\gamma}, U$ imply a *PS-values matrix* $V(\boldsymbol{\gamma}, U)$. This matrix contains the values for alternatives by which voters decide how to vote. A voter $i$'s *PS-value* for $a$ weighs their own utility $u_i(a)$ to a $(1 - \gamma_i)$ extent, and $a$'s social welfare $\mathrm{sw}(a, U)$ to a $\gamma_i$ extent:

$$v_i(a, \boldsymbol{\gamma}, U) = (1 - \gamma_i)u_i(a) + \gamma_i \mathrm{sw}(a, U)/n. \tag{1}$$

Note that $\mathrm{sw}(a, U)/n$ is interpreted as voters' *average utility* for $a$. Per this equation, the mathematical interpretation of a voter's public spirit level is the weight they place on the average utility versus their own in this convex combination.

**Rankings.** A *ranking* $\pi$ is a permutation of $[m]$. Voter $i$ expresses their preferences over alternatives as a strict, complete ranking $\pi_i$. We denote that $i$ ranks $a$ ahead of $b$ by $a >_{\pi_i} b$. We say that $\pi_i(j)$ is the alternative that voter $i$ ranks in the $j$-th position.

**Preference profiles.** A *preference profile* $\boldsymbol{\pi}$ is the $n$-tuple of all $n$ voters' rankings: $\boldsymbol{\pi} := (\pi_i : i \in [n])$. We let $\Pi$ be the set of all preference profiles. To compare how two alternatives' relative positions compare within a profile $\boldsymbol{\pi}$, we denote the number of voters in $\boldsymbol{\pi}$ who prefer $a$ to $b$ as $|\{i : a >_{\pi_i} b\}|$. A *pairwise election* between $a$ and $b$ in $\boldsymbol{\pi}$ compares $|\{i : a >_{\pi_i} b\}|$ and $|\{i : b >_{\pi_i} a\}|$; we say that *a pairwise-dominates $b$* if it wins this pairwise election, i.e., if $|\{i : a >_{\pi_i} b\}| > n/2$, and we add *weakly* if the inequality is weak. We say that $a$ is a *Condorcet winner* in $\boldsymbol{\pi}$ if $a$ pairwise-dominates all $b \neq a$ (noting that not all profiles have a Condorcet winner).

**Translating instances to preference profiles.** In any instance $(\boldsymbol{\gamma}, U)$, its associated PS-values matrix $V(\boldsymbol{\gamma}, U)$ naturally implies a preference profile in which alternatives are ordered in decreasing order of PS-value; formally, for any voter $i$,

$$v_i(a, \boldsymbol{\gamma}, U) > v_i(b, \boldsymbol{\gamma}, U) \implies a >_{\pi_i} b. \tag{2}$$

We do not specify the ranking implied when $v_i(a, \boldsymbol{\gamma}, U) = v_i(b, \boldsymbol{\gamma}, U)$; rather, we allow either ranking to be *consistent* with such PS-values, and thus, there can be *multiple* profiles consistent with the same $V(\boldsymbol{\gamma}, U)$. We let $\Pi_{V(\boldsymbol{\gamma}, U)}$ be the set of all profiles consistent with $V(\boldsymbol{\gamma}, U)$.

## 2.2 Voting Rules

A preference profile maps to a winning alternative via a (resolute) *voting rule* $f : \Pi \to [m]$. Then, $f(\boldsymbol{\pi}) = a$ means that on profile $\boldsymbol{\pi}$, rule $f$ chooses $a$ as the winner. We study two main classes of voting rules, *uncovered set rules* and *positional scoring rules*, defined below.[4] All of our examples will be strict, so we need not specify tie-breaking methods.

**Uncovered Set Rules.** The *uncovered set* of a given profile $\boldsymbol{\pi}$ is the set of all alternatives $a$ such that there is no $b$ that pairwise-dominates both $a$ and all alternatives pairwise-dominated by $a$. *Uncovered set rules* are all voting rules whose winner lies in the uncovered set, for all profiles. From this class, we primarily study the well-known rule COPELAND, where the score of an alternative is the number of alternatives it pairwise-dominates, and an alternative with maximum score is the COPELAND winner. We also study SLATER, which selects the ranking that is inconsistent with the outcomes of as few pairwise elections as possible.

**Positional Scoring Rules.** Positional scoring rules are defined by a score vector $\mathbf{s}$ of weakly decreasing scores $s_1 \geq \cdots \geq s_m$, where (without loss of generality) $s_1 = 1$ and $s_m = 0$. The winner by positional scoring rule $f_{\mathbf{s}}$ is the alternative that receives the most points, where $a$ receives $s_j$ points for every voter that ranks it $j$th. We will study three standard positional scoring rules, PLURALITY with score vector $\mathbf{s} = (1, 0, \ldots, 0)$, BORDA with score vector $\mathbf{s} = (1, 1 - 1/m-1, 1 - 2/m-1, \ldots, 1/m-1, 0)$, and VETO with score vector $\mathbf{s} = (1, \ldots, 1, 0)$. We will also define a new positional scoring rule PIECEWISE in Section 3, which will achieve better distortion than any of the previous three.

---

[4]The rules we study are standard, defined in, e.g., Conitzer [2006] (SLATER) and Xia and Conitzer [2010] (all others).

**Other rules and axioms.** We characterize one additional rule, MAXIMIN, which chooses the alternative with the lowest minimax score, defined for $a$ as the magnitude of $a$'s most severe pairwise domination, i.e., $\max_{\tilde{a} \neq a} |\{i : \tilde{a} \succ_{\pi_i} a\}|$. We also sometimes discuss the axiom *Condorcet consistency*, where $f$ is Condorcet consistent if it selects the Condorcet winner in all profiles in which one exists. Of the rules we study, COPELAND, SLATER, and MAXIMIN are Condorcet consistent.

## 2.3 Distortion of Voting Rules

The distortion of a voting rule $f$ in an instance $(\boldsymbol{\gamma}, U)$, called $\text{dist}(f, \boldsymbol{\gamma}, U)$, is the ratio between the respective welfares of the highest-welfare alternative $a^*$ and the winner $f(\boldsymbol{\pi})$. As is standard, we use *distortion*, called $\text{dist}(f, \boldsymbol{\gamma})$, to mean the *worst-case* such ratio over all $U$ (here, for a fixed $\boldsymbol{\gamma}$).

$$\text{dist}(f, \boldsymbol{\gamma}, U) := \sup_{\boldsymbol{\pi} \in \Pi_{V(\boldsymbol{\gamma}, U)}} \frac{\text{sw}(a^*, U)}{\text{sw}(f(\boldsymbol{\pi}), U)}, \quad \text{and} \quad \text{dist}(f, \boldsymbol{\gamma}) := \sup_{U \in \mathbb{R}_{\geq 0}^{n \times m}} \text{dist}(f, \boldsymbol{\gamma}, U).$$

## 3 DISTORTION BOUNDS FOR VOTING RULES

We now analyze the distortion of several voting rules under the condition that $\gamma_{min}$, the minimum level of public spirit among voters, is positive. First, in Section 3.1, we prove our key lemma, which founds our analysis of specific voting rules and gives intuition for why public spirit *should* limit the distortion. In Section 3.2, we will apply this lemma in various forms to upper-bound on the distortion of several standard voting rules. Section 3.3 contains our lower bounds for these rules, which match in almost all cases. We summarize these bounds in Table 1. Most include exact constants; the few asymptotic results we give are asymptotic in $m$, as is standard in the distortion literature.

| Rule | Upper bounds | | Lower bounds | |
|---|---|---|---|---|
| Uncovered set rules | $(2z_{\gamma_{min}} + 1)^2$ | (Thm. 3.3) | | |
|   COPELAND | $(2z_{\gamma_{min}} + 1)^2$ | | $(2z_\gamma + 1)^2$ | (Prop. 3.9) |
|   SLATER | $(2z_{\gamma_{min}} + 1)^2$ | | $(2z_\gamma + 1)^2$ | (Prop. 3.10) |
| Positional scoring rules | | | $\Omega(\sqrt{m})$ | (Thm. 3.11) |
|   PLURALITY | $mz_{\gamma_{min}} + 1$ | (Prop. 3.5) | $mz_\gamma + 1$ | (Prop. 3.15) |
|   BORDA | $mz_{\gamma_{min}} + 1$ | (Prop. 3.6) | $(m-1)z_\gamma + 1$ | (Prop. 3.13) |
|   VETO | | | infinite | (Prop. 3.14) |
|   PIECEWISE | $O(m^{2/3})$ | (Prop. 3.7) | $\Omega(\sqrt{m})$ | |
| MAXIMIN | $mz_{\gamma_{min}} + 1$ | (Prop. 3.8) | $(m-1)z_\gamma + 1$ | (Prop. 3.16) |

Table 1. Bounds on the distortion of voting rules. Upper bounds hold for all $\boldsymbol{\gamma}$; lower bounds hold for all uniform $\boldsymbol{\gamma} = \gamma \mathbf{1}$. As shorthand, we let $z_\gamma = {}^{(1-\gamma)}/_\gamma$. Gray-text results are inherited from more general results.

## 3.1 Key Lemma

LEMMA 3.1. *For all $U$, all alternatives $a, b$ with $\text{sw}(a, U) > 0$, all $\boldsymbol{\gamma}$ with $\gamma_{min} > 0$, and all $\boldsymbol{\pi} \in \Pi_{V(\boldsymbol{\gamma}, U)}$,*

$$\frac{\text{sw}(b, U)}{\text{sw}(a, U)} \leq \frac{1 - \gamma_{min}}{\gamma_{min}} \cdot \frac{n}{|\{i : a \succ_{\pi_i} b\}|} + 1.$$

Conceptually, Lemma 3.1 states that for arbitrary alternatives $a, b$, the more voters who rank $a$ ahead of $b$, the less the welfare of $b$ can exceed that of $a$ (assuming $\gamma_{min} > 0$). The intuition for the proof, below, is that any voter $i$ who ranks $a \succ_{\pi_i} b$ must have utility for $a$ that exceeds $b$ sufficiently to close the countervailing gap $\text{sw}(b, U) - \text{sw}(a, U)$, which is weighted by $\gamma_i$ in $i$'s PS-value. This fact implies a lower bound on $i$'s utility for $a$, which grows in $\gamma_i$; summing over all voters $i$, we get a lower bound on $\text{sw}(a, U)$ relative to $\text{sw}(b, U)$, which grows stronger in $\gamma_{min}$.

PROOF. Fix a $U$, $\boldsymbol{\gamma}$, and let $\boldsymbol{\pi} \in \Pi_{V(\boldsymbol{\gamma}, U)}$. Let $N_{a>b}$ be the set of voters in $\boldsymbol{\pi}$ who ranks $a$ ahead of $b$, and let $i \in N_{a>b}$. The fact that $a \succ_{\pi_i} b$ means that $v_i(a, \boldsymbol{\gamma}, U) \geq v_i(b, \boldsymbol{\gamma}, U)$, implying that

$$(1 - \gamma_i)u_i(a) + \gamma_i \frac{\text{sw}(a, U)}{n} = v_i(a, \boldsymbol{\gamma}, U) \geq v_i(b, \boldsymbol{\gamma}, U) = (1 - \gamma_i)u_i(b) + \gamma_i \frac{\text{sw}(b, U)}{n} \geq \gamma_i \frac{\text{sw}(b, U)}{n}.$$

Now, dividing both sides by $\gamma_i$ and then adding up both sides over all $i \in N_{a>b}$:

$$\sum_{i \in N_{a>b}} \left( \frac{1 - \gamma_i}{\gamma_i} u_i(a) + \frac{\text{sw}(a, U)}{n} \right) \geq \sum_{i \in N_{a>b}} \frac{\text{sw}(b, U)}{n}.$$

Using that $\frac{1-\gamma_i}{\gamma_i}$ decreasing in $\gamma_i$ and making simplifications,

$$\implies |N_{a>b}|/n \cdot \text{sw}(a, U) + \frac{1 - \gamma_{min}}{\gamma_{min}} \sum_{i \in N_{a>b}} u_i(a) \geq |N_{a>b}|/n \cdot \text{sw}(b, U).$$

Finally, we use that $\sum_{i \in N_{a>b}} u_i(a) \leq \text{sw}(a, U)$ to conclude the claim. □

In the next sections, we will apply this lemma to upper bound the distortion of various voting rules. Although we apply it in different ways across voting rules, the key idea is always the same: so long as enough voters rank the election winner $a'$ ahead of the highest-welfare alternative $a^*$ (or an alternative with social welfare comparable to $a^*$), then $\text{sw}(a^*, U)$ cannot exceed $\text{sw}(a', U)$ by more than a bounded amount, bounding the distortion. Intuitively, for "reasonable" voting rules, the number of voters who prefer $a'$ to some such alternative *should* be lower-bounded — otherwise, $a'$ would not be the winner. We will formalize this intuition as we prove our upper bounds.

## 3.2 Upper bounds

### 3.2.1 Uncovered Set Rules.
We will now show that, when $\gamma_{min}$ is separated from 0, all uncovered set rules — most notably including COPELAND and SLATER — have *constant* distortion. To prove this, we apply Lemma 3.1 in two different ways: in the first case, we use it to directly compare $a'$, the winner, and $a^*$. In the second and more interesting case, we apply the lemma twice, first to compare $a'$ with some intermediate alternative $a$, and then to compare $a$ with $a^*$. The choice of this intermediate alternative $a$ arises from a known[5] property of the uncovered set (Lemma 3.2):

LEMMA 3.2 (Moulin 1986). If $a'$ is in the uncovered set then for all $a \neq a'$, $a'$ either weakly pairwise-dominates $a$, or there exists some $a''$ such that $a'$ weakly pairwise-dominates $a''$ and $a''$ weakly pairwise-dominates $a$.

THEOREM 3.3. For all uncovered set rules $f$ and all $\boldsymbol{\gamma}$ with $\gamma_{min} > 0$,

$$\text{dist}(f, \boldsymbol{\gamma}) \leq \left( \frac{2(1 - \gamma_{min})}{\gamma_{min}} + 1 \right)^2.$$

---

[5]Our framing slightly adapts the classic result [Moulin, 1986] to permit pairwise ties. We remark that this result was also used to prove the constant distortion of uncovered set rules under metric preferences [Anshelevich et al., 2018, Thm. 5].

Proof. Let $f$ be an uncovered set rule, and fix arbitrary $U$, $\gamma$ and $\pi \in \Pi_{V(\gamma,U)}$. Let $a^*$ be the highest-welfare alternative in $U$, and let $a'$ be the winner by $f$, i.e., $a' = f(\pi)$. Then, we know $a'$ is in the uncovered set. If $a'$ weakly pairwise-dominates $a^*$, then $|\{i : a' \succ_{\pi_i} a^*\}|/n \geq 1/2$ and by applying Lemma 3.1 with $a' = a, a^* = b$, we immediately obtain an upper bound stronger than the claim. Else, by Lemma 3.2, there exists some $a$ such that $a'$ weakly pairwise-dominates $a$, and $a$ weakly pairwise-dominates $a^*$. Fix this $a$. Then, by Lemma 3.1, both $\mathrm{sw}(a^*)/\mathrm{sw}(a)$ and $\mathrm{sw}(a)/\mathrm{sw}(a')$ are at most $2(1-\gamma_{min})/\gamma_{min} + 1$. Multiplying these inequalities implies the claim. □

### 3.2.2 Positional scoring rules.
In giving upper bounds on the distortion of positional scoring rules, we will establish an upper bound on the distortion of *all* voting rules — one which will turn out to be tight for not only key positional scoring rules, but also some Condorcet consistent rules (e.g., Maximin, as analyzed in Section 3.2.3). This upper bound will be a corollary of Lemma 3.1, derived by using the lemma to compare the social welfares of $a'$ directly with $a^*$.

Formally, we deduce this corollary by plugging in $a = f(\pi)$ (for any $\pi \in \Pi_{V(\gamma,U)}$) and $b = a^*$. Then, for a given $f$, we need only to bound the quantity $|\{i : f(\pi) \succ_{\pi_i} a^*\}|/n$. We thus define the parameter $\kappa_f(m)$, the minimum fraction of voters who must rank the winner $f(\pi)$ ahead of *any* other given alternative, in *any* profile $\pi$.

$$\kappa_f(m) := \min_{\pi} \min_{a \neq f(\pi)} |\{i : f(\pi) \succ_{\pi_i} a\}|/n. \tag{3}$$

Although we express this quantity as a function of $m$, for brevity, we will often write it as $\kappa_f$. For a fixed $f$, we then have by definition that $|\{i : f(\pi) \succ_{\pi_i} a^*\}|/n \geq \kappa_f$ for all instances $(\gamma, U)$ and corresponding $\pi \in \Pi_{V(\gamma,U)}$, as needed. From this we conclude the following corollary of Lemma 3.1, which we emphasize is an upper bound on the distortion of *any* voting rule $f$:

Corollary 3.4 (Universal Upper Bound). For all rules $f$ and all $\gamma$ with $\gamma_{min} > 0$,

$$\mathrm{dist}(f, \gamma) \leq \frac{1 - \gamma_{min}}{\gamma_{min} \cdot \kappa_f} + 1.$$

To apply this corollary to upper bound the distortion of a specific $f$, we must simply lower bound $\kappa_f$. One useful observation, before doing so, is that for *all* $f$, $\kappa_f \leq 1/m$; thus, Corollary 3.4 can be used to prove linear distortion at best.[6]

Now, we prove upper bounds on the standard positional scoring rules Borda and Plurality by characterizing their respective $\kappa_f$ values and applying Corollary 3.4:

Proposition 3.5. $\kappa_{\text{Plurality}} = 1/m$, so for all $\gamma$ with $\gamma_{min} > 0$, $\mathrm{dist}(\text{Plurality}, \gamma) \leq m\frac{1-\gamma_{min}}{\gamma_{min}} + 1$.

Proposition 3.6. $\kappa_{\text{Borda}} = 1/m$, so for all $\gamma$ with $\gamma_{min} > 0$, $\mathrm{dist}(\text{Borda}, \gamma) \leq m\frac{1-\gamma_{min}}{\gamma_{min}} + 1$.

For Veto, we cannot apply the same approach, because $\kappa_{\text{Veto}}$ is $1/n$ — i.e., there exists an instance in which just one voter must rank the winner ahead of any other alternative — and thus the upper bound given by Corollary 3.4 is unbounded in $n$. It will turn out that, as Corollary 3.4 would suggest, the distortion of Veto is truly unbounded, shown via an instance in which the Veto-winner is almost never ranked ahead of the highest-welfare alternative.

---

[6]To see why $\kappa_f \leq 1/m$ for all $f$, divide $[n]$ into $m$ equal-sized groups $G_1, ..., G_m$. Then, for each group $G_k$, suppose the voters have rankings $k \succ k+1 \succ ... \succ m \succ 1 \succ ...k-1$. In this case, every alternative's worst pairwise defeat is to be ranked behind another alternative by an $(m-1)/m$ voters. Hence, $\kappa_f \leq 1/m$.

So far, we have not found a positional scoring rule that has sub-linear distortion, prompting the question: does one exist? We answer this question in the affirmative with Piecewise, a voting rule we newly define. It can be seen as a hybrid of Plurality and Borda, defined by a score vector with $m^{2/3}$ non-zero entries: $\mathbf{s} = (1, 1 - 1/m^{2/3}, 1 - 2/m^{2/3}, \ldots, 1/m^{2/3}, 0, \ldots, 0)$. We now show that, when $\gamma_{min}$ is any nonzero constant, Piecewise suffers at most $O(m^{2/3})$ distortion. Here, we depart from the approach of directly applying Corollary 3.4 (as we must in order to obtain a sub-linear bound).

Proposition 3.7. For all $\boldsymbol{\gamma}$ with (fixed) $\gamma_{min} > 0$, dist(Piecewise, $\boldsymbol{\gamma}$) $\in O(m^{2/3})$.

The proof of this proposition, found in Appendix A.4, again applies our key lemma, but in a more intricate fashion than in the preceding bounds. Similarly to the proof of Theorem 3.3, the argument considers one case comparing the Piecewise winner $a'$ directly to $a^*$, and another comparing $a'$ to some intermediate alternative(s) other than $a^*$. The first case is invoked in profiles where at least half of voters rank $a^*$ in the first $m^{2/3}$ positions; then, normalizing $a^*$'s social welfare to be constant, $a'$ must have social welfare $\Omega(m^{-2/3})$ in order to win the election. In the second case, over half the voters must rank $a^*$ in the *last* $m - m^{2/3}$ positions, implying that each of these voters must rank at least $m^{2/3}$ many alternatives ahead of $a^*$. In order for $a'$ to win the election over these other alternatives, $a'$ must again have social welfare $\Omega(m^{-2/3})$.

*3.2.3 Maximin.* Given that $\kappa_f$ is not meaningfully lower-bounded for Copeland and Slater (indeed, per the instance giving Proposition 3.9, it can be arbitrarily small), one might think that this is the case for all Condorcet consistent rules. On the contrary, here we show that $\kappa_{\text{Maximin}} = 1/m$, and thus Corollary 3.4 gives a useful distortion upper bound for Maximin — in fact, it will turn out that this upper bound is tight. The proof of this proposition is found in Appendix A.5.

Proposition 3.8. $\kappa_{\text{Maximin}} = 1/m$, so for all $\boldsymbol{\gamma}$ with $\gamma_{min} > 0$, dist(Maximin, $\boldsymbol{\gamma}$) $\leq m \frac{1 - \gamma_{min}}{\gamma_{min}} + 1$.

## 3.3 Lower bounds

We give matching lower bounds for all voting rules analyzed in Section 3.2 except Piecewise. However, a lower bound on Piecewise of $\Omega(\sqrt{m})$ (thus leaving an asymptotic gap of $m^{1/6}$) is implied by Theorem 3.11, which shows that even when voters are public-spirited, *all* positional scoring rules must suffer at least $\Omega(\sqrt{m})$ distortion. The proofs of all our lower bounds proceed by fixing an arbitrary uniform PS-vector $\boldsymbol{\gamma} = \gamma \mathbf{1}$, and then constructing a utility matrix $U$ whose entries depend on $\gamma$, in which the election winner $a'$ has far lower social welfare than $a^*$.

*3.3.1 Uncovered set rules.*

Proposition 3.9. For all uniform $\boldsymbol{\gamma} = \gamma \mathbf{1}$, $\gamma \in [0, 1]$, dist(Copeland, $\boldsymbol{\gamma}$) $\geq \left( \frac{2(1-\gamma)}{\gamma} + 1 \right)^2$.

Proposition 3.10. For all uniform $\boldsymbol{\gamma} = \gamma \mathbf{1}$, $\gamma \in [0, 1]$, dist(Slater, $\boldsymbol{\gamma}$) $\geq \left( \frac{2(1-\gamma)}{\gamma} + 1 \right)^2$.

The proofs of these propositions are found in Appendices A.6 and A.7, respectively. Both use the same instance, constructed so that $a'$ pairwise-dominates every alternative except $a^*$, and $a^*$ pairwise-dominates all but two alternatives, $a_1, a_2 \neq a'$. (We use two such alternatives here only to ensure that $a^*$ is *not* contained in the uncovered set, and thus the winner $a'$ is unique. Proving the bound requires reasoning about $a_1$ *or* $a_2$; here, we explain the bound via $a_1$.) Normalizing the average utility of $a^*$ to be 1, observe that because at least half of voters rank $a_1$ ahead of $a^*$, $a_1$ must have average utility at least $2(1-\gamma)/\gamma$. In turn, because at least half of voters rank $a'$ ahead of $a_1$, $a'$ must have average utility of at least $(2(1-\gamma)/\gamma)^2$. Then, the $U$ that minimizes $a'$'s social welfare relative to $a^*$ while also realizing the above profile makes all these inequalities tight, giving the lower bound.

*3.3.2 Positional scoring rules.* First, in Theorem 3.11, we show that whenever $\gamma_{min} < 1$, *all* positional scoring rules must have distortion at least $\Omega(\sqrt{m})$. Note that this result implies a fundamental separation between positional scoring rules and uncovered set rules, which per Theorem 3.3 have at most constant distortion for fixed values of $\gamma_{min} > 0$.

THEOREM 3.11. *For all positional scoring rules $f$ and uniform $\boldsymbol{\gamma} = \gamma\mathbf{1}$ with (fixed) $\gamma \in [0, 1)$,*

$$\text{dist}(f, \boldsymbol{\gamma}) \in \Omega(\sqrt{m}).$$

The key observation underlying this lower bound, proven formally in Appendix A.8, is that in any positional scoring rule's score vector, there exists some position $t$ amongst the first $\sqrt{m}$ entries in the score vector — that is, $t \in \{1, \dots, \sqrt{m}\}$ — such that the gap $s_t - s_{t+1}$ between the scores for positions $t$ and $t + 1$ is at most $1/\sqrt{m}$ (this is simply by averaging). Then, for fixed $\gamma$ and corresponding PS-vector $\boldsymbol{\gamma} = \gamma\mathbf{1}$, one can use this fact to construct an instance $(\boldsymbol{\gamma}, U)$ which realizes order-$\sqrt{m}$ distortion. The construction works as follows: Divide voters into two groups, a small group of size $O(1/\sqrt{m})$, and the remainder of the electorate. Let all voters in the larger group $a^*$ in the $t$-th position and the winner $a'$ in the $(t + 1)$-st position. In the small group, $a'$ is ranked first and $a^*$ is ranked last, thereby compensating for $a'$'s 'scoring' deficit in the larger group and allowing it to win the election. Because $a'$ is so rarely ranked ahead of $a^*$ in this profile, it can be realized by a utility matrix in which $a^*$ has constant average utility, while all voters have utility $O(1/\sqrt{m})$ for the winner $a'$, resulting in a distortion of order $O(\sqrt{m})$.

It turns out that many positional scoring rules have distortion far exceeding $\Omega(\sqrt{m})$ distortion; this is true, for instance, for all voting rules with a small value of $\Delta_f := s_1 - s_2$, the gap in scores of the first two ranking positions:

LEMMA 3.12. *For all positional scoring rules $f$ and uniform $\boldsymbol{\gamma} = \gamma\mathbf{1}$, $\gamma \in [0, 1]$, $\text{dist}(f, \boldsymbol{\gamma}) \geq \frac{1-\gamma}{\gamma\Delta_f} + 1$.*

In the proof of this proposition, found in Appendix A.9, we again construct an instance in which as few voters as possible rank the winner $a'$ ahead of $a^*$. To illustrate why smaller $\Delta_f$ permits fewer voters to rank $a'$ ahead of $a^*$, we will describe this construction. Divide voters into two groups: voters in the first group rank $a'$ first and $a^*$ last, and voters in the second group rank $a^*$ and $a'$ adjacently over the first two positions. In order for $a'$ to win this election, the first group must contain at least $\Delta_f$ voters; moreover, only these voters must have non-negligible utility for $a'$. Note that the use of the gap over the *first two positions* is essential: if we placed $a^* > a'$ over a smaller adjacent gap elsewhere, $a'$ would be ranked below several other alternatives by many voters, and we could no longer guarantee that it wins the election.

We can now directly apply Lemma 3.12 to lower bound the distortion of BORDA and VETO, using that $\Delta_{\text{BORDA}} = 1/(m-1)$ and $\Delta_{\text{VETO}} = 0$.

PROPOSITION 3.13. *For all uniform $\boldsymbol{\gamma} = \gamma\mathbf{1}$, $\gamma \in [0, 1]$, $\text{dist}(\text{BORDA}, \boldsymbol{\gamma}) \geq (m-1) \cdot \frac{1-\gamma}{\gamma} + 1$.*

PROPOSITION 3.14. *For all uniform $\boldsymbol{\gamma} = \gamma\mathbf{1}$, $\gamma \in [0, 1]$, $\text{dist}(\text{VETO}, \boldsymbol{\gamma}) = \infty$.*

For PLURALITY, $\Delta_{\text{PLURALITY}} = 1$, so Lemma 3.12 does not give a useful lower bound. However, we can get a tight lower bound using a similar construction: We let a $1/m + \epsilon$ fraction of voters rank $a'$ first, and all other voters rank $a'$ last. Only the former group of voters must have non-negligible utility for $a'$, while all other alternatives can receive non-negligible utility from the much larger second group of voters, yielding linear distortion. The full proof is found in Appendix A.10.

PROPOSITION 3.15. *For all uniform $\boldsymbol{\gamma} = \gamma\mathbf{1}$, $\gamma \in [0, 1]$, $\text{dist}(\text{PLURALITY}, \boldsymbol{\gamma}) \geq m \cdot \frac{1-\gamma}{\gamma} + 1$.*

*3.3.3 MAXIMIN.* Finally, we show that our upper bound on MAXIMIN's distortion was, indeed, tight.

PROPOSITION 3.16. For all uniform $\boldsymbol{\gamma} = \gamma \mathbf{1}$, dist(MAXIMIN, $\boldsymbol{\gamma}$) $\geq (m-1) \cdot \frac{1-\gamma}{\gamma} + 1$.

The formal proof of this proposition is found in Appendix A.11. The construction is somewhat involved, but it intuitively works as follows: voters are divided into two groups. In Group 1, containing a $1/(m-1)$ fraction of voters, the election winner $a'$ is ranked first; in Group 2, composed of the remaining voters, $a'$ is ranked last. The relative ranking of alternatives other than $a'$ is 'cyclical'—that is, all voters order them identically, up to a shift. There are $m-1$ possible such shifts, and each shifted ranking occupies a $1/(m-1)$ fraction of the voters. In this profile, $a'$'s greatest pairwise defeat is by $(m-2)/(m-1)$ fraction of voters, and the cyclical treatment of all other alternatives ensures that each suffers a pairwise defeat at least as severe as $a'$, making $a'$ the winner. This profile can be realized with a utility matrix in which all alternatives besides $a'$ get utility 1 from all voters in Group 2, while $a'$ only gets utility from Group 1.

# 4 PS-MONOTONICITY

Given that increasing voters' public spirit can only promote higher-welfare alternatives in their rankings, it seems natural that distortion should decrease as voters' public spirit increases. We refer to this general property of voting rules—decreasing distortion with increasing public spirit—as *public-spirit monotonicity* (for short, *PS-monotonicity*). Our upper (and matching lower) bounds from Section 3 already hint at a weak form of PS-monotonicity, as they are decreasing in $\gamma_{min}$.

In this section, we pursue stronger forms of PS-monotonicity, which ask for monotonicity not just in $\gamma_{min}$, but in voters' individual levels of public spirit. To this end, we define and analyze three notions of PS-monotonicity, from weakest to strongest. We first study *uniform PS-monotonicity*, which requires that distortion decreases as public spirit increases uniformly across voters. We find that this property holds for *all voting rules*—i.e., it is a fundamental property of the model. We next study a much stronger notion, *nonuniform PS-monotonicity*, which requires that the distortion decreases as voters' public spirit increases heterogeneously. We show that this notion holds for all voting rules when $m \leq 3$, and it holds for arbitrary $m$ for COPELAND and PLURALITY.

These first two notions examine monotonicity in the *worst-case* distortion. Even more optimistically, one might hope that public spirit would decrease the distortion on an *instance-wise* basis: i.e. in a fixed instance, if all voters' public spirit levels weakly increase, the welfare of the chosen outcome should only increase. We refer to this property as *instance-wise PS-monotonicity*. Unfortunately, we prove via classical voting axioms that no reasonable voting rule satisfies this notion: specifically, any weakly unanimous voting rule that satisfies instance-wise PS-monotonicity must be a dictatorship.

## 4.1 Uniform PS-monotonicity

*Definition 4.1 (Uniform PS-monotonicity).* A voting rule $f$ exhibits uniform PS-monotonicity if, for all $\gamma' \geq \gamma$ and associated uniform $\boldsymbol{\gamma} = \gamma \mathbf{1}, \boldsymbol{\gamma}' = \gamma' \mathbf{1}$, dist($f, \boldsymbol{\gamma}'$) $\leq$ dist($f, \boldsymbol{\gamma}$).

THEOREM 4.2. All voting rules are uniform PS-monotonic.

PROOF. We will prove this theorem by showing that, given arbitrary $U$ and $\gamma_{big} \geq \gamma_{small}$, we can find $\tilde{U}$ such that dist($f, \gamma_{big}, U$) = dist($f, \gamma_{small}, \tilde{U}$): roughly, under a lower level of public spirit, there exists a utility matrix with distortion at least as high. In fact, this distortion-preserving $\tilde{U}$ will simply be $U$ with some carefully-chosen amount of public spirit applied:

$$\tilde{U} := V(\gamma^*, U), \quad \text{where} \quad \gamma^* := \frac{\gamma_{\text{big}} - \gamma_{\text{small}}}{1 - \gamma_{\text{small}}}. \tag{4}$$

We begin by considering two $n \times m$ matrices: $U$ (which we can interpret as a matrix) and $W_U$, whose columns contain the column sums of $U$:

$$W_U = \begin{bmatrix} \text{sw}(a_1, U)/n & \dots & \text{sw}(a_m, U)/n \\ \vdots & & \vdots \\ \text{sw}(a_1, U)/n & \dots & \text{sw}(a_m, U)/n \end{bmatrix}.$$

We think of applying an arbitrary $\gamma$ to $U$ as a linear transformation on $U$, where varying $\gamma$ from 0 to 1 interpolates between the matrices $U$ and $W_U$: applying $\gamma = 0$ returns $U$, applying $\gamma = 1$ returns $W_U$, and there is an infinite sequence of matrices in between ranging over $\gamma \in [0, 1]$, where the $\gamma$-th matrix is equal to a convex combination of $U$ and $W_U$ — that is, $V(\gamma, U) = (1 - \gamma)U + \gamma W_U$.

A key property of this transformation is that it is *column-sum-preserving*, so all matrices in this sequence have the same column sums; that is, for all $\gamma \in [0, 1]$, $W_{V(\gamma, U)} = W_U$. We use this fact to make the general observation that applying public spirit $\gamma_1$ and then $\gamma_2$ in succession is the same as applying $\gamma_1 + \gamma_2 - \gamma_1 \gamma_2$ public spirit all at once:

LEMMA 4.3. *For arbitrary $U$ and arbitrary $\gamma_1, \gamma_2 \in [0, 1]$, $V(\gamma_2, V(\gamma_1, U)) = V(\gamma_1 + \gamma_2 - \gamma_1 \gamma_2, U)$.*

*Proof of Lemma 4.3:*

$$\begin{aligned} V(\gamma_2, V(\gamma_1, U)) &= \gamma_2 W_{V(\gamma_1, U)} + (1 - \gamma_2)V(\gamma_1, U) \\ &= \gamma_2 W_{V(\gamma_1, U)} + (1 - \gamma_2)\big((1 - \gamma_1)U + \gamma_1 W_U\big) \\ &= \gamma_2 W_U + (1 - \gamma_2)((1 - \gamma_1)U + \gamma_1 W_U) \\ &= (1 - \gamma_1)(1 - \gamma_2)U + \big(\gamma_1 + \gamma_2 - \gamma_1 \gamma_2\big)W_U \\ &= V(\gamma_1 + \gamma_2 - \gamma_1 \gamma_2, U) \qquad\qquad\qquad \square \end{aligned}$$

Because applying public spirit is column-sum-preserving, we can set $\tilde{U}$ to *any* matrix $V(\gamma, U)$, $\gamma \in [0, 1]$ and be certain that $\tilde{U}$ will give the same welfares to all alternatives as $U$. We will carefully choose this $\gamma = \gamma^*$ according to Lemma 4.3: $\gamma^* = \gamma_1$, $\gamma_{\text{small}} = \gamma_2$, and $\gamma_{\text{big}} = \gamma_1 + \gamma_2 - \gamma_1 \gamma_2$, which means setting $\gamma^*$ as in Equation (4). This setting of $\gamma^*$ then ensures that the rankings are preserved:

$$V(\gamma_{\text{small}}, V(\gamma^*, U)) = V(\gamma_{\text{big}}, U) \implies \Pi_{V(\gamma_{\text{small}}, \tilde{U})} = \Pi_{V(\gamma_{\text{big}}, U)}.$$

Thus, across $(U, \gamma_{\text{big}})$ and $(\tilde{U}, \gamma_{\text{small}})$, the rankings (and therefore the winner) and social welfares are identical. The distortion must then be the same across the instances, proving the claim.     $\square$

## 4.2 Nonuniform PS-monotonicity

Here, we define the ordering of vectors in the standard way: $\gamma' \geq \gamma$ iff $\gamma_i' \geq \gamma_i$ for all $i \in [n]$.

*Definition 4.4 (Nonuniform PS-monotonicity).* A voting rule $f$ exhibits nonuniform PS-monotonicity if for all $\gamma, \gamma'$ where $\gamma' \geq \gamma$, $\text{dist}(f, \gamma') \leq \text{dist}(f, \gamma)$.

First, we show that nonuniform PS-monotonicity holds for *all* voting rules when $m \leq 3$.

PROPOSITION 4.5. *If $m \leq 3$, then all voting rules exhibit nonuniform monotonicity.*

We defer the proof of this proposition to Appendix B.1, as it is fairly involved. The main intuition behind the proof is as follows: If $U$ is a utility matrix, $\boldsymbol{\gamma}$ is some PS-vector and $\tilde{\boldsymbol{\gamma}}$ arises from *lowering* an entry in $\boldsymbol{\gamma}$, then we can explicitly construct another utility matrix $\tilde{U}$ such that the profile(s) implied by $(\boldsymbol{\gamma}, U)$ and $(\tilde{\boldsymbol{\gamma}}, \tilde{U})$ are identical (i.e., $\Pi_{V(\boldsymbol{\gamma}, U)} = \Pi_{V(\tilde{\boldsymbol{\gamma}}, \tilde{U})}$), and the social welfares of all alternatives are preserved (i.e., $\mathrm{sw}(a, U) = \mathrm{sw}(a, \tilde{U})$ for all $a$). Across these instances, the election winner, and thus the distortion, must be the same.

The construction used to show Proposition 4.5 is already considerably complicated when $m = 3$; proving the claim for all (or a broad class of) voting rules when $m \geq 4$ remains an interesting open problem. However, we do affirmatively resolve this question for two specific voting rules, showing that COPELAND and PLURALITY both satisfy nonuniform PS-monotonicity for arbitrary $m$.

PROPOSITION 4.6. COPELAND *is nonuniform PS-monotonic.*

PROPOSITION 4.7. PLURALITY *is nonuniform PS-monotonic.*

These propositions are proven in Appendices B.2 and B.3, respectively. Although the constructions used to analyze COPELAND and PLURALITY are different, both reflect the argument from Proposition 4.5: given $U, \boldsymbol{\gamma}, \tilde{\boldsymbol{\gamma}}$ where $\tilde{\boldsymbol{\gamma}} \leq \boldsymbol{\gamma}$, we construct a $\tilde{U}$ such that the election winner and welfares are preserved instances. Note that these arguments can be simpler than the proof of Proposition 4.5 because, given that we are not reasoning about *all* voting rules, preserving these features across instances does not necessitate preserving the full preference profile. As such, in the analysis of COPELAND, $\tilde{U}$ just preserves the relevant aspects of the uncovered set; in the analysis of PLURALITY, $\tilde{U}$ just preserves the first-ranked alternatives.

## 4.3 Instance-wise PS-Monotonicity

*Definition 4.8 (Instance-wise PS-monotonicity).* A voting rule $f$ is *instance-wise PS-monotonic* iff, for all $U$ and all $\boldsymbol{\gamma}, \boldsymbol{\gamma}'$ where $\boldsymbol{\gamma}' \geq \boldsymbol{\gamma}$, $\mathrm{dist}(f, \boldsymbol{\gamma}', U) \leq \mathrm{dist}(f, \boldsymbol{\gamma}, U)$.

Unfortunately, Theorem 4.11 shows that no reasonable — i.e., *weakly unanimous* (Definition 4.9) and *non-dictatorial* (Definition 4.10) — voting rule satisfies this property. Although the proof is involved, the intuition is simple: consider three alternatives in order of decreasing welfare, $a, b, c$. Suppose $a$ wins initially, but after increasing voters' public spirit, all voters promote $b$ over $c$ but no other relative rankings change. For any monotonic and otherwise reasonable voting rule, $b$ — whose welfare is lower than $a$'s — must in some cases be able to become the winner.

*Definition 4.9 (weakly unanimous).* A voting rule $f$ is *weakly unanimous* iff for every profile $\boldsymbol{\pi}$, if there is a pair of alternatives $a, b$ such that $a \succ_{\pi_i} b$ for all voters $i$, then $f(\boldsymbol{\pi}) \neq a$.

*Definition 4.10 (dictatorship).* Voter $i$ is a *dictator* with respect to $f$ if $f$ always selects $i$'s top choice: for every profile $\boldsymbol{\pi}$, $f(\boldsymbol{\pi}) = a$ iff for all $a' \neq a$, $a \succ_{\pi_i} a'$. $f$ is a *dictatorship* if it has a dictator.

THEOREM 4.11. *If $m \geq 3$ and $f$ is weakly unanimous and instance-wise monotonic, $f$ is a dictatorship.*

We prove Theorem 4.11 at the end of this subsection by showing that instance-specific PS-montonicity implies an increasingly strong series of voting axioms. We build up this system of axiomatic implications until they meet the preconditions of a known result by Muller and Satterthwaite [1977] implying that $f$ is a dictatorship. Below, we step through each of these axiomatic implications, defining the relevant axioms as we go.

First, Lemmas 4.13 and 4.15 (proven in Appendix B.4 and Appendix B.5) show that for all weakly unanimous $f$, instance-wise PS-monotonicity implies *monotonicity* (Definition 4.12), the standard voting axiom, and *swap invariance* (Definition 4.14), which we newly define.

*Definition 4.12 (monotonic).* A voting rule rule $f$ is *monotonic* iff, for every profile $\boldsymbol{\pi}$ such that $f(\boldsymbol{\pi}) = a$, and for every $i \in [n]$, if $\boldsymbol{\pi}'$ is identical to $\boldsymbol{\pi}$ except that in ranking $\pi_i'$, $a$ is promoted (with one adjacent swap) compared in $\pi_i$, then $f(\boldsymbol{\pi}') = a$.

LEMMA 4.13. *If $f$ is weakly unanimous and instance-wise PS-monotonic, then it is monotonic.*

*Definition 4.14 (swap invariant).* A voting rule $f$ satisfies *swap invariance* iff, for every profile $\boldsymbol{\pi}$ such that $f(\boldsymbol{\pi}) = a$, every $i \in [n]$, and every pair of alternatives $b, c \in [m]$ where $b, c \neq a$, if $\boldsymbol{\pi}'$ is identical to $\boldsymbol{\pi}$ except $b$ and $c$ are adjacently swapped in $\pi_i'$, then $f(\boldsymbol{\pi}') = a$.

LEMMA 4.15. *If $f$ weakly unanimous and monotonic, then if $f$ is instance-wise PS-monotonic, it must also be swap-invariant.*

Next, Lemma 4.17 (proven in Appendix B.6) shows that together, monotonicity and swap invariance imply a stronger notion of monotonicity known as *Maskin monotonicity* (Definition 4.16).

*Definition 4.16 (Maskin-monotonic).* A voting rule $f$ is Maskin-monotonic iff, for every preference profile $\boldsymbol{\pi}$ such that $f(\boldsymbol{\pi}) = a$, if $\boldsymbol{\pi}'$ is another profile such that $a >_{\pi_i'} b$ whenever $a >_{\pi_i} b$ for every voter $i$ and every alternative $b$, then $f(\boldsymbol{\pi}') = a$.

LEMMA 4.17. *If $f$ is monotonic and swap-invariant, then it is Maskin-monotonic.*

Finally, we apply Theorem 4.18, a known result by Muller and Satterthwaite, which shows that any voting rules that is weakly unanimous and Maskin-monotonic must also be a dictatorship.

THEOREM 4.18 ([Muller and Satterthwaite, 1977]). *When $m \geq 3$, if $f$ is weakly unanimous and Maskin-monotonic, it is also dictatorial.*

We prove Theorem 4.11 by applying these lemmas in sequence.

PROOF OF THEOREM 4.11.

$f$ is *weakly unanimous* and *instance-wise PS-monotonic* $\implies f$ is *monotonic*     (Lemma 4.13)

$f$ is *weakly unanimous*, *monotonic*, and *instance-wise PS-monotonic*

$$\implies f \text{ is } \textit{swap-invariant} \quad \text{(Lemma 4.15)}$$

$f$ is *monotonic* and *swap-invariant*      $\implies f$ is *Maskin-monotonic*     (Lemma 4.17)

$f$ is *weakly unanimous* and *instance-wise PS-monotonic*

$$\implies f \text{ is } \textit{weakly unanimous} \text{ and } \textit{Maskin-monotonic}$$

$$\implies f \text{ is a dictatorship.} \quad \text{(Theorem 4.18)} \quad \square$$

## 5 ROBUSTNESS OF DISTORTION BOUNDS

So far, we have considered the distortion of voting rules under two ideal conditions, which we will now relax: (a) $\gamma_{min}$, the minimum public spirit level, is bounded away from zero, and (b) voters act according to precise and internally-consistent values of the model inputs $u_i(a)$, $\gamma_i$, and $\text{sw}(a, U)$.

We will show that the distortion is asymptotically maintained — and degrades smoothly by constant factors — as we relax these conditions (up to an extent, for (a)).

When proving robustness to violation of (a), we essentially work within our model; to study deviations from (b), we meaningfully generalize our model to encompass a variety of errors. Our arguments for both types of robustness follow the same structure, paralleling the main upper bound results from Sections 3.1 and 3.2 in the robust setting. In particular, for both (a) and (b), we first prove a "robust" version of Lemma 3.1, and then deduce corresponding "robust" distortion upper bounds via the same arguments used to deduce our original upper bounds from Lemma 3.1.

## 5.1 Robustness to a Non-Public-Spirited Contingent

Here, we show that our upper bounds from Section 3 continue to hold up to constants as long as the number of non-public-spirited voters $i$, i.e. with $\gamma_i = 0$, is not too large. We begin by proving a "robust" version of Lemma 3.1, with respect to this form of robustness:

LEMMA 5.1. Let $U$ be any utility matrix, and let $\boldsymbol{\gamma}$ be such that $\gamma_{min} > 0$. Then, for any $c < 1$, any alternatives $b, a$ with $\mathrm{sw}(a, U) > 0$ and any $\tilde{\boldsymbol{\gamma}}$ which arises from setting the public spirit of at most any $c \cdot |\{a \succ_{\pi_i} b\}|$ voters in $\boldsymbol{\gamma}$ to zero, it holds that

$$\frac{\mathrm{sw}(b, U)}{\mathrm{sw}(a, U)} \leq \frac{1 - \gamma_{min}}{\gamma_{min}} \cdot \frac{n}{|\{i : a \succ_{\pi_i} b\}|(1 - c)} + 1.$$

PROOF. Let us denote the set of voters who both have at least $\gamma_{min}$ public spirit and rank $a$ ahead of $b$ by $\tilde{N}_{a>b} := |\{i : a \succ_{\pi_i} b \text{ and } \gamma_i \geq \gamma_{min}\}|$. Then, we can follow the same arguments as in the proof of Lemma 3.1 with $\tilde{N}_{a>b}$ in place of $N_{a>b}$ to obtain the inequality

$$\frac{|\tilde{N}_{a>b}|}{n}\mathrm{sw}(a, U) + \frac{1 - \gamma_{min}}{\gamma_{min}}\mathrm{sw}(a, U) \geq \frac{|\tilde{N}_{a>b}|}{n}\mathrm{sw}(b, U).$$

Dividing both sides by $\mathrm{sw}(a, U) \cdot |\tilde{N}_{a>b}|/n$ yields

$$\frac{\mathrm{sw}(b, U)}{\mathrm{sw}(a, U)} \leq \frac{1 - \gamma_{min}}{\gamma_{min}} \frac{n}{|\tilde{N}_{a>b}|} + 1.$$

By assumption, $|\tilde{N}_{a>b}| \geq (1 - c)|\{i : a \succ_{\pi_i} b\}|$, and the claim follows. □

Then, since $\kappa_f$ lower bounds the fraction of agents who must rank ahead the winner (which we think of as $a$ in the lemma above) ahead of the maximum welfare alternative (which we think of as $b$), Lemma 5.1 immediately implies the following corollary, as Lemma 3.1 implied Corollary 3.4.

COROLLARY 5.2. Let $f$ be any voting rule, and let $\boldsymbol{\gamma}$ with $\gamma_{min} > 0$. Then, for any $c < 1$ and any $\tilde{\boldsymbol{\gamma}}$ created by setting the public spirit of at most $c\kappa_f \cdot n$ many voters in $\boldsymbol{\gamma}$ to zero,

$$\mathrm{dist}(f, \tilde{\boldsymbol{\gamma}}) \leq \frac{1 - \gamma_{min}}{\gamma_{min}(1 - c)\kappa_f} + 1.$$

Similarly, for Uncovered Set Rules, Lemma 5.1 implies the following corollary (analogously to Lemma 3.1 implying Theorem 3.3).

COROLLARY 5.3. Let $f$ be an uncovered set rule, and let $\boldsymbol{\gamma}$ with $\gamma_{min} > 0$. Then, for any $c < 1/2$ and for any $\tilde{\boldsymbol{\gamma}}$ created by setting the public spirit of a $c$-fraction of voters in $\boldsymbol{\gamma}$ to zero,

$$\mathrm{dist}(f, \tilde{\boldsymbol{\gamma}}) \leq \left( \frac{1 - \gamma_{min}}{\gamma_{min}(1/2 - c)} + 1 \right)^2.$$

## 5.2 Robustness to inaccurate or internally-inconsistent voter behavior

Our model assumes that voters know (or can come to know) their own utilities and the respective welfares of all alternatives, to which they then uniformly apply some level of public spirit. However, voters almost certainly do not maintain precise internal values of $\gamma_i$ and $u_i(a)$, sw$(a, U)$ for all $a$, and then vote by tabulating their PS-values. In fact, it is dubious whether a voter, if asked, could even assign useful numeric values to these quantities. As a result, the best we can probably hope for in practice is that voters have some internal *sense* of these quantities, which may be subject to errors, biases, and internal inconsistencies.

This motivates our extension of our upper bounds to the case where voters may deviate from our model with respect to *any input* to Equation (1). First, we allow voters to misestimate their utilities, and likewise the social welfares, by some bounded multiplicative error. Since we can always rescale utilities by a multiplicative factor without changing the voting outcome or the distortion, we can without loss of generality only consider the case where voters *overestimate* these quantities. Second, voters may apply different levels of public spirit to different alternatives. These are not "errors", per se, because voters' levels of public spirit do not factor into the utilitiarian social welfare (our benchmark) and thus do not necessarily have a ground-truth value. These deviations can rather be interpreted as internal inconsistencies — or even natural behaviors — where voters are partial to the nature of certain alternatives' social benefit over that of others.

To formalize these errors, we assume voter $i$ applies multiplicative errors $\delta_i(a) \geq 1$ to their utility for $a$ and $\eta_i(a) \geq 1$ to the social welfare of $a$. We define $\delta^* = \max_{i \in [n], a \in [m]} \delta_i(a)$ and $\eta^* = \max_{i \in [n], a \in [m]} \eta_i(a)$ as the maximum such errors across all voters and alternatives, and we let $\boldsymbol{\delta} \in [1, \delta^*]^{n \times m}$, $\boldsymbol{\eta} \in [1, \eta^*]^{n \times m}$ be the matrices of these errors across voters and alternatives. We also assume $i$ applies public spirit level $\gamma_i(a)$ to each alternative $a$, and we let the *PS-matrix* $\Gamma \in [0, 1]^{n \times m}$ be the matrix of these $\gamma$ over all voters and alternatives. We now let $\gamma_{min} = \min_{i \in [n], a \in [m]} \gamma_i(a)$ be the minimum level of public spirit in $\Gamma$.

Incorporating these deviations, voter $i$'s *effective* PS-value is then

$$\tilde{v}_i(a, \Gamma, U, \boldsymbol{\delta}, \boldsymbol{\eta}) := (1 - \gamma_i(a)) \cdot \delta_i(a) u_i(a) + \gamma_i(a) \cdot \eta_i(a) \text{sw}(a, U)/n.$$

Correspondingly, we let $\tilde{V}(\Gamma, U, \boldsymbol{\delta}, \boldsymbol{\eta})$ be the matrix of all voters' effective PS-values. Finally, we define distortion under such errors, bounded above by $\delta^*, \eta^*$ respectively, as

$$\text{dist}^{\delta^*, \eta^*}(f, \Gamma) := \sup_{U \in \mathbb{R}_{\geq 0}^{n \times m}, \ \boldsymbol{\delta} \in [1, \delta^*]^{n \times m}, \ \boldsymbol{\eta} \in [1, \eta^*]^{n \times m}} \ \sup_{\boldsymbol{\pi} \in \Pi_{\tilde{V}(\Gamma, U, \boldsymbol{\delta}, \boldsymbol{\eta})}} \frac{\text{sw}(a^*, U)}{\text{sw}(f(\boldsymbol{\pi}), U)}$$

*A priori*, it seems that the distortion of a voting rule might not be at all robust to such errors, because even a minimal deviation could cause a pivotal switch in two alternatives, changing the winner and causing a jump in distortion. Surprisingly, however, we find that we can give distortion upper bounds on any voting rule that increase smoothly in $\delta^*$ and incur merely an additive term of $\eta^*/\gamma_{min}$. At a high level, this holds because Lemma 3.1 must still upper-bound the ratio of the *estimated* social welfares of the winner and $a^*$, which in turn bounds the ratio of the *true* welfares, given that the estimates are not too far off. We formalize this intuition below in a generalized "robust" version of Lemma 3.1 that incorporates these errors.

LEMMA 5.4. Fix utility matrix $U$, $\delta^*, \eta^*$, errors $\boldsymbol{\delta} \in [1, \delta^*]^{n \times m}$ and $\boldsymbol{\eta} \in [1, \eta^*]^{n \times m}$, and a PS-matrix $\Gamma$ with $\gamma_{min} > 0$. Then, for any alternatives $a, b$ with sw$(a, U) > 0$ and any $\boldsymbol{\pi} \in \Pi_{\tilde{V}(\Gamma, U, \boldsymbol{\delta}, \boldsymbol{\eta})}$,

$$\frac{\text{sw}(b, U)}{\text{sw}(a, U)} \leq \frac{\delta^* \cdot (1 - \gamma_{min})}{\gamma_{min}} \cdot \frac{n}{|\{i : a \succ_{\pi_i} b\}|} + \frac{\eta^*}{\gamma_{min}}$$

Proof. We take the same approach as in the proof of Lemma 3.1, this time accounting for all deviations. For any voter $i$ ranking $a > b$, and thus having $\tilde{v}_i(a, \Gamma, U, \boldsymbol{\delta}, \boldsymbol{\eta}) \geq \tilde{v}_i(b, \Gamma, U, \boldsymbol{\delta}, \boldsymbol{\eta})$,

$$
(1 - \gamma_i(a)) \cdot \delta^* u_i(a) + \gamma_i(a) \frac{\eta^* \operatorname{sw}(a, U)}{n} \geq (1 - \gamma_i(a)) \cdot \delta_i(a) u_i(a) + \gamma_i(a) \cdot \eta_i(a) \frac{\operatorname{sw}(a, U)}{n}
$$

$$
\geq (1 - \gamma_i(b)) \cdot \delta_i(b) u_i(b) + \gamma_i(b) \cdot \eta_i(b) \frac{\operatorname{sw}(b, U)}{n}
$$

$$
\geq \gamma_i(b) \frac{\operatorname{sw}(b, U)}{n}.
$$

Then, following through the same rearrangements as in the proof Lemma 3.1 and summing over $N_{a>b}$ (shorthand for $\{i : a \succ_{\pi_i} b\}$), we conclude the proof:

$$
\eta^* \cdot \operatorname{sw}(a, U) \frac{|N_{a>b}|}{n} \frac{\gamma_i(a)}{\gamma_i(b)} + \delta^* \cdot \frac{1 - \gamma_i(a)}{\gamma_i(b)} \operatorname{sw}(a, U) \geq \frac{|N_{a>b}|}{n} \operatorname{sw}(b, U)
$$

$$
\implies \frac{\operatorname{sw}(b, U)}{\operatorname{sw}(a, U)} \leq \delta^* \cdot \frac{1 - \gamma_i(a)}{\gamma_i(b)} \frac{n}{|N_{a>b}|} + \eta^* \cdot \frac{\gamma_i(a)}{\gamma_i(b)} \leq \frac{\delta^* (1 - \gamma_{min})}{\gamma_{min}} \frac{n}{|N_{a>b}|} + \frac{\eta^*}{\gamma_{min}}. \qquad \square
$$

Now, we conclude the robust versions of our original distortion upper bounds. Lemma 5.4 implies Corollary 5.5 just as Lemma 3.1 implied Corollary 3.4. Similarly, Lemma 5.4 implies Corollary 5.6 just as Lemma 3.1 implied Theorem 3.3.

Corollary 5.5. For all voting rules $f$, all $\delta^*, \eta^* \geq 1$ and PS-matrices $\Gamma \in [0, 1]^{n \times m}$ with $\gamma_{min} > 0$,

$$
\operatorname{dist}^{\delta^*, \eta^*}(f, \Gamma) \leq \frac{\delta^* (1 - \gamma_{min})}{\gamma_{min} \kappa_f} + \frac{\eta^*}{\gamma_{min}}.
$$

Corollary 5.6. For all uncovered set rules $f$, all $\delta^*, \eta^* \geq 1$ and PS-matrices $\Gamma \in [0, 1]^{n \times m}$ with $\gamma_{min} > 0$,

$$
\operatorname{dist}^{\delta^*, \eta^*}(f, \Gamma) \leq \left( \frac{2\delta^* (1 - \gamma_{min})}{\gamma_{min}} + \frac{\eta^*}{\gamma_{min}} \right)^2.
$$

**A remark about tightness.** Most of the upper bounds derived from Lemma 3.1 were tight for constant PS-vectors $\boldsymbol{\gamma} = \gamma \mathbf{1}$ (Section 3.3). Thus, one may wonder whether the upper bounds in this subsection are likewise tight for constant PS-matrices. This question merits formal theoretical treatment, because one must construct a separate lower bound for each voting rule, as in Section 3.3. However it does seem that tightness should hold via the following simple construction: let $a'$ be the election winner. Then, construct a profile in which, for all voters $i$, $\delta_i(a') = \delta^*$ and $\eta_i(a') = \eta^*$ and $\delta_i(a) = \eta_i(a) = 1$ for all other $a \neq a'$. Intuitively, this construction allows $a'$ to win the election with the smallest true utility possible, and should yield lower bounds corresponding to those in Section 3.3 as follows: if a lower bound on the standard model is, for some functions $h, g$, of the form $h(g(m) \cdot (1 - \gamma)/\gamma + 1)$, then it should be $h(g(m) \cdot \delta^*(1 - \gamma)/\gamma + \eta^*)$ in the generalized model.

# 6 DISCUSSION

A key contribution of our work is to establish *cultivating voters' public spirit* as a new approach to increasing the welfare of democratic decision-making — an approach which can be operationalized via publicly-palatable interventions like deliberation. In the introduction, we discussed why increasing the welfare of voting outcomes is a pressing goal; however, regardless of how pressing one believes this goal to be, our results suggest that in many senses, interventions that promote public spirited voting can only help.

Of course, these results arise from a theoretical model, so their practical implications depend on how our model may capture — or fail to capture — reality. On this note, our robustness results in Section 5.2 cover a wide range of plausible behavioral deviations: they allow voters to, e.g., assess their utilities on different scales, overestimate their own utilities compared to others', have biases that cause them to systematically under-weight the interests of certain groups, apply different levels of public spirit to different alternatives, or even more coarsely, just maintain a ranking over alternatives rather than any sense of these quantities (this corresponds to arbitrary errors in utilities and social welfares). Our results in Section 5.1 also allow for participants who exhibit *no* public spirit; however, a key issue we sidestep is the case where some participants not only lack public spirit, but are actually adversarial to the process. We address this in future work below.

## 6.1  Future work

In addition to the theoretical directions identified below, we remark that our work motivates further experiments studying how voters' public spirit changes over the course of deliberation. In turn, with a more detailed understanding of the structure of voters' deviations from our model, one can get more fine-grained robustness bounds than we achieve in Section 5.2.

**Strategic voters among public spirited voters.** As always, in our setting there is the potential for manipulation — perhaps more so here because some voters are prioritizing the collective rather than acting in rational self-interest. The possibility of some voters being strategic opens several questions, such as: 'Does public spirit among most voters make the voting process more or less robust to a few manipulators?' and 'Given that the presence of strategic voters might pose a risk to others, how might voters who would otherwise intend to be public-spirited respond?'

**Sufficient conditions for (approximate) instance-wise monotonicity.** While in many respects, our results suggest that increased public spirit is beneficial, Theorem 4.11 shows an extremely fundamental impossibility: that in general, public spirit may not help on an instance-by-instance basis. This begs the question: can we establish sufficient conditions on instances — ideally which are roughly detectable in practice — under which we *can* be certain that increasing the public spirit will improve outcomes? Moreover, even if we cannot hope for *exact* monotonicity, can we show approximate notions, e.g., in which the social welfare increases up to bounded fluctuations?

**Extensions to other notions of social welfare.** In this paper, we assume that public-spirited voters determine how positively an alternative impacts society according to its *utilitarian* social welfare. However, voters might just as easily take an *egalitarian* perspective, thus quantifying an alternative's social welfare by how it affects the person it benefits the *least*. Even further, there is no guarantee that public spirited voters apply the *same* priorities when assessing the social welfare. These points open questions such as, if voters are public-spirited *but quantify the social good via different objectives*, does public spirit still increase the welfare of the outcome?

**Other collective decision mechanisms.** Our results identify public-spirited voting behavior within democratic contexts as a powerful, practically-motivated beyond-worst-case assumption. We have demonstrated this specifically for deterministic voting mechanisms where voters express preferences as complete rankings. However, there are many other well-studied collective decision mechanisms — e.g., randomized voting rules, approval voting, multi-winner elections, liquid democracy, participatory budgeting — that could potentially benefit from public spirit, too. To initiate the study of public spirit in other mechanisms, we remark that all the aforementioned mechanisms can be analyzed in the same utilitarian social welfare framework: one needs only to specify a model of how voters translate their underlying utilities into ballot responses — analogous to our Equations (1) and (2) — that allows voters to weigh their own interests against the common good.

# REFERENCES

E. Anshelevich, O. Bhardwaj, E. Elkind, J. Postl, and P. Skowron. 2018. Approximating Optimal Social Choice under Metric Preferences. *Artificial Intelligence* 264 (2018), 27–51.

E. Anshelevich, A. Filos-Ratsikas, N. Shah, and A. A. Voudouris. 2021. Distortion in Social Choice Problems: The First 15 Years and Beyond. arXiv:2103.00911.

M. M. Bechtel and R. Liesch. 2020. Reforms and redistribution: Disentangling the egoistic and sociotropic origins of voter preferences. *Public Opinion Quarterly* 84, 1 (2020), 1–23.

G. S. Becker. 1976. Altruism, egoism, and genetic fitness: Economics and sociobiology. *Journal of economic Literature* 14, 3 (1976), 817–826.

I. Caragiannis and A. D. Procaccia. 2011. Voting Almost Maximizes Social Welfare Despite Limited Communication. *Artificial Intelligence* 175, 9–10 (2011), 1655–1671.

P.-A. Chen, B. de Keijzer, D. Kempe, and G. Schäfer. 2014. Altruism and Its Impact on the Price of Anarchy. *ACM Transactions on Economics and Computation* 2, 4 (2014), article 17.

V. Conitzer. 2006. Computing Slater Rankings Using Similarities among Candidates. In *Proceedings of the 21st AAAI Conference on Artificial Intelligence (AAAI)*. 613–619.

Ernst Fehr and Klaus M Schmidt. 1999. A theory of fairness, competition, and cooperation. *The quarterly journal of economics* 114, 3 (1999), 817–868.

J. Gastil, C. Bacci, and M. Dollinger. 2010. Is Deliberation Neutral? Patterns of Attitude Change During the Deliberative Polls. *Journal of public deliberation* 6, 2 (2010).

V. Gkatzelis, D. Halpern, and N. Shah. 2020. Resolving the Optimal Metric Distortion Conjecture. In *Proceedings of the 61st Symposium on Foundations of Computer Science (FOCS)*. 1427–1438.

K. Grönlund, K. Herne, and M. Setälä. 2017. Empathy in a Citizen Deliberation Experiment. *Scandinavian Political Studies* 40, 4 (2017), 457–480.

The Irish Citizen's Assembly Project. 2019. *Irish citizens' assembly (2019–2020)*. http://www.citizenassembly.ie/work/

O. E. Kangas. 1997. Self-Interest and the Common Good: The Impact of Norms, Selfishness and Context in Social Policy Opinions. *The Journal of Socio-Economics* 26, 5 (1997), 475–494.

C. F. Karpowitz and T. Mendelberg. 2011. An Experimental Approach to Citizen Deliberation. *Cambridge Handbook of Experimental Political Science* (2011), 258–272.

C. Kendall and J. Matsusaka. 2021. *The Common Good and Voter Polarization.* Technical Report. Mimeo, University of Southern California.

D. R. Kinder and D. R. Kiewiet. 1981. Sociotropic politics: the American case. *British Journal of Political Science* 11, 2 (1981), 129–161.

F. E. Kizilkaya and D. Kempe. 2022. Plurality Veto: A Simple Voting Rule Achieving Optimal Metric Distortion. In *Proceedings of the 31st International Joint Conference on Artificial Intelligence (IJCAI)*. 349–355.

J. Ledyard. 1997. Public Goods: A Survey of Experimental Research. In *Handbook of Experimental Economics*, J. Kagel and A. Roth (Eds.). Princeton University Press.

A. Lindbeck and J. W. Weibull. 1988. Altruism and time consistency: the economics of fait accompli. *Journal of Political Economy* 96, 6 (1988), 1165–1182.

T. Mendelberg. 2002. The Deliberative Citizen: Theory and Evidence. *Political Decision Making, Deliberation and Participation* 6, 1 (2002), 151–193.

H. Moulin. 1986. Choosing From a Tournament. *Social Choice and Welfare* 3, 4 (1986), 271–291.

E. Muller and M. A. Satterthwaite. 1977. The Equivalence of Strong Positive Association and Strategy-Proofness. *Journal of Economic Theory* 14, 2 (1977), 412–418.

Participedia. 2022. . https://participedia.net

A. D. Procaccia and J. S. Rosenschein. 2006. The Distortion of Cardinal Preferences in Voting. In *Proceedings of the 10th International Workshop on Cooperative Information Agents (CIA)*. 317–331.

R. Wang, J. S. Fishkin, and R. C. Luskin. 2020. Does Deliberation Increase Public-Spiritedness? *Social Science Quarterly* 101, 6 (2020), 2163–2182.

L Xia and V Conitzer. 2010. Determining possible and necessary winners under common voting rules given partial orders. A longer unpublished version of [38].

I. Zettler, B. E. Hilbig, and J. Haubrich. 2011. Altruism at the Ballots: Predicting Political Attitudes and Behavior. *Journal of Research in Personality* 45, 1 (2011), 130–133.

## A  SUPPLEMENTAL MATERIALS FROM SECTION 3

By convention, throughout the appendices $a'$ denotes the winner of the election, i.e. $a' = f(\pi)$, and $a^*$ denotes the highest-welfare alternative.

### A.1  Explanation of instance diagrams

In this section of the appendix, we will present the utility matrices of counterexample instances (usually for proving lower bounds) via diagrams. Below, we show the anatomy of such a diagram:

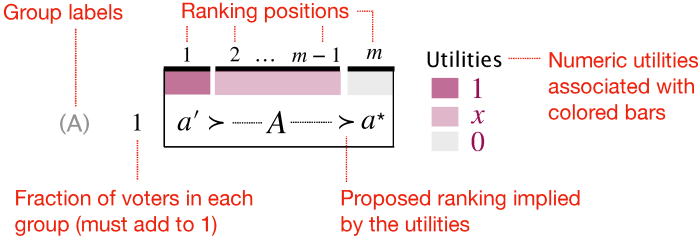

*Voters.* Most diagrams will have multiple rows, but this one has just a single row, reflecting the fact that this utility matrix has only one *group* of voters, labeled as group (A) on the left. All members of a given group have the same utilities for all alternatives, and thus the same ranking over alternatives. On the left of the box is the number 1, indicating that all voters (a 1-fraction) belong to group (A).

*Alternatives.* The alternatives are listed in the white region of the box. In this instance, there are $m$ alternatives: $a', a^*$, and all alternatives in $A$, which represents a bloc of alternatives that are interchangeable in the instance, i.e., treated identically by all voters.

*Utilities.* We encode voters' utilities for alternatives with colored bars corresponding to the alternative below them, where darker colors correspond to higher utilities. The utility value associated with each color is on the right hand side of the diagram in the key labeled 'Utilities'. Sometimes, this key will contain variables like $x$, which we will set carefully in the proof, as they are functions of $\gamma$. For example, in the diagram above, every voter in group (A) has utility 1 for alternative $a'$, $x$ for all $a \in A$, and 0 for $a^*$. In these examples, we will occasionally set utilities to be larger than 1 to make the math clearer because the scaling is more convenient.

*Rankings.* Finally, these diagrams encode the rankings that we propose are implied by the utilities. These rankings are denoted by the list of alternatives in the box, separated by $>$ symbols to denote that they are ordered. For instance, the ranking proposed in the above instance is $a' > A > a^*$, i.e., all voters in group (A) rank $a'$ first, $a^*$ last, and all other alternatives in between. Of course, the fact that these rankings are realized by the given utilities requires proving, which we will do when we prove our lower bounds. The ranking positions are given above the box.

Regarding the rankings of alternatives in blocs like $A$, we will make various assumptions about how the alternatives within $A$ are ranked, via arbitrarily small perturbations of the utilities of those alternatives.[7]

---

[7]We will usually assume that the alternatives in $A$ are cycled symmetrically across voters' rankings (using arbitrarily small epsilons to tie-break), but sometimes we will instead assume that these alternatives are always ranked consistently. Either way, we can do this tie-breaking without affecting the distortion.

## A.2 Proof of Proposition 3.5

PROPOSITION 3.5. $\kappa_{\text{PLURALITY}} = 1/m$, so for all $\gamma$ with $\gamma_{min} > 0$, $\text{dist}(\text{PLURALITY}, \gamma) \leq m^{\frac{1-\gamma_{min}}{\gamma_{min}}} + 1$.

PROOF. In light of Corollary 3.4, proving the claim amounts to proving $\kappa_{\text{PLURALITY}}(m) \geq 1/m$. Let $f = \text{PLURALITY}$. For the sake of contradiction, suppose there exists a profile $\pi$ and an alternative $a$ such that $|\{i|f(\pi) \succ_{\pi_i} a\}|/n < 1/m$. For shorthand, let $a' = f(\pi)$. Then, $a'$ must be ranked first by less than a $1/m$ fraction of the voters in $\pi$, meaning $a'$ receives strictly less than $n/m$ points. There are $n$ total points awarded across alternatives, so by averaging, there must be an alternative $a \neq a'$ that receives strictly more than $1/m$ points, implying that $f(\pi) \neq a'$, a contradiction. □

## A.3 Proof of Proposition 3.6

PROPOSITION 3.6. $\kappa_{\text{BORDA}} = 1/m$, so for all $\gamma$ with $\gamma_{min} > 0$, $\text{dist}(\text{BORDA}, \gamma) \leq m^{\frac{1-\gamma_{min}}{\gamma_{min}}} + 1$.

PROOF. Proving this claim amounts to proving that $\kappa_{\text{BORDA}}(m) \geq 1/m$. Let $f = \text{BORDA}$. For the sake of contradiction, suppose there exists a profile $\pi$ and an alternative $a$ such that $|\{i : f(\pi) \succ_{\pi_i} a\}|/n < 1/m$. For shorthand, let $a' = f(\pi)$.

Now, divide voters into two groups: those who rank $a' \succ a$ (an $x < 1/m$ fraction of the voters), and those who rank $a \succ a'$ (the remaining $1 - x$ fraction of voters). Among all voters in the first group, the point gap between $a'$ and $a$ is at most 1, corresponding to $a'$ ranked first and $a$ last. For all voters in the second group, the point gap between $a'$ and $a$ is at most $-1/(m-1)$, i.e., $a$ receives at least $1/(m-1)$ more points than $a'$ from each of these voters' rankings. Then, denoting the respective point totals by $P(a')$ and $P(a)$,

$$P(a') - P(a) \leq x \cdot 1 + (1-x) \cdot \frac{-1}{m-1} < \frac{1}{m} + \left(1 - \frac{1}{m}\right) \cdot \frac{-1}{m-1} = 0.$$

Therefore, $a'$ must receive less points than $a$ and cannot be the BORDA winner, a contradiction. □

## A.4 Proof of Proposition 3.7

PROPOSITION 3.7. For all $\gamma$ with (fixed) $\gamma_{min} > 0$, $\text{dist}(\text{PIECEWISE}, \gamma) \in O(m^{2/3})$.

PROOF. Let $U \in \mathbb{R}_{\geq 0}^{n \times m}$, fix arbitrary $\gamma$ with $\gamma_{min} > 0$ (as in the hypothesis), and let $\pi \in \Pi_{V(\gamma, U)}$. Let $a' = \text{PIECEWISE}(\pi)$ and $a^*$ denote the winner and the highest-welfare alternative, respectively. Without loss of generality, let us assume that the average utility of $a^*$ is $\text{sw}(a^*)/n = 1$. We treat separately the scenarios where the lower bound on the social welfare of $a'$ comes from $a'$ *having to beat $a^*$* (Case 1), and when it comes from *having to beat some other alternative* (Case 2).

**Case 1:** Suppose at least half of voters rank $a^*$ in the first $m^{2/3}$ positions. Let us call this subset of voters

$$N^* = \{i : (\pi_i)^{-1}(a^*) \leq m^{2/3}\},$$

satisfying $|N^*| \geq n/2$. If $a'$ ranks ahead of $a^*$ in more than half of $N^*$, then Lemma 3.1 immediately gives a *constant* distortion bound. If on the other hand $a'$ ranks behind $a^*$ in more than half of $N^*$, and since $a^*$ is located in the first $m^{2/3}$ positions where the spacing between consecutive positions is $s_t - s_{t-1} = m^{-2/3}$, $a'$ amasses a point deficit of at least

$$m^{-2/3} \cdot \frac{|N^*|}{2} \geq m^{-2/3} \cdot \frac{n}{4},$$

relative to $a^*$. Thus, in order to beat $a^*$ overall, $a'$ must rank ahead of $a^*$ at least $m^{-2/3} \cdot n/4$ times. Therefore, using Lemma 3.1, we obtain a distortion bound of the order $O(m^{2/3})$:

$$\frac{\mathrm{sw}(a^*, U)}{\mathrm{sw}(a', U)} \leq \frac{1 - \gamma_{min}}{\gamma_{min}} \cdot 4m^{2/3} + 1.$$

**Case 2:** Now suppose $a^*$ is ranked in the first $m^{2/3}$ positions by *less* than $1/2$ of the voters. Again let $N^*$ be again the voters where $a^*$ ranks in the first $m^{2/3}$ positions; we have that $|(N^*)^c| \geq n/2$ (using $(\cdot)^c$ to denote the complement). Now, for each alternative $a$, define the frequency with which $a$ occurs in the first $m^{2/3}$ positions amongst $(N^*)^c$ by

$$F_a = \frac{|\{i \in (N^c)^* : (\pi_i)^{-1}(a) \leq m^{2/3}\}|}{n}.$$

Since $|(N^*)^c| \geq n/2$, the *average* frequency of occurrence in the first $m^{2/3}$ positions must satisfy

$$\frac{1}{m} \sum_{a \in [m]} F_a \geq \frac{nm^{2/3}}{2mn} = \frac{m^{-1/3}}{2}. \tag{5}$$

Now, we need a further case distinction, based on *how many alternatives* have, roughly speaking, above-average frequency of occurrence in the first $m^{2/3}$ positions. To this end, let $\bar{A}$ be the set of alternatives that have $F_a \geq m^{-1/3}/4$:

$$\bar{A} := \{a \in A : F_a \geq m^{-1/3}/4\}.$$

**Case 2a:** Suppose $|\bar{A}| > m^{2/3}$. Let us now lower bound the average utility of alternatives $a \in \bar{A}$. First, since agents $i \in (N^*)^c$ rank $a^*$ in a lower position than $m^{2/3}$, the set featuring in the definition of $F_a$ is contained as follows

$$\{i : (N^c)^* : (\pi_i)^{-1}(a) \leq m^{2/3}\} \subseteq \{i : a \succ_{\pi_i} a^*\}.$$

Therefore, we may use Lemma 3.1 to estimate

$$\frac{n}{\mathrm{sw}(a, U)} \leq \frac{1 - \gamma_{min}}{\gamma_{min}} \frac{n}{\{i : a \succ_{\pi_i} a^*\}} + 1 \leq \frac{1 - \gamma_{min}}{\gamma_{min}} \frac{1}{F_a} + 1 \leq \frac{1 - \gamma_{min}}{\gamma_{min}} 4m^{1/3} + 1, \tag{6}$$

which leads to the lower bound

$$\frac{\mathrm{sw}(a, U)}{n} \geq \left(\frac{1 - \gamma_{min}}{\gamma_{min}} 4m^{1/3} + 1\right)^{-1} =: \bar{w}, \text{ satisfying } \bar{w} = \Omega(m^{-1/3}).$$

Next, we deduce from this a lower bound on the social welfare of $a'$. Since there are in total $nm^{2/3}/2$ points awarded in the election, $a'$ has to score at least $m^{-1/3}/2$ points per voter (on average) to win. Thus, $a'$ has to rank in the first $m^{2/3}$ positions at least $n/(2m^{1/3})$ many times – denote this set of voters by

$$N' := \{i : (\pi_i)^{-1}(a') \leq m^{2/3}\}, \text{ satisfying } |N'|/n \geq m^{-1/3}/2.$$

Since $|\bar{A}| > m^{2/3}$, every time that $a'$ ranks in the first $m^{2/3}$ positions, it has to rank ahead of an alternative $a \in \bar{A}$, whose average utility is lower bounded by $\bar{w}$. Therefore, arguing as in Lemma 3.1, we obtain that

$$\frac{\bar{w}}{\mathrm{sw}(a', U)/n} \leq \frac{1 - \gamma_{min}}{\gamma_{min}} \frac{n}{|N'|} + 1,$$

which implies

$$\frac{\mathrm{sw}(a', U)}{n} \geq \bar{w} \left(\frac{1 - \gamma_{min}}{\gamma_{min}} \frac{n}{|N'|} + 1\right)^{-1} \geq \bar{w} \left(\frac{1 - \gamma_{min}}{\gamma_{min}} 2m^{1/3} + 1\right)^{-1} = \Omega(m^{-2/3}).$$

**Case 2b:** Now suppose $|\bar{A}| \leq m^{2/3}$. Using (5), we then obtain that

$$\frac{m^{-1/3}}{2} \leq \frac{1}{m} \Big( \sum_{a \in \bar{A}} F_a + \sum_{a \notin \bar{A}} F_a \Big)$$

$$= \frac{|\bar{A}|}{m} \cdot \frac{1}{|\bar{A}|} \sum_{a \in \bar{A}} F_a + \frac{|\bar{A}^c|}{m} \cdot \frac{1}{|\bar{A}^c|} \sum_{a \notin \bar{A}} \frac{m^{-1/3}}{4}$$

$$\leq \frac{|\bar{A}|}{m} \cdot \frac{1}{|\bar{A}|} \sum_{a \in \bar{A}} F_a + \frac{m^{-1/3}}{4}.$$

Rearranging and using that $|\bar{A}| \leq m^{2/3}$, we obtain that

$$\frac{m^{-1/3}}{4} \leq \frac{|\bar{A}|}{m} \cdot \frac{1}{|\bar{A}|} \sum_{a \in \bar{A}} F_a \leq \frac{m^{2/3}}{m} \cdot \frac{1}{|\bar{A}|} \sum_{a \in \bar{A}} F_a = m^{-1/3} \frac{1}{|\bar{A}|} \sum_{a \in \bar{A}} F_a$$

$$\implies \frac{1}{|\bar{A}|} \sum_{a \in \bar{A}} F_a \geq \frac{1}{4}.$$

It follows that there must exist at least one alternative $\bar{a} \in \bar{A}$ such that $F_{\bar{a}} \geq 1/4$. Since at least $n/4$ voters rank $\bar{a}$ ahead of $a^*$, Lemma 3.1 implies that

$$\frac{n}{\mathrm{sw}(\bar{a}, U)} \leq 4 \frac{1 - \gamma_{min}}{\gamma_{min}} + 1,$$

and thus the average utility of $\bar{a}$ is lower bounded by a constant, $\mathrm{sw}(\bar{a}, U)/n = \Omega(1)$. We may now complete the proof by arguing as in Case 1: Indeed, each time $a'$ ranks behind $\bar{a}$, it incurs a scoring deficit of $m^{-2/3}$. It thus must rank ahead of $\bar{a}$ at least $\Omega(nm^{-2/3})$ times, which, via Lemma 3.1, gives the desired lower bound $\mathrm{sw}(a', U)/n = \Omega(m^{-2/3})$.                                                                                     □

## A.5    Proof of Proposition 3.8

The goal of this section is to show the following upper bound for the distortion of Maximin.

Proposition 3.8. $\kappa_{\mathrm{Maximin}} = 1/m$, so for all $\boldsymbol{\gamma}$ with $\gamma_{min} > 0$, $\mathrm{dist}(\mathrm{Maximin}, \boldsymbol{\gamma}) \leq m \frac{1 - \gamma_{min}}{\gamma_{min}} + 1$.

Our high-level proof strategy is to show that $\kappa_{\mathrm{Maximin}} = 1/m$. Then, the proposition follows immediately from an application of Corollary 3.4. Since every voting rule satisfies $\kappa_f \leq 1/m$, we only have to show that $\kappa_{\mathrm{Maximin}} \geq 1/m$, which is directly implied by the following lemma.

Lemma A.1. For every $\boldsymbol{\pi}$ preference profile, there exists some alternative $\bar{a} \in [m]$ such that

$$\min_{a \neq \bar{a}} \{ i : \bar{a} \succ_{\pi_i} a \} \geq n/m.$$

In particular, the Maximin winner $a'$ (which is the alternative with the smallest maximum pairwise loss) must also satisfy

$$\min_{a \neq a'} |\{ i : a' \succ_{\pi_i} a \}| \geq n/m.$$

Consequently, it also holds that $\kappa_{\mathrm{Maximin}} \geq 1/m$.

Proof. We define a sequence of alternatives $(a_j : j \geq 1)$ as follows. Start with an arbitrary alternative $a_1$. Given $a_j$, we let $a_{j+1}$ be the alternative which pairwise-dominates $a_j$ by the most,

$$a_{j+1} := \arg \max_{a \in [m] \setminus \{a_j\}} |\{ i : a \succ_{\pi_i} a_j \}|.$$

In this process, if we encounter an alternative that has previously been part of the sequence, i.e. $a_{j+1} = a_k$ for some $k \leq j$, then we exit the recursive procedure, and draw a cycle $(a_k, \ldots, a_{j+1})$. Then, the longest such cycle we can create is of length $m + 1$. Since $a_1$ was arbitrary, we may without loss of generality assume that the constructed cycle starts at $k = 1$, and has length $L$, i.e. the cycle is $(a_1, \ldots, a_L)$ with $a_1 = a_L$. Now, let $N_j \subseteq [n]$ denote the set of voters who rank $a_{j+1} > a_j$, i.e., who contribute to $a_j$'s worst pairwise defeat. We now make the following claim.

**Claim:** There exists some $j^* \in [L]$ such that $|N_{j^*}| \leq \frac{L-2}{L-1} n$.

To prove the claim, we first note that there cannot exist any voter $i$ such that

$$a_1 >_{\pi_i} \ldots >_{\pi_i} a_L >_{\pi_i} a_1,$$

since this ranking would be cyclical. It follows that

$$\bigcap_{j=1}^{L-1} N_j = \emptyset.$$

Now, assume for the sake of contradiction that for all $j = 1, \ldots, L$ it holds that $|N_j| > \frac{L-2}{L-1} n$. Then, this implies that.

$$\left| \bigcap_{j=1}^{J} N_j \right| > \frac{L-1-J}{L-1} n, \text{ for all } J = 1, \ldots, L-1.$$

Intuitively, we are saying that if all $N_j$ individually comprise nearly the entire set of voters, their intersection must be somewhat large. Now, looking in particular at the case where $J = L - 1$, the above inequality implies that $|\bigcap_{j=1}^{L-1} N_j| > 0$, which contradicts that the intersection of all $N_j : j \in [L-1]$ must be empty, as above. We conclude that the claim is true.

Since $L - 1 \leq m$, the preceding claim implies that there exists some $a_{j^*}$ whose worst defeat is by less than $\frac{L-1}{L} \leq \frac{m-1}{m}$ fraction of voters, i.e.,

$$\max_{a \neq a_{j^*}} \frac{|\{i : a >_{\pi_i} a_{j^*}\}|}{n} \leq \frac{m-1}{m}.$$

This proves the first assertion of the proposition, that is, by setting $\bar{a} = a_{j^*}$, we obtain the desired alternative for which at least $n/m$ voters must rank $\bar{a} > a$.

Since $a'$ is the Maximin winner, we further obtain that

$$\min_{a \neq a'} |\{i : a' >_{\pi_i} a\}| \geq \min_{a \neq a_{j^*}} |\{i : a >_{\pi_i} a_{j^*}\}| \geq \frac{n}{m}.$$

$\square$

## A.6 Proof of Proposition 3.9

PROPOSITION 3.9. *For all uniform* $\boldsymbol{\gamma} = \gamma \mathbf{1}$, $\gamma \in [0, 1]$, $\text{dist}(\text{COPELAND}, \boldsymbol{\gamma}) \geq \left( \frac{2(1-\gamma)}{\gamma} + 1 \right)^2$.

PROOF. The claim is true when $\gamma = 1$ trivially, so we will consider $\gamma < 1$. Let $U$ be the utility matrix described by the diagram (see Appendix A.1 for a primer on reading these diagrams), where $\epsilon > 0$ and $W$ is some sufficiently large value that depends on $\gamma$ (but not $\epsilon$).

$$w = W, \quad x = \frac{\gamma/2}{1 - \gamma/2}, \quad y = \left( \frac{\gamma/2}{1 - \gamma/2} \right)^2.$$

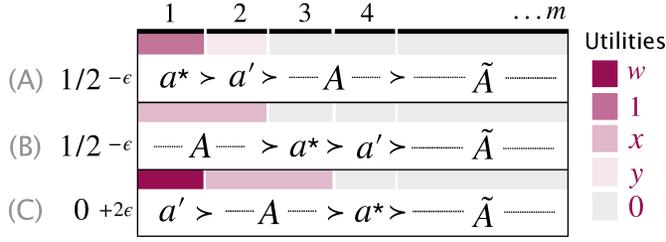

Observe that $1 > x > y > 0$, and the average utilities of alternatives are the following, where here and throughout this analysis, we will gray out $\epsilon$ terms, as they can be made arbitrarily small.

Now, establishing the average utilities: $\mathrm{sw}(a^*, U)/n = 1/2 - \epsilon$; $\mathrm{sw}(a', U)/n = y/2 + \epsilon(2w - y/2)$; for all $a \in A$, $\mathrm{sw}(a, U)/n = x(1/2 + \epsilon) = x/2 + x\epsilon$; and for all $a \in \tilde{A}$, $\mathrm{sw}(a, U)/n = 0$.

***Claim 1.*** *The utilities imply the claimed rankings.* First, observe that by virtue of having zero social welfare, the alternatives in $\tilde{A}$ are always ranked last. We will consider only the other relative rankings throughout this analysis. We confirm each group's ranking left to right by comparing the values of $v_i(a, \boldsymbol{\gamma}, U)$ (Equation (2)), derived below.

Let $i \in$ Group (A) and $a \in A$. Then, $a^* >_{\pi_i} a' >_{\pi_i} a$:

$$v_i(a^*, \boldsymbol{\gamma}, U) = (1 - \gamma) + \gamma(1/2 - \epsilon) = 1 - \gamma/2 - \gamma\epsilon$$

$$v_i(a', \boldsymbol{\gamma}, U) = (1 - \gamma)y + \gamma(y(1/2 - \epsilon) + 2\epsilon W) = y(1 - \gamma/2) + \epsilon\gamma(2W - y)$$

$$v_i(a, \boldsymbol{\gamma}, U) = \gamma x(1/2 + \epsilon) = \gamma/2 \cdot \left( \frac{\gamma/2}{1 - \gamma/2} \right) + \epsilon\gamma x = y(1 - \gamma/2) + \epsilon\gamma x$$

Let $i \in$ Group (B) and $a \in A$. Then, $a >_{\pi_i} a^* >_{\pi_i} a'$:

$$v_i(a, \boldsymbol{\gamma}, U) = x(1 - \gamma + \gamma(1/2 + \epsilon)) = x(1 - \gamma/2) + \gamma x\epsilon = \gamma/2 + \gamma x\epsilon$$

$$v_i(a^*, \boldsymbol{\gamma}, U) = \gamma(1/2 - \epsilon) = \gamma/2 - \gamma\epsilon$$

$$v_i(a', \boldsymbol{\gamma}, U) = \gamma(y(1/2 - \epsilon) + 2\epsilon W) = \gamma y/2 + \epsilon\gamma(2W - y)$$

Let $i$ be in Group (C), and $a \in A$. Then, $a' >_{\pi_i} a >_{\pi_i} a^*$:

$$v_i(a', \boldsymbol{\gamma}, U) = (1 - \gamma)W + \gamma(y(1/2 - \epsilon) + 2W\epsilon) = (1 - \gamma)W + \gamma y/2 + \epsilon\gamma(2W - y)$$

$$v_i(a, \boldsymbol{\gamma}, U) = x(1 - \gamma + \gamma(1/2 + \epsilon)) = \left( \frac{\gamma/2}{1 - \gamma/2} \right)(1 - \gamma/2) + \epsilon\gamma x = \gamma/2 + \epsilon\gamma x$$

$$v_i(a^*, \boldsymbol{\gamma}, U) = \gamma(1/2 - \epsilon) = \gamma/2 - \epsilon\gamma$$

***Claim 2.*** $a'$ *is the* Copeland *winner.* To do this analysis quickly, we draw the pairwise majority graph for this instance, where an arrow $a \to \tilde{a}$ indicates that $a$ pairwise-dominates $\tilde{a}$:

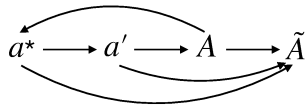

Because we assume that items are symmetrically within $A$, and similarly within $\tilde{A}$, $a'$ is the unique COPELAND winner:[8]

- $a'$ gets $m-2$ points by strictly pairwise defeating 2 items in $A$ and $m-4$ items in $\tilde{A}$.

- $a^*$ gets $m-3$ points by strictly pairwise defeating $a'$ and $m-4$ items in $\tilde{A}$.

- all $a \in A$ get $m-3$ points by strictly pairwise defeating $a^*$ and $m-4$ items in $\tilde{A}$.

- all $a \in \tilde{A}$ get 0 points.

*Distortion.* It follows that the distortion in this instance, provided the proposed rankings are realized, approaches the following quantity as $\epsilon \to 0$:

$$\frac{\mathrm{sw}(a^*, U)}{\mathrm{sw}(a', U)} \xrightarrow{\epsilon \to 0} \frac{1/2}{y/2} = \left(\frac{1-\gamma/2}{\gamma/2}\right)^2 = \left(\frac{2-\gamma}{\gamma}\right)^2 = \left(\frac{2(1-\gamma)}{\gamma}+1\right)^2 \qquad \square$$

## A.7 Proof of Proposition 3.10

PROPOSITION 3.10. *For all uniform $\boldsymbol{\gamma} = \gamma\mathbf{1}$, $\gamma \in [0,1]$, $\mathrm{dist}(\textsc{Slater}, \boldsymbol{\gamma}) \geq \left(\frac{2(1-\gamma)}{\gamma}+1\right)^2$.*

PROOF. We can lower-bound SLATER's distortion identically to COPELAND's, as in Proposition 3.9, via the same instance (with slightly different treatment of the alternatives in $A, \tilde{A}$). In particular, where before we cycled alternatives symmetrically in these set, now assume that items are always ordered the same way within $A$, and similarly within $\tilde{A}$. In particular, let $\pi_A, \pi_{\tilde{A}}$ be these consistent sub-rankings. Fix this instance $\boldsymbol{\gamma}, U$. Then, $a'$ is the unique SLATER winner, by the argument below. Note that this is all we need to prove identical distortion to Proposition 3.9, because we have already confirmed that the rankings in this instance are realized by the utilities, as well as the distortion itself, in the proof of Proposition 3.9.

First, we will pare down the possible slater rankings. Observe that because items within $A, \tilde{A}$ are always ranked as $\pi_A, \pi_{\tilde{A}}$ in $\boldsymbol{\pi}^{\gamma, U}$, the slater ranking must also rank them in this order to minimize pairwise disagreements. Similarly, the slater ranking will always rank everything in $\tilde{A}$ in the last $m-4$ slots, as those items are always in those slots in $\boldsymbol{\pi}^{\gamma, U}$.

That leaves us with the possible slater rankings listed below, using $\pi_A, \pi_{\tilde{A}}$ to denote all alternatives in those sets in their fixed ordering. Note that $A$ contains 2 alternatives and $\tilde{A}$ contains $m-4$ alternatives. For each ranking, we tally its disagreements with the pairwise majority graph.

- $a' \succ \pi_A \succ a^* \succ \pi_{\tilde{A}}$ disagrees with 1

- $a' \succ a^* \succ \pi_A \succ \pi_{\tilde{A}}$ disagrees with 3

- $a^* \succ a' \succ \pi_A \succ \pi_{\tilde{A}}$ disagrees with 2

- $a^* \succ \pi_A \succ a' \succ \pi_{\tilde{A}}$ disagrees with 4

- $\pi_A \succ a^* \succ a' \succ \pi_{\tilde{A}}$ disagrees with 2

- $\pi_A \succ a' \succ a^* \succ \pi_{\tilde{A}}$ disagrees with 3

The slater ranking is the first one, so the winner is $a'$. $\qquad \square$

---

[8]Here, we additionally assume that $n$ is even (a similar instance, with a third identical alternative added to the set $A$ to form a Condorcet cycle within $A$, would work for odd $n$, see Appendix B.2 for a similar construction.).

### A.8   Proof of Theorem 3.11

THEOREM 3.11. *For all positional scoring rules $f$ and uniform $\boldsymbol{\gamma} = \gamma\mathbf{1}$ with (fixed) $\gamma \in [0, 1)$,*

$$\mathrm{dist}(f, \boldsymbol{\gamma}) \in \Omega(\sqrt{m}).$$

PROOF. Let $\mathbf{s} = (s_1, \ldots, s_m)$ denote the (decreasing) scoring vector of $f$, and recall that $s_1 = 1$, $s_m = 0$. Then, there must exist some position $t \in \{1, \ldots, \sqrt{m}\}$ such that $s_t - s_{t+1} \leq 1/\sqrt{m}$. We then construct a utility matrix $U$ as pictured in the diagram below (see Appendix A.1 for a primer on reading these diagrams), where

$$x = \frac{1}{1 - \gamma} \quad \text{and} \quad y = C'/\sqrt{m}$$

and $C, C'$ are constant to be chosen later.

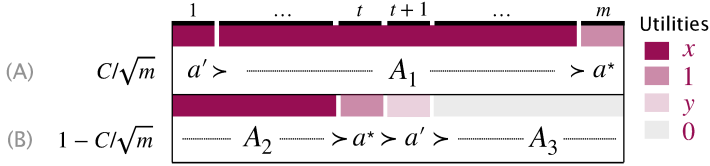

For the ranking of group (A), we assume that $A_1$ contains alternatives $1, \ldots, m-2$ occupy the ranks in *cyclically*, i.e. that any given alternative $a = 1, \ldots, m-2$ occupies any rank $r \in \{2, \ldots, m-1\}$ in a $1/(m-2)$ fraction of group (A) (this is permitted since $a = 1, \ldots, m-2$ are treated symmetrically, so we may choose the preference orderings between them arbitrarily when the PS-values are tied.) Similarly, in group (B) we may assume that the alternatives $1, \ldots, m-2$ are cycled through the $m-2$ occupied by $A_2 \cup A_3$ – this way their welfares are equal, $\mathrm{sw}(1, U) = \cdots = \mathrm{sw}(m-2, U)$, and the PS-values are hence always tied between the alternatives within positions $A_2$, and within positions $A_3$.

We now argue that the above utilities induce the the rankings profile shown in the diagram. To verify the rankings in group (A), we first note that

$$v_i(a^*, \boldsymbol{\gamma}, U) = 1 \leq (1 - \gamma)u_i(a') \leq v_i(a', \boldsymbol{\gamma}, U).$$

Moreover, for any $1 \leq a \leq m-2$, since $t \leq \sqrt{m}$ we have

$$\frac{\mathrm{sw}(a, U)}{n} \leq \frac{C}{(1 - \gamma)\sqrt{m}} + \frac{t - 1}{(1 - \gamma)(m - 2)} \leq \frac{C + 2}{(1 - \gamma)\sqrt{m}}.$$

while for $a'$ we have, for any $m$ large enough such that $C/\sqrt{m} \leq 1/2$,

$$\frac{\mathrm{sw}(a', U)}{n} = \frac{C}{(1 - \gamma)\sqrt{m}} + \left(1 - \frac{C}{\sqrt{m}}\right)\frac{C'}{\sqrt{m}} \geq \frac{C}{(1 - \gamma)\sqrt{m}} + \frac{C'}{2\sqrt{m}}.$$

Thus, for $C'$ chosen large enough (depending on $\gamma, C$), we obtain that for $a = 1, \ldots, m-2$, $\mathrm{sw}(a', U) \geq \mathrm{sw}(a, U)$. It follows that also $v_i(a', \boldsymbol{\gamma}, U) \geq v_i(a, \boldsymbol{\gamma}, U)$, and the rankings of group (A) are confirmed.

We now verify the rankings in group (B). For alternatives $a$ in positions $A_2$, we have

$$v_i(a^*, \boldsymbol{\gamma}, U) = 1 \leq (1 - \gamma)u_i(a) \leq v_i(a, \boldsymbol{\gamma}, U),$$

so that they indeed rank ahead of $a^*$. Since $C'$ was chosen above such that $\mathrm{sw}(a', U) \geq \mathrm{sw}(a, U)$ (for all $a = 1, \ldots, m-2$), $a'$ is indeed ranked ahead of $A_3$, and $\mathrm{sw}(a', U) = O(1/\sqrt{m}) = o(\mathrm{sw}(a^*, U))$, we conclude that $a'$ is indeed ranked in the $t+1$-st position. Thus the positions in group (B) are confirmed, too.

It remains to verify that in the ranking profile from the diagram, $a'$ is indeed the positional scoring rule winner. For $i \in [n], a \in [m]$, let $\pi_i^{-1}(a)$ denote the position that voter $i$ ranks alternative $a$ in. Then, we may write the point totals as

$$P(a) := \sum_{i \in [n]} s_{\pi_i^{-1}(a)}, \ a \in [m]$$

Firstly, $a'$ beats $a^*$, since

$$\frac{1}{n}(P(a') - P(a^*)) = \frac{C}{\sqrt{m}} - \left(1 - \frac{C}{\sqrt{m}}\right)(s_t - s_{t+1}) \geq \frac{C-1}{\sqrt{m}} > 0,$$

as long as we choose $C > 1$. Secondly, to see that $a'$ beats $1, \ldots, m-2$, we prove that $P(a') > P(1)$ (which suffices because $P(1) = \cdots = P(m-2)$). Note that the fraction of times alternative 1 occupies any position $l \in \{1, \ldots, t+1\}$ is bounded by

$$\frac{|\{i : \pi_i^{-1}(1) \leq t+1\}|}{n} = \frac{C}{\sqrt{m}}\frac{t}{m-2} + \left(1 - \frac{C}{\sqrt{m}}\right)\frac{t-1}{m} \leq \frac{t}{m-2} \lesssim \frac{1}{\sqrt{m}},$$

where we again used that $t \leq \sqrt{m}$. Since $a'$ ranks first a $C/\sqrt{m}$ fraction of times, and otherwise occupies the $(t+1)$-th place, we may enforce that $P(a') > P(1)$, by choosing $C > 0$ large enough. □

## A.9 Proof of Lemma 3.12

**Lemma 3.12.** For all positional scoring rules $f$ and uniform $\boldsymbol{\gamma} = \gamma\mathbf{1}, \gamma \in [0,1]$, $\text{dist}(f, \boldsymbol{\gamma}) \geq \frac{1-\gamma}{\gamma\Delta_f} + 1$.

**Proof.** The claim is true when $\gamma = 1$, because given that all positional scoring rules $f$ are unanimous, $\text{dist}_1(f) = 1$. For the remainder of the proof, we will thus consider $\gamma < 1$.

Fix an arbitrary positional scoring rule $f$ with gap $\Delta_f$, defined as the gap between the scores given to the first two positions (i.e., $s_1 - s_2$). Fix some $\gamma \in [0, 1)$, and let $\boldsymbol{\gamma} = \gamma\mathbf{1}$. Now, consider the instance $(\boldsymbol{\gamma}, U)$ depicted in the diagram below, where $U$ is as shown in the following diagram with $\epsilon > 0$ and

$$x = \frac{\gamma(1-\Delta_f)}{1-\gamma+\gamma\Delta_f} \iff x(1-\gamma+\gamma\Delta_f) = \gamma(1-\Delta_f)$$

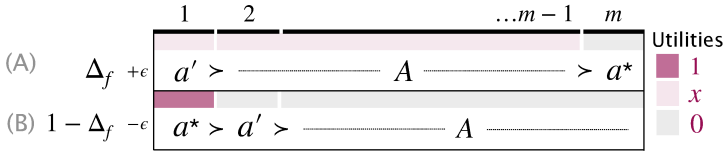

Fig. 1. $A$ contains all alternatives other than $a', a^*$, cycled symmetrically over rankings, and all $\pm\epsilon$ are used for tie-breaking only.

First, we prove two necessary claims, and then analyze the distortion given that $a'$ is the winner by $f$.

***Claim 1:*** *the utilities imply the proposed rankings.* Since $a'$ always has PS-values which are always greater or equal than that of any alternative in $A$, we may always rank $a'$ ahead of all alternatives in $A$, and thus the relative rankings of $a', A$ are correct. Now, we verify the relative orderings of $a^*$ and all other alternatives in both groups:

$$x = \frac{\gamma(1-\Delta_f)}{1-\gamma+\gamma\Delta_f} \iff x(1-\gamma+\gamma\Delta_f) = \gamma(1-\Delta_f)$$

$$\implies (1-\gamma+\gamma(\Delta_f+\epsilon))x > \gamma(1-\Delta_f-\epsilon)$$

$$\iff v_i(a,\boldsymbol{\gamma},U) > v_i(a^*,\boldsymbol{\gamma},U) \text{ for all } a \neq a^*, i \in \text{group (A)}.$$

We now analyze group (B)'s ranking. Since $u_i(a') = u_i(a)$ for all $i \in [n]$ and $a \in A$, it suffices to check that

$$\gamma \text{sw}(a',U) = v_i(a',\boldsymbol{\gamma},U) \leq v_i(a^*,\boldsymbol{\gamma},U) \text{ for all } i \in \text{group (B)}.$$

Since $u_i(a') = 0$ in group (B), it suffices to verify that $\text{sw}(a',U) \leq \text{sw}(a^*,U)$:

$$\text{sw}(a',U) \leq \text{sw}(a^*,U) \iff x(\Delta_f+\epsilon) \leq (1-\Delta_f-\epsilon)$$

$$\iff \frac{\gamma(\Delta_f+\epsilon)}{1-\gamma+\gamma\Delta_f}(1-\Delta_f) \leq (1-\Delta_f-\epsilon)$$

$$\iff \gamma(1-\Delta_f)(\Delta_f+\epsilon) \leq \gamma(1-\Delta_f-\epsilon)(\Delta_f+1/\gamma-1)$$

Since we assumed that $\gamma < 1$, clearly we may choose $\epsilon > 0$ small enough such that the inequality in the last line holds true. This confirms the rankings in group (B).

***Claim 2:*** *$a'$ is the winner per the proposed rankings.* $a'$ is always ranked ahead of all $a \in A$, so $a'$ must receive a higher score than all these alternatives. $a'$ also receives more points than $a^*$: $a'$ receives $\Delta_f + \epsilon + (1-\Delta_f-\epsilon)(1-\Delta_f) > 1-\Delta_f$ points, which is larger than the $1-\Delta_f-\epsilon$ points received by $a^*$.

Now, to analyze the distortion we let $\epsilon \to 0$:

$$\text{dist}_{\boldsymbol{\gamma}}(f) \geq \frac{\text{sw}(a^*,U)}{\text{sw}(a',U)} \xrightarrow{\epsilon\to 0} \frac{1-\Delta_f}{\Delta_f x} = \frac{1-\gamma}{\gamma\Delta_f}+1.$$

$\square$

## A.10 Proof of Proposition 3.15

PROPOSITION 3.15. *For all uniform $\boldsymbol{\gamma} = \gamma\mathbf{1}$, $\gamma \in [0,1]$, $\text{dist}(\text{PLURALITY},\boldsymbol{\gamma}) \geq m \cdot \frac{1-\gamma}{\gamma}+1$.*

PROOF. Fix an arbitrary uniform $\boldsymbol{\gamma} = \mathbf{1}\gamma$ and let $U$ be the utility matrix depicted in the following diagram, where all alternatives in $A$ are cycled symmetrically, and

$$x = \frac{\gamma(m-1)/m}{1-\gamma+\gamma/m}$$

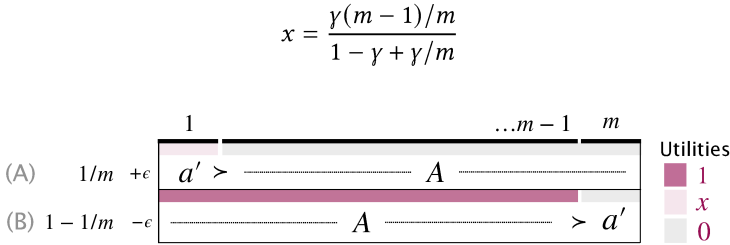

The average utilities of the alternatives are then the following: $\text{sw}(a',U)/n = x(1/m+\epsilon) = x/m_{+x\epsilon}$, and for all $a \in A$, $\text{sw}(a,U)/n = (m-1)/m_{-\epsilon}$.

***Claim 1.*** *The proposed rankings are realized by the utilities.* We confirm each ranking left to right by comparing voters' PS-values, per Equation (2).

Let $i \in$ group (A) and $a \in A$. Then,

$$v_i(a', \boldsymbol{\gamma}, U) = x(1 - \gamma + \gamma(1/m + \epsilon)) = x(1 - \gamma + \gamma/m) + x\epsilon = \left(\frac{\gamma(m-1)/m}{1 - \gamma + \gamma/m}\right)(1 - \gamma + \gamma/m) + x\epsilon$$
$$= \gamma(m-1)/m + x\epsilon$$
$$v_i(a, \boldsymbol{\gamma}, U) = \gamma(1 - 1/m - \epsilon) = \gamma(m-1)/m - \gamma\epsilon$$

Let $i \in$ group (B) and $a \in A$. Then,

$$v_i(a, \boldsymbol{\gamma}, U) = 1 - \gamma + \gamma(1 - 1/m - \epsilon) = 1 - \gamma/m - \gamma\epsilon$$
$$v_i(a', \boldsymbol{\gamma}, U) = \gamma(1/m + \epsilon)x = \gamma x/m + \gamma x\epsilon = \gamma/m \cdot \left(\frac{\gamma(m-1)/m}{1 - \gamma + \gamma/m}\right) + \gamma x\epsilon$$
$$= \gamma/m \cdot \left(\frac{1}{1 - \gamma + \gamma/m} - 1\right) + \gamma x\epsilon$$
$$= \frac{\gamma/m}{1 - \gamma + \gamma/m} - \gamma/m + \gamma x\epsilon$$
$$< 1 - \gamma/m - \gamma\epsilon$$

Where the last step holds for sufficiently small $\epsilon$, and $\gamma/m \le 1 - \gamma/m$ holds when $m \ge 2$.

***Claim 2.*** $a'$ *is the winner.* $a'$ is the PLURALITY winner because it is ranked first a $1/m + \epsilon$ fraction of the time, while all other alternatives $a \in A$ are ranked first a $1/m - \epsilon/(m-1)$ fraction of the time.

By Claims 1 and 2, the distortion in this instance approaches the following as $\epsilon \to 0$ (where $a$ is an arbitrary alternative in $A$):

$$\frac{\mathrm{sw}(a, U)}{\mathrm{sw}(a', U)} = \frac{(m-1)/m}{x/m} = (m-1) \cdot \frac{1 - \gamma + \gamma/m}{\gamma(m-1)/m} = \frac{1-\gamma}{\gamma}m + 1. \qquad \square$$

## A.11  Proof of Proposition 3.16

PROPOSITION 3.16. *For all uniform* $\boldsymbol{\gamma} = \gamma\mathbf{1}$, $\mathrm{dist}(\text{MAXIMIN}, \boldsymbol{\gamma}) \ge (m-1) \cdot \frac{1-\gamma}{\gamma} + 1$.

PROOF. We first specify a preference profile $\boldsymbol{\pi}$ with $m$ alternatives in which $a'$ is the winner, i.e., $\text{MAXIMIN}(\boldsymbol{\pi}) = a'$; we will later show that $\boldsymbol{\pi}$ can be realized by suitable utilities.

We split the population into two groups, A and B:

- **Group A** is of size $n/(m-1)$, and voters $i$ in group A rank

$$a' >_i \text{ all other } m-1 \text{ alternatives.}$$

- **Group B** contains the rest of the voters, i.e. is of size $n(m-2)/(m-1)$. In this group, voters $i$ have ranking of the form

$$\text{all other } m-1 \text{ alternatives} >_{\pi_i} a'.$$

In these rankings, we assume that the $m - 1$ non-winning alternatives (call them $1 \ldots m - 1$) are ranked cyclically – that is, each group is further divided into $m - 1$ subgroups of equal size, where voters $i$ in the respective $k$-th subgroups rank

$$k >_{\pi_i} k + 1 >_{\pi_i} \cdots >_{\pi_i} m - 1 >_{\pi_i} 1 >_{\pi_i} \cdots >_{\pi_i} k - 1.$$

We now verify that indeed $a' = \text{MAXIMIN}(\boldsymbol{\pi})$. Firstly, $a'$ performs equally well in all comparisons with other alternatives, i.e.

$$\max_{a \neq a'} |\{i : a >_{\pi_i} a'\}| = |\{i : 1 >_{\pi_i} a'\}| = n - \frac{n}{m-1} = n\frac{m-2}{m-1}.$$

On the other hand, for each of the remaining alternatives $k = 1, \ldots, m - 1$, their worst defeat comes from the preceding alternative $k - 1$ (for $k = 1$, this alternative is $m - 1$) – in particular, the cyclical rankings in both Group 1 and 2 immediately imply that

$$\max_{a \neq k} |\{i : a >_{\pi_i} k\}| = |\{i : k - 1 >_{\pi_i} k\}| = n\frac{m-2}{m-1} \geq \max_{a \neq a'} |\{i : a >_{\pi_i} a'\}|,$$

confirming that $a'$ wins the election.

We now specify the utilities as follows.

- In **Group A**, voters $i$ have $u_i(a') = \frac{\gamma(m-2)}{(1-\gamma)(m-1)+\gamma}$ and $u_i(a) = 0$ for all remaining alternatives.

- In **Group B**, voters $i$ have $u_i(a') = 0$ and $u_i(a) = 1$ for all remaining alternatives $a = 1, \ldots, m - 1$.

The cyclical rankings amongst $a = 1, \ldots, m - 1$ can be realized since we have treated those alternatives symmetrically, so that $v_i(a, \boldsymbol{\gamma}, U)$ are tied for all $i \in [n]$ and $a = 1, \ldots, m - 1$. The ranking of voters in group $B$ is confirmed by comparison of social welfares and utilities. $a'$ is ranked ahead of all other $a$ for all $i$ in Group A by the following reasoning:

$$\begin{aligned}
v_i(a', \boldsymbol{\gamma}, U) &= (1 - \gamma)u_i(a') + \text{sw}(a', U) \\
&= (1 - \gamma)\frac{\gamma(m-2)}{(1-\gamma)(m-1)+\gamma} + \frac{\gamma}{m-1}\frac{\gamma(m-2)}{(1-\gamma)(m-1)+\gamma} \\
&= (1 - \gamma + \frac{\gamma}{m-1})\frac{\gamma(m-2)}{(m-1)[(1-\gamma)+\gamma/(m-1)]} \\
&= \gamma\frac{m-2}{m-1} = v_i(a, \boldsymbol{\gamma}, U).
\end{aligned}$$

Since we may assume worst-case tie breaking, we may rank $a'$ ahead of $a$. Note that in this profile, all the alternatives $a = 1, \ldots, m - 1$ have equal social welfare. Fixing any such $a$, the distortion in this instance is

$$\frac{\text{sw}(a, U)}{\text{sw}(a', U)} = \frac{m-2}{m-1} \cdot \frac{m-1}{\frac{\gamma(m-2)}{(1-\gamma)(m-1)+\gamma}} = \frac{(1-\gamma)(m-1)+\gamma}{\gamma}. \qquad \square$$

# B  SUPPLEMENTAL MATERIAL FOR SECTION 4

In this appendix, we will often apply the following lemma:

LEMMA B.1. *For any utility matrix, decreasing a voter $i$'s level of public spirit cannot result in them promoting a higher-welfare alternative over a lower-welfare alternative in their ranking.*

PROOF. Fix an arbitrary utility matrix $U$, arbitrary voter $i$, and $\boldsymbol{\gamma}, \tilde{\boldsymbol{\gamma}}$ which differ in that $\tilde{\gamma}_i < \gamma_i$ (they may also differ in other ways — it is irrelevant to this proof). Fix corresponding profiles $\boldsymbol{\pi} \in \Pi_{V(\boldsymbol{\gamma}, U)}$ and $\tilde{\boldsymbol{\pi}} \in \Pi_{V(\tilde{\boldsymbol{\gamma}}, U)}$. Let $a, a'$ be an arbitrary pair of alternatives such that $a' >_{\pi_i} a$ and $\mathrm{sw}(a, U) \geq \mathrm{sw}(a', U)$ (if no such pair exists, because $i$'s alternatives are ranked in decreasing order of welfare, and thus we are done because $i$ cannot promote a higher-welfare alternative over a lower-welfare alternative). We will show that $a$ cannot be promoted over $a'$ from $\pi_i$ to $\tilde{\pi}_i$ — that is, $a' >_{\tilde{\pi}_i} a$, thereby showing the claim.

First, observe that because $a$ has greater social welfare than $a'$, $i$ must have higher utility for $a'$ than $a$ to create their relative ranking in $\pi_i$:

$$a' >_{\pi_i} a \implies u_i(a') > u_i(a).$$

Then, by $\tilde{\gamma}_i < \gamma_i$, $\mathrm{sw}(a', U) - \mathrm{sw}(a, U) < 0$ and $u_i(a') - u_i(a) > 0$,

$$
\begin{aligned}
v_i(a', \tilde{\boldsymbol{\gamma}}, U) - v_i(a, \tilde{\boldsymbol{\gamma}}, U) &= (1 - \tilde{\gamma}_i)(u_i(a') - u_i(a)) + \tilde{\gamma}_i(\mathrm{sw}(a', U) - \mathrm{sw}(a, U)) \\
&> (1 - \gamma_i)(u_i(a') - u_i(a)) + \gamma_i(\mathrm{sw}(a', U) - \mathrm{sw}(a, U)) \\
&= v_i(a', \boldsymbol{\gamma}, U) - v_i(a, \boldsymbol{\gamma}, U) > 0.
\end{aligned}
$$

The inequality deduced above concludes the proof: $v_i(a', \tilde{\boldsymbol{\gamma}}, U) - v_i(a, \tilde{\boldsymbol{\gamma}}, U) > 0 \implies a' >_{\tilde{\pi}_i} a$.  □

## B.1  Proof of Proposition 4.5

PROPOSITION 4.5. *If $m \leq 3$, then all voting rules exhibit nonuniform monotonicity.*

We prove this for $m = 2$ and $m = 3$ separately, though the arguments use the same overall strategy. We present the proof of the $m = 2$ case more gently as a warm-up, to illustrate the high-level approach; the proof of $m = 3$ requires more careful handling of additional technicalities.

**Proposition 4.5(a).** *When $m = 2$, all voting rules exhibit nonuniform monotonicity.*

PROOF. Fix an arbitrary resolute voting rule $f$, and suppose our two alternatives are $a, b$. To show the claim, it suffices to show that, starting with an instance $\boldsymbol{\gamma}, U$ and given a $\tilde{\boldsymbol{\gamma}}$ which only differs from $\boldsymbol{\gamma}$ in that $\tilde{\gamma}_1 < \gamma_1$ (i.e., only a single voter's public spirit is decreased), we can find some $\tilde{U}$ with the following two properties:

- *Property 1:* $\mathrm{sw}(a, U) = \mathrm{sw}(a, \tilde{U})$ and $\mathrm{sw}(b, U) = \mathrm{sw}(b, \tilde{U})$

- *Property 2:* $\Pi_{V(\boldsymbol{\gamma}, U)} \subseteq \Pi_{V(\tilde{\boldsymbol{\gamma}}, \tilde{U})}$.

Together these properties imply that $\mathrm{dist}(f, \boldsymbol{\gamma}, U) \leq \mathrm{dist}(f, \tilde{\boldsymbol{\gamma}}, \tilde{U})$.

*Construction of $\tilde{U}$.* Note that for all $i > 1$, we immediately have that $v_i(a, \boldsymbol{\gamma}, U) = v_i(a, \tilde{\boldsymbol{\gamma}}, U)$ and $v_i(b, \boldsymbol{\gamma}, U) = v_i(b, \tilde{\boldsymbol{\gamma}}, U)$. Then, if it is already the case voter 1's values under $U$ match ordinally across $\boldsymbol{\gamma}, \tilde{\boldsymbol{\gamma}}$ — that is $v_1(a, \boldsymbol{\gamma}, U) \geq v_1(b, \boldsymbol{\gamma}, U)$ and $v_1(a, \tilde{\boldsymbol{\gamma}}, U) \geq v_1(b, \tilde{\boldsymbol{\gamma}}, U)$, or $v_1(b, \boldsymbol{\gamma}, U) \geq v_1(a, \boldsymbol{\gamma}, U)$ and $v_1(b, \tilde{\boldsymbol{\gamma}}, U) \geq v_1(a, \tilde{\boldsymbol{\gamma}}, U)$ — then we are done: set $\tilde{U} = U$, and we automatically get properties 1 and 2.

Else, we have that $v_1(a, \boldsymbol{\gamma}, U) \geq v_1(b, \boldsymbol{\gamma}, U)$ and $v_1(b, \tilde{\boldsymbol{\gamma}}, U) \geq v_1(a, \tilde{\boldsymbol{\gamma}}, U)$, where moreover, one of these inequalities is strict. Then, we have the following facts:

FACT B.2. By Lemma B.1 and $\tilde{\gamma}_1 < \gamma_1$, $\mathsf{sw}(b, U) < \mathsf{sw}(a, U)$.

FACT B.3. By the fact that $v_1(b, \tilde{\boldsymbol{\gamma}}, U) > v_1(a, \tilde{\boldsymbol{\gamma}}, U)$ and Fact B.2, $u_1(b) > u_1(a)$.

Let $N'$ be the set of all voters $i$ for whom $u_i(a) > u_i(b)$. Note that by Fact B.2, $v_i(a, \boldsymbol{\gamma}, U) > v_i(b, \boldsymbol{\gamma}, U)$ for all $i \in N'$. We now show Equation (7), which states that in order for $\mathsf{sw}(a, U) \geq \mathsf{sw}(b, U)$, the gap between voters' utilities for $a$ and $b$ in $N'$ must at least compensate for the gap between voter 1's utilities for $b$ and $a$:

$$0 \leq \mathsf{sw}(a, U) - \mathsf{sw}(b, U) = -(u_1(b) - u_1(a)) + \sum_{i \in N \setminus \{1\}} (u_i(a) - u_i(b))$$
$$\leq -(u_1(b) - u_1(a)) + \sum_{i \in N'} (u_i(a) - u_i(b)).$$

and we conclude

$$\sum_{i \in N'} (u_i(a) - u_i(b)) \geq u_1(b) - u_1(a). \tag{7}$$

Then, by Equation (7), there must exist some vector of non-negative real numbers $\boldsymbol{\delta} = (\delta_i : i \in N')$ such that

$$0 \leq \delta_i \leq u_i(a) - u_i(b) \text{ for all } i \in N' \qquad \text{and} \qquad \sum_{i \in N'} \delta_i \geq u_1(b) - u_1(a).$$

Fix this vector $\boldsymbol{\delta}$, and use it to construct $\tilde{U}$ in the following way: first, for all voters $i$, set $\tilde{u}_i(b) = u_i(b)$. Then, set voters' utilities for $a$ as follows:

- $\tilde{u}_1(a) = u_1(a) + \sum_{i \in N'} \delta_i$,

- for all $i \in N'$, $\tilde{u}_i(a) = u_i(a) - \delta_i$, and

- for all other $i$, $\tilde{u}_i(a) = u_i(a)$.

By inspection, per this construction we have *Property 1*: that $\mathsf{sw}(a, U) = \mathsf{sw}(a, \tilde{U})$ and $\mathsf{sw}(b, U) = \mathsf{sw}(b, \tilde{U})$.

Finally, we show *Property 2*, that $\Pi_{V(\boldsymbol{\gamma}, U)} \subseteq \Pi_{V(\tilde{\boldsymbol{\gamma}}, \tilde{U})}$. First for all voters $i \in N' \cup \{1\}$, we have by the construction above that $\tilde{u}_i(a) \geq \tilde{u}_i(b)$; By Property 1, we also have that $\mathsf{sw}(a, \tilde{U}) > \mathsf{sw}(b, \tilde{U})$. Thus, $v_i(a, \tilde{\boldsymbol{\gamma}}, \tilde{U}) > v_i(b, \tilde{\boldsymbol{\gamma}}, \tilde{U})$. This is consistent with the fact that $v_i(a, \boldsymbol{\gamma}, U) > v_i(b, \boldsymbol{\gamma}, U)$ for all $i \in N' \cup \{1\}$, as fixed earlier in the proof. For all remaining voters $i \notin N' \cup \{1\}$, we did not change their utilities from $U$ to $\tilde{U}$, so we have that $v_i(a, \tilde{\boldsymbol{\gamma}}, \tilde{U}) = v_i(a, \boldsymbol{\gamma}, U)$ and $v_i(b, \tilde{\boldsymbol{\gamma}}, \tilde{U}) = v_i(b, \boldsymbol{\gamma}, U)$. We conclude that all PS-values are ordinally consistent for all voters across $V(\boldsymbol{\gamma}, U)$ and $V(\tilde{\boldsymbol{\gamma}}, \tilde{U})$, and thus $\Pi_{V(\boldsymbol{\gamma}, U)} \subseteq \Pi_{V(\tilde{\boldsymbol{\gamma}}, \tilde{U})}$, concluding the proof. □

**Proposition 4.5(b).** *When $m = 3$, all voting rules exhibit nonuniform monotonicity.*

PROOF. Fix an arbitrary $f$. Fix $U$ and $\tilde{\boldsymbol{\gamma}} < \boldsymbol{\gamma}$ where $\tilde{\gamma}_1 < \gamma_1$ and $\tilde{\gamma}_i = \gamma_i$ for all $i > 1$. We will prove the claim by showing that we can find some other utility matrix $\tilde{U}$ so that $\mathsf{dist}(f, \boldsymbol{\gamma}, U) \leq \mathsf{dist}(f, \tilde{\boldsymbol{\gamma}}, \tilde{U})$.

For notational convenience, for any instance $(\boldsymbol{\gamma}, U)$ we will write $\boldsymbol{\pi}^{\boldsymbol{\gamma}, U}$ to denote a profile compatible with $(\boldsymbol{\gamma}, U)$. Fix an arbitrary $\boldsymbol{\pi}^{\boldsymbol{\gamma}, U}$, and fix another profile $\boldsymbol{\pi}^{\tilde{\boldsymbol{\gamma}}, U}$ with the same tie-breaking when PS-values are equal. Note that these two profiles may differ only in voter 1's ranking (and if they don't, we can set $\tilde{U} = U$ and we are done). This proof will be conceptually similar to that of Proposition 4.5(a), except that instead of correcting one pairwise ranking, we must correct multiple in succession.

Define the swap$(\pi, a, b)$ function as one that intakes a ranking and two alternatives that are ranked adjacently in $\pi$, and outputs the ranking in which they are swapped; e.g., swap$(b \succ a, b, a) = a \succ b$. Now, define a sequence of unique pairwise swaps of alternatives adjacent in $\pi_1^{\tilde{\gamma}, U}$ such that, if made, would transform $\pi_1^{\tilde{\gamma}, U}$ into $\pi_1^{\gamma, U}$. Let this sequence be $(a_1, b_1), (a_2, b_2), \ldots, (a_T, b_T)$ where, by convention, we are swapping $b_t \succ a_t \rightarrow a_t \succ b_t$. That is, if we apply swap successively to $\pi_1^{\tilde{\gamma}, U}$ for alternatives $a_1, b_1 \ldots a_T, b_T$, we will get $\pi_1^{\gamma, U}$.

By Lemma B.4 (below), we can define a sequence of utility matrices $\tilde{U}_1, \tilde{U}_2, \ldots \tilde{U}_T$ such that

- **Property 1:** $\pi_i^{\tilde{U}_t, \tilde{\gamma}} = \pi_i^{\gamma, U}$ for all $i \neq 1, t \in [T]$
  (the rankings of all voters other than 1 are preserved in step $t$)

- **Property 2** $\pi_1^{\tilde{U}_{t+1}, \tilde{\gamma}} = \text{swap}(\pi_1^{\tilde{U}_t, \tilde{\gamma}}, b_{t+1}, a_{t+1})$ for all $t \in [T-1]$
  (so 1's pairwise mis-ordering of $a_{t+1}$ and $b_{t+1}$ is corrected in the $t + 1$-st step)

- **Property 3:** sw$(a, U) = $ sw$(a, \tilde{U}_t)$ for all $a \in [m], t \in [T]$
  (the welfares are preserved in step $t$)

It follows that $\boldsymbol{\pi}^{\tilde{U}_T, \tilde{\gamma}} = \boldsymbol{\pi}^{\gamma, U}$ and sw$(a, U) = $ sw$(a, \tilde{U}_T)$, together implying that

$$\text{dist}(f, \boldsymbol{\gamma}, U) = \text{dist}(f, \tilde{\gamma}, \tilde{U}_T),$$

concluding the proof.                                                                                                 □

LEMMA B.4. Let $m = 3$. Fix arbitrary $U$ and $\gamma > \tilde{\gamma}$, where $\tilde{\gamma}_1 < \gamma_1$ but all other voters' entries are identical. Let alternatives $a, b$ be such that $a \succ_{\pi_1^{\gamma, U}} b$ and $b \succ_{\pi_1^{\tilde{\gamma}, U}} a$, where $a$ and $b$ are ranked adjacently in $\pi_1^{\tilde{\gamma}, U}$. Then, there exists a $\tilde{U}$ such that:

- **Property 1** $\pi_i^{\tilde{\gamma}, \tilde{U}} = \pi_i^{\gamma, U}$ for all $i \neq 1$
  (the rankings of all voters other than 1 are preserved)

- **Property 2** $\pi_1^{\tilde{\gamma}, \tilde{U}} = \text{swap}(\pi_1^{\tilde{\gamma}, U}, b, a)$
  (so 1's pairwise mis-ordering of $a$ and $b$ is corrected)

- **Property 3:** sw$(a, U) = $ sw$(a, \tilde{U})$ for all $a \in [m]$
  (the welfares are preserved)

PROOF. We begin by establishing a series of facts:

FACT B.5. By the fact that $a, b$ were in the list of pairwise swaps, $a \succ_{\pi_1^{\gamma, U}} b$ and $b \succ_{\pi_1^{\tilde{\gamma}, U}} a$.

FACT B.6. By Lemma B.1, the fact that $b \succ_{\pi_1^{\tilde{\gamma}, U}} a$, and $\tilde{\gamma}_1 < \gamma_1$,

$$\text{sw}(b, U) \leq \text{sw}(a, U).$$

FACT B.7. By Fact B.5 and Fact B.6,[9] we have that

$$u_1(b) > u_1(a).$$

Now, define our set of voters $N'$ as in the $m = 2$ proof, i.e., as the set of all voters $i \in [n]$ such that $u_i(a) > u_i(b)$ (and thus, given Fact B.6, $a \succ_{\pi_i^{U,\tilde{\gamma}}} b$). Then, we know that by the same argument as before, using Facts B.6 and B.7, that

$$\sum_{i \in N'} (u_i(a) - u_i(b)) > u_1(b) - u_1(a). \qquad (8)$$

Now, we have that for all $i \in N'$, we have that $a \succ_{\pi_i^{U,\tilde{\gamma}}} b$. Let $c$ be the third, remaining alternative that is not equal to $a$ or $b$. Then, a voter $i \in N'$ can have one of three possible rankings in $\pi^{U,\tilde{\gamma}}$:

$$(1) \; a \succ b \succ c, \quad (2) \; c \succ a \succ b, \text{ or } \quad (3) \; a \succ c \succ b.$$

We will now prove three claims, one per ranking, which will lay the foundations for our later construction of $\tilde{U}$. We use $N'_{(1)}$ to mean the set of voters in $N'$ with ranking (1), and likewise for (2) and (3).

**Claim 1:** For all voters $i \in N'_{(1)}$ and for all $\delta_i^1 \in [0, u_i(a) - u_i(b)]$,

$$v_i(a, \tilde{\boldsymbol{\gamma}}, U) \geq (1 - \tilde{\gamma}_i)(u_i(b) + \delta_i) + \tilde{\gamma}_i \mathsf{sw}(b, U) \geq v_i(c, \tilde{\boldsymbol{\gamma}}, U).$$

The first inequality holds by Fact B.6 combined with $\delta_i$ being defined in $[u_i(b), u_i(a)]$. The second inequality is implied by the fact that $v_i(b, \tilde{\boldsymbol{\gamma}}, U) \geq v_i(c, \tilde{\boldsymbol{\gamma}}, U)$, inferred from the fact that $b \succ_{\pi_i^{U,\tilde{\gamma}}} c$ (i.e., $i$ ranks $b$ ahead of $c$).

**Claim 2:** For all voters $i \in N'_{(2)}$, and for all $\delta_i \in [0, u_i(a) - u_i(b)]$,

$$v_i(c, \tilde{\boldsymbol{\gamma}}, U) \geq (1 - \tilde{\gamma}_i)(u_i(a) - \delta_i) + \tilde{\gamma}_i \mathsf{sw}(a, U) \geq v_i(b, \tilde{\boldsymbol{\gamma}}, U).$$

*Proof of Claim 2:* The proof is essentially the same as that of Claim 1: The first inequality is implied by the fact that $v_i(c, \tilde{\boldsymbol{\gamma}}, U) \geq v_i(a, \tilde{\boldsymbol{\gamma}}, U)$, inferred from the fact that $i$ ranks $c$ ahead of $a$, and the second inequality holds by Fact B.6 combined with $\delta_i$ being defined in $[u_i(b), u_i(a)]$.

**Claim 3:** For all voters $i \in N'_{(3)}$, there exists some $u^*$ in the following interval

$$\left[ u_i(c) + \frac{\tilde{\gamma}_i(\mathsf{sw}(c, U) - \mathsf{sw}(a, U))}{(1 - \tilde{\gamma}_i)n}, \; u_i(c) + \frac{\tilde{\gamma}_i(\mathsf{sw}(c, U) - \mathsf{sw}(b, U))}{(1 - \tilde{\gamma}_i)n} \right],$$

such that $u^*$ is also in the interval $[u_i(b), u_i(a)]$ and satisfies

$$(1 - \tilde{\gamma}_i)u^* + \tilde{\gamma}_i \mathsf{sw}(a)/n \geq u_i^{\tilde{\gamma}}(c) \geq (1 - \tilde{\gamma}_i)u^* + \tilde{\gamma}_i \mathsf{sw}(b)/n. \qquad (9)$$

*Proof of Claim 3:* By Fact B.6, the upper end of the interval is indeed at least the lower end, so there can exist a $u^*$, as this is a non-empty region of the real line. Second, fixing any $u^*$ in this interval, the chain of inequalities in (9) is proven by simply rearranging the given fact that $u^*$ is in the provided interval. Finally, the given interval must overlap the interval $[u_i(b), u_i(a)]$, so we can choose some $u^*$ within both intervals. We show in two steps. First, the upper end of the interval is weakly larger than $u_i(b)$:

$$v_i(b, \tilde{\boldsymbol{\gamma}}, U) \leq v_i(c, \tilde{\boldsymbol{\gamma}}, U) \iff (1 - \tilde{\gamma}_i)u_i(b) + \tilde{\gamma}_i \mathsf{sw}(b, U)/n \leq (1 - \tilde{\gamma}_i)u_i(c) + \tilde{\gamma}_i \mathsf{sw}(c, U)/n$$

$$\iff u_i(c) + \frac{\tilde{\gamma}_i(\mathsf{sw}(c, U) - \mathsf{sw}(b, U))}{(1 - \tilde{\gamma}_i)n} \geq u_i(b).$$

---

[9]This is strict for the same reason as in the $m = 2$ case.

And the lower end of the interval is at most $u_i(a)$:

$$v_i(a, \tilde{\boldsymbol{\gamma}}, U) \leq v_i(c, \tilde{\boldsymbol{\gamma}}, U) \iff (1 - \tilde{\gamma}_i)u_i(a) + \tilde{\gamma}_i \mathsf{sw}(a, U)/n \geq (1 - \tilde{\gamma}_i)u_i(c) + \tilde{\gamma}_i \mathsf{sw}(c, U)/n$$

$$\iff u_i(c) + \frac{\tilde{\gamma}_i(\mathsf{sw}(c, U) - \mathsf{sw}(b, U))}{(1 - \tilde{\gamma}_i)n} \leq u_i(a).$$

*End of proof of Claim 3.*

**Claim 4 (Corollary of Claim 3).** For arbitrary $u^*$ satisfying the conditions of Claim 3, for all $\delta_i^{3,a} \in [0, u_i(a) - u^*]$, $i \in N'_{(3)}$ and all $\delta_i^{3,b} \in [0, u^* - u_i(b)]$, $i \in N'_{(3)}$, we have that

$$(1 - \tilde{\gamma}_i)(u_i(a) - \delta_i^{3,a}) + \tilde{\gamma}_i \mathsf{sw}(a)/n \geq (1 - \tilde{\gamma}_i)u^* + \tilde{\gamma}_i \mathsf{sw}(a)/n \tag{9}$$

$$\geq v_i(c, \tilde{\boldsymbol{\gamma}}, U)$$

$$\geq (1 - \tilde{\gamma}_i)u^* + \tilde{\gamma}_i \mathsf{sw}(b)/n \tag{9}$$

$$\geq (1 - \tilde{\gamma}_i)(u_i(b) + \delta^{3,b}) + \tilde{\gamma}_i \mathsf{sw}(a)/n.$$

*Choosing the $\boldsymbol{\delta}$s.* Taking the $\delta_i^{1,a}$, $\delta_i^{2,b}$, $\delta_i^{3,a}$ and $\delta_i^{3,b}$ and their domains from Claims 1, 2, and 4, we have that

$$0 \leq \sum_{i \in N'_{(1)}} \delta_i^{1,a} + \sum_{i \in N'_{(2)}} \delta_i^{2,b} + \sum_{i \in N'_{(3)}} (\delta_i^{3,a} + \delta_i^{3,b})$$

$$\leq \sum_{i \in N'_{(1)} \cup N'_{(2)}} (u_i(a) - u_i(b)) + \sum_{i \in N'_{(3)}} (u_i(a) - u^*) + (u^* - u_i(b))$$

$$= \sum_{i \in N'} u_i(a) - u_i(b)$$

$$> u_1(b) - u_1(a). \tag{8}$$

Thus, for any constant $t \in [0, u_1(b) - u_1(a)]$, there must exist settings of these deltas so that their sum over $i \in N'$ is equal to $t$. We will choose $\delta^*$ values $\delta_i^{*1,a}$ for all $i \in N_{(1)}$, $\delta_i^{*2,b}$ for all $i \in N_{(2)}$, $\delta_i^{*3,a}$ and $\delta_i^{*3,b}$ for all $i \in N_{(3)}$, so that they add up to

$$t^* = u_1(b) - u_1(a) - \frac{\tilde{\gamma}}{1 - \tilde{\gamma}}(\mathsf{sw}(a, U) - \mathsf{sw}(b, U))/n \tag{10}$$

Note that this value falls in the permitted range as it is clearly at most $u_1(b) - u_1(a)$, and it is at least 0 by a simple rearrangement of the known inequality $v_i(b, \tilde{\boldsymbol{\gamma}}, U) \geq v_i(a, \tilde{\boldsymbol{\gamma}}, U)$.

*Construction of $\tilde{U}$.*

- For all $i \notin N' \cup \{1\}$, set $i$'s utilities in $\tilde{U}$ as in $U$, i.e., $\tilde{u}_i(a) = u_i(a)$ and likewise for $b$ and $c$.

- For all $i \in N'$ set $\tilde{u}_i(c) = u_i(c)$, and

– for $i \in N'_{(1)}$, set

$$\tilde{u}_i(a) = u_i(a) - \delta_i^{*1,a},$$
$$\tilde{u}_i(b) = u_i(b)$$

– for $i \in N'_{(2)}$, set

$$\tilde{u}_i(a) = u_i(a)$$
$$\tilde{u}_i(b) = u_i(b) + \delta_i^{*2,b}$$

– for $i \in N'_{(3)}$, set

$$\tilde{u}_i(a) = u_i(a) - \delta_i^{*3,a}$$
$$\tilde{u}_i(b) = u_i(b) + \delta_i^{*3,b}.$$

- For voter 1, set $\tilde{u}_1(c) = u_1(c)$, and set

$$\tilde{u}_1(a) = u_1(a) + \sum_{i \in N'_{(1)}} \delta_i^{*1,a} + \sum_{i \in N'_{(3)}} \delta_i^{*3,a}$$
$$\tilde{u}_1(b) = u_1(b) - \sum_{i \in N'_{(2)}} \delta_i^{*2,b} - \sum_{i \in N'_{(3)}} \delta_i^{*3,b}.$$

By construction, all utilities are nonnegative.

$\tilde{U}$ *satisfies Property 3.* We want to show that $\mathrm{sw}(a, \tilde{U}) = \mathrm{sw}(a, U)$, and likewise for alternatives $b$ and $c$. This is true for $c$ by inspection, as for all $i \in [n]$, $\tilde{u}_i(c) = u_i(c)$. For $a$ and $b$, the argument is also by inspection, noting that the utility added or subtracted among the $N'$ group for either alternative is exactly compensated by the change to voter 1's utility for that alternative.

$\tilde{U}$ *satisfies Property 1.* We need to conclude that by our construction, all voters' other than 1's rankings were preserved, i.e., $\pi_i^{\tilde{\gamma},U} = \pi_i^{\tilde{\gamma},\tilde{U}}$ for all $i \neq 1$. We will confirm this by group:

- For all $i \notin N' \cup \{1\}$, this holds simply by the fact that $\tilde{U}$ satisfies Property 3, $\tilde{\gamma}_i = \gamma_i$, and $\tilde{u}_i(a) = u_i(a)$, $\tilde{u}_i(b) = u_i(b)$, and $\tilde{u}_i(c) = u_i(c)$.

- For all $i \in N'$, this follows from claims 1, 2, and 3 and the fact that we set the $\delta$s as specified according to the conditions of those claims.

$\tilde{U}$ *satisfies Property 2.* This is implied by the fact that $v_1(a, \tilde{\boldsymbol{\gamma}}, \tilde{U}) = v_1(b, \tilde{\boldsymbol{\gamma}}, \tilde{U})$, which we will prove now. First, we will following equality using (10):

$$\tilde{u}_1(b) - \tilde{u}_1(a) = u_1(b) - u_1(a) - t^* = \frac{\tilde{\gamma}}{1 - \tilde{\gamma}}(\mathrm{sw}(a, U) - \mathrm{sw}(b, U))/n.$$

Then, applying this equality,

$$v_1(b, \tilde{\boldsymbol{\gamma}}, \tilde{U}) - v_1(a, \tilde{\boldsymbol{\gamma}}, \tilde{U}) = (1 - \tilde{\gamma}_1)(\tilde{u}_1(b) - \tilde{u}_1(a)) + \tilde{\gamma}_1(\mathrm{sw}(b, U) - \mathrm{sw}(a, U))/n$$
$$= (1 - \tilde{\gamma}_1) \cdot \frac{\tilde{\gamma}_1}{1 - \tilde{\gamma}_1}(\mathrm{sw}(a, U) - \mathrm{sw}(b, U)/n + \tilde{\gamma}_1(\mathrm{sw}(b, U) - \mathrm{sw}(a, U))/n$$
$$= 0. \qquad \square$$

## B.2 Proof of Proposition 4.6

Proposition 4.6. Copeland is nonuniform PS-monotonic.

Proof. Let $f = $ Copeland. Since the case $m \leq 3$ is covered by Proposition 4.5, we may assume here that $m \geq 4$. For notational convenience, for any instance $(\boldsymbol{\gamma}, U)$ we will write $\boldsymbol{\pi}^{\boldsymbol{\gamma}, U}$ to denote a profile compatible with $(\boldsymbol{\gamma}, U)$.

It suffices to show that when a single voter's public spirit level is decreased, the worst-case distortion weakly increases. Suppose this voter is voter 1, and that their public spirit is decreased from $\gamma_1$ to $\tilde{\gamma}_1$, corresponding to a change from PS-vector $\boldsymbol{\gamma}$ to $\tilde{\boldsymbol{\gamma}}$ (all else kept the same). To prove monotonicity, it suffices to prove that for an arbitrary utility matrix $U$, we can find a utility matrix $\tilde{U}$ such that the winner $a'$ remains the same (i.e., $a' = $ Copeland$(\boldsymbol{\pi}^{\boldsymbol{\gamma}, U}) = $ Copeland$(\boldsymbol{\pi}^{\tilde{\boldsymbol{\gamma}}, \tilde{U}})$), and such that sw$(a', U) = $ sw$(a', \tilde{U})$, sw$(a^*, U) = $ sw$(a^*, \tilde{U})$. We make a case distinction now on whether $a'$ pairwise-dominates $a^*$ in $\boldsymbol{\pi}^{\boldsymbol{\gamma}, U}$.

Case 1: If $a'$ strictly pairwise-dominates $a^*$ in $\boldsymbol{\pi}^{\boldsymbol{\gamma}, U}$, then define $\tilde{U}$ such that for all $i \in [n]$,

- $\tilde{u}_i(a) = 0$ for all $a \notin \{a', a^*\}$
- $\tilde{u}_i(a) = u_i(a)$ for all $a \in \{a', a^*\}$

Now, we argue that dist(Copeland, $\boldsymbol{\gamma}, U$) = dist(Copeland, $\tilde{\boldsymbol{\gamma}}, \tilde{U}$):

Observation 1. The welfares of $a', a^*$ are preserved across $U, \tilde{U}$, i.e., sw$(a', U) = $ sw$(a', \tilde{U})$, sw$(a^*, U) = $ sw$(a^*, \tilde{U})$.

Observation 2. for all voters $i \neq 1$, $i$ has the same relative ordering of $a', a^*$ in $\boldsymbol{\pi}^{\boldsymbol{\gamma}, U}$ and $\boldsymbol{\pi}^{\tilde{\boldsymbol{\gamma}}, \tilde{U}}$. This is because from $\boldsymbol{\gamma}, U$ to $\tilde{\boldsymbol{\gamma}}, \tilde{U}$, $a', a^*$'s average utilities don't change, $i$'s utilities for $a', a^*$ don't change, and $\gamma_i$ doesn't change, meaning that $v_i(a', \boldsymbol{\gamma}, U) = v_i(a', \tilde{\boldsymbol{\gamma}}, \tilde{U})$ and $v_i(a^*, \boldsymbol{\gamma}, U) = v_i(a^*, \tilde{\boldsymbol{\gamma}}, \tilde{U})$.

Observation 3. In $\boldsymbol{\pi}^{\tilde{\boldsymbol{\gamma}}, \tilde{U}}$, $a'$ and $a^*$ pairwise-dominate all $a \notin \{a', a^*\}$. This is because all voters must rank $a', a^*$ in the first two positions and all the other alternatives in positions $3 \dots m$, by virtue of the fact that we can wlog assume that some voter has nonzero utility for $a^*$ (else the distortion will be 0), and thus some voter has nonzero utility for $a'$ (since it is sometimes ranked ahead of $a'$). In contrast, all other alternatives have average utility 0, and thus must be ranked behind $a', a^*$.

Observation 4. $a'$ pairwise-dominates $a^*$ in $\boldsymbol{\pi}^{\tilde{\boldsymbol{\gamma}}, \tilde{U}}$. If $a^* >_{\pi_1^{\boldsymbol{\gamma}, U}} a'$, then either voter 1's ranking is preserved, or $a' >_{\pi_i^{\tilde{\boldsymbol{\gamma}}, \tilde{U}}} a^*$, which can only strengthen $a'$'s pairwise domination of $a^*$. Conversely, if $a' >_{\boldsymbol{\pi}^{\boldsymbol{\gamma}, U}} a^*$, $a^*$ cannot overtake $a'$ by Lemma B.1.

These four observations, taken together, imply that in $\boldsymbol{\pi}^{\tilde{U}, \tilde{\boldsymbol{\gamma}}}$, $a'$ still pairwise-dominates $a^*$, and moreover, both $a'$ and $a^*$ pairwise-dominate everything else. We conclude that the uncovered set is $\{a'\}$, and thus $a'$ is the unique winner. By Observation 1, this directly implies that the distortion is preserved across $(\boldsymbol{\gamma}, U)$ and $(\tilde{\boldsymbol{\gamma}}, \tilde{U})$.

Case 2: Now, suppose $a'$ does not strictly dominate $a^*$. We may without loss of generality assume that $a^*$ is not a Copeland winner — indeed, if it were, then for this $U$ we would have dist(Copeland, $\boldsymbol{\gamma}, U$) = 1, in which case the distortion can only increase when voter 1's PS-level is dropped.

When $a^*$ is not a Copeland winner, it has a strictly lower Copeland score than $a'$, and thus there must exist some alternative $b$ such that $a'$ *strictly* pairwise-dominates $b$ and $b$ weakly pairwise dominates $a^*$. We now construct $\tilde{U}$ from $U$ in three steps. In the first step, for all alternatives $a \notin \{a', a^*, b\}$ and all voters $i \in [n]$, we set $\tilde{u}_i(a) = 0$. For all voters $i \neq 1$, we set their utilities in $\tilde{U}$ for $a', a^*, b$ to be the same as in $U$.

In the second step, we set the utilities for $a^*, a', b$ for voter $i = 1$, depending on the following case distinction.

- Suppose $\mathrm{sw}(b, U) > \mathrm{sw}(a', U)$.

  - In this case, the social welfares are ordered $\mathrm{sw}(a^*, U) \geq \mathrm{sw}(b, U) > \mathrm{sw}(a', U)$, while the above pairwise wins are

    $$a' \xrightarrow{\text{strictly}} b \xrightarrow{\text{weakly}} a^*.$$

    Since dropping $\gamma_1$ can only promote lower-welfare alternatives, we keep the same utilities for voter 1, and these pairwise wins will continue to hold.

  - Now, suppose $\mathrm{sw}(b, U) \leq \mathrm{sw}(a', U)$.

    * In this case, we know that dropping $\gamma_1$ can lead to the following promotions in 1's ranking: $b$ over $a'$, $b$ over $a^*$, or $a'$ over $a^*$. The last one doesn't concern us, as the promotion of $a'$ only helps $a'$ win, and the second-last one does not concern us because it will just strengthen the existing pairwise win of $b$ versus $a^*$. Thus, as long as the first promotion doesn't occur, we keep the same utilities as before.

    * If $b$ is promoted over $a'$, we drop its utility to $\tilde{u}_1(b) = 0$ for voter 1 (then, it will not be promoted, leaving only the option of promoting $a'$ over $a^*$). Then, if there exists someone who ranks $b$ ahead of $a'$, we add this utility to someone who ranks $b$ ahead of $a'$. If the person ranks $b$ ahead of $a^*$, this preserves their exact ranking; if they rank $b$ behind $a^*$, this may result in a strengthening of the pairwise defeat of $a^*$ by $b$, which does not change the Copeland winner. Else, if there is no one who ranks $b$ ahead of $a'$, then $b$ dominating $a'$ pairwise is not possible by changing any single person's ranking, so add this utility arbitrarily.

Finally, in the third step, we add identical copies of the 'intermediate' alternative $b$, to make $a'$ the unique Copeland winner. Again, we need a case distinction.

- *$n$ is even.* We take an 'empty' alternative $\bar{b} \in [m] \setminus \{a', a^*, b\}$ for which we previously set the utilities to 0, and re-set its utilities to be identical to $b$. We moreover choose the preference profile where any individual's preference between $a', a^*, \bar{b}$ is identical to the preference between $a', a^*, \bar{b}$ (i.e. $b, \bar{b}$ are always neighbouring in any $\pi_i$), and that $b, \bar{b}$ are in a tie (i.e. $|\{i : b \succ b'\}| = n/2$). In this constellation, $a'$ at least pairwise beats $b$ and $\bar{b}$ ($\geq 2$ points), $b$ and $\bar{b}$ at best pairwise beat $a^*$ ($\leq 1$ point), and $a^*$ at best beats $a'$ ($\leq 1$ point), so $a'$ is the winner.

- *$n$ is odd and $m \geq 5$.* Since we are unable to create pairwise ties when $n$ is odd, we have to treat this case separately. Let us assume first that $m \geq 5$. Then, we have at least two 'empty' alternatives for which we previously set the utilities to 0; let us call these $\bar{b}, \tilde{b} \in [m] \setminus \{a', a^*, b\}$. We then re-set the utilities for $\bar{b}, \tilde{b}$ to be identical to $b$, such that they are ranked relative to $a', a^*$ the same as $b$ by any individual. We moreover order $b, \bar{b}, \tilde{b}$ in so that they form a Condorcet cycle, and

$$b \xrightarrow{\text{strictly}} \bar{b} \xrightarrow{\text{strictly}} \tilde{b} \xrightarrow{\text{strictly}} b.$$

Note that we may do so freely, since all three alternatives are identical.

In this scenario, the Copeland scores are

– $a'$ gets 3 points (for beating $b, \bar{b}, \tilde{b}$),

– $a^*$ gets 1 point (for beating $a'$),

– $b, \bar{b}, \tilde{b}$ get 2 points,

whence $a'$ wins.

- **$n$ is odd and $m = 4$.** The previous arguments held for the Copeland rule with arbitrary tie-breaking between alternatives with identical Copeland score. In the specific case of $n$ being odd and $m = 4$, we need to make a slight refinement to our definition of distortion, namely that the distortion is a supremum over the whole Copeland set $CS(\pi^{\gamma,U})$ for any $(\gamma, U)$-compatible profile $\pi^{\gamma,U}$.

$$\text{dist}(\text{Copeland}, \gamma, U) = \sup_{a \in CS(\pi^{\gamma,U})} \frac{\text{sw}(a^*, U)}{\text{sw}(a, U)}.$$

It then suffices to ensure that $a'$ is *one* of the Copeland winners under $(\tilde{U}, \tilde{\gamma})$, not the unique one. Let the four alternatives be called $a', a^*, b, \bar{b}$. Since (i) $n$ is odd, (ii) we assumed that $a'$ does not strictly pairwise dominate $a^*$ and since we assumed that $a^*$ is not a Copeland winner, we can deduce that

– $a'$ has exactly Copeland score 2 (for beating $b, \bar{b}$.)

– $a^*$ has exactly Copeland score 1 (for beating $a'$.)

– There exist exactly two elements in the Copeland set (alternatives with score 2), suppose that $b$ is this element.

– Note that this $b$ is an admissible choice in the second step, We assume that it was chosen in the second step.

After the second step, we may here create a $\bar{b}$ identical to $b$, and suppose that $b$ pairwise beats $\bar{b}$. Then the Copeland set will again consist of the same alternatives $\{a', b\}$. Since the welfare of $b$ was preserved in the second step, the proof is now complete. □

## B.3 Proof of Proposition 4.7

Proposition 4.7. Plurality is nonuniform PS-monotonic.

Proof. For notational convenience, for any instance $(\gamma, U)$ we will write $\pi^{\gamma,U}$ to denote any profile compatible with $(\gamma, U)$.

It suffices to prove that when a single voter's public spirit level is decreased, the worst-case distortion increases. Suppose this voter is voter $i = 1$, and that $\gamma_1$ is changed from some value $\gamma_1 = \rho$ (Scenario 1) is changed to some lower value $\gamma_1 = \tilde{\rho} < \rho$ (Scenario 2). Let us denote by $\gamma$ the original PS-vector (with $\gamma_1 = \rho$), and by $\tilde{\gamma}$ the one which arises from lowering $\gamma_1$ to $\tilde{\rho}$. To prove monotonicity, it suffices to prove that for any utility matrix $U \in \mathbb{R}^{n \times m}$, we can find a utility matrix $\tilde{U}$ such that

(1) the winner remains the same, $a' = f(\pi^{\gamma,U}) = f(\pi^{\tilde{\gamma},\tilde{U}})$,

(2) the social welfares of $a', a^*$ are preserved, i.e.

$$\text{sw}(a', U) = \text{sw}(a', \tilde{U}), \ \text{sw}(a^*, U) = \text{sw}(a^*, \tilde{U}).$$

If voter 1's first-ranked alternative remains unchanged, there is nothing to prove, so let us assume the first-ranked alternative does change – let us denote by $\pi_1^{\gamma, U}(1) = a$ the alternatives which receive voter 1's vote in scenario 1 such that voter 1's rankings are of the form

- Scenario 1: $a \succ$ all other alternatives,

- Scenario 2: alternatives $A_1 \succ a \succ$ alternatives $A_2$.

Since the second ranking arises from the first ranking by lowering $\tilde{\gamma}$, only *alternatives with lower welfare* can be promoted over $a$, i.e. $A_1$ consists of alternatives with welfare below $\text{sw}(a, U)$.

We construct $\tilde{U}$ from $U$ in two steps. First, we set voter 1's utility for all alternatives in $A_1$ to zero. Since all those alternatives have lower welfare than $a$, this will restore $a$ as voter 1's first-ranked alternative. Since the highest-welfare alternative $a^*$ cannot have not been promoted over $a$, i.e. $a^* \in A_2$, its welfare remains unchanged.

This second step is to restore *some* of the welfares of alternatives in $A_1$ which were affected by the previous step. Specifically, let $\bar{a} \in A_1$. If there is a non-empty set $N_{\bar{a}} \subseteq [n]$, $|N_{\bar{a}}| \geq 1$ of voters (in Scenario 1) who rank $\bar{a}$ first, we add an $u_1(\bar{a})/|N_j|$ amount of utility to all the voters in $N_{\bar{a}}$,

$$\tilde{u}_i(\bar{a}) = u_i(\bar{a}) + \frac{u_1(\bar{a})}{|N_{\bar{a}}|}, \ \forall i \in N_{\bar{a}}.$$

If on the other hand $\bar{a}$ is ranked first by no voter, we do not intervene.

We claim that these two steps combined restore the first-ranked alternatives of all voters, and thus the winner of the election. To see this, we notice the following.

- **Welfares.** For any $\bar{a} \in A_1$ with $N_{\bar{a}} \neq \emptyset$, $\text{sw}(\bar{a}, U) = \text{sw}(\bar{a}, \tilde{U})$. The other alternatives $\bar{a} \in A_1$ with $N_{\bar{a}} = \emptyset$ may have lower welfare $\text{sw}(\bar{a}, \tilde{U}) \leq \text{sw}(\bar{a}, U)$. The welfares of alternatives in $A_2$, in particular of $a^* \in A_2$, remain unchanged.

- **Voters with first-choice in $A_1$.** If a voter first-ranks some alternative $\bar{a} \in \tilde{A}_1$ in Scenario 1 $(\gamma, U)$, then they still do so in Scenario 2 $(\tilde{\gamma}, \tilde{U})$, since they have added utility for $\bar{a}$ while the welfares of all other alternatives are either the same or lower.

- **Voters with first-choice in $\{a\} \cup A_2$.** Suppose a voter first-ranks some $\bar{a} \in \{a\} \cup A_2$ under $(\gamma, U)$. Then, since both their utility and welfare for $\bar{a}$ are the same under $(\tilde{\gamma}, \tilde{U})$ while the welfares of other alternatives can only have decreased, they continue to first-rank $\bar{a}$ under $(\tilde{\gamma}, U)$.

This concludes the proof.                                                                                 □

## B.4  Proof of Lemma 4.13

LEMMA 4.13. *If $f$ is weakly unanimous and instance-wise PS-monotonic, then it is monotonic.*

PROOF. Suppose that $f$ is weakly unanimous but not monotonic; we will show that it is not instance-wise PS-monotonic. Fix a pair of profiles $\pi, \pi'$ in which monotonicity is violated, i.e., where there exists some voter $i^* \in [n]$ such that $a$ is promoted via an adjacent swap in $\pi'_{i^*}$ compared to $\pi_{i^*}$, but $f(\pi) = a$ and $f(\pi') = b$. Let $\bar{a}$ be the alternative over which $a$ is promoted from $\pi_{i^*}$ to $\pi'_{i^*}$.

Given that $f(\pi) = a$ and the fact that $f$ is weakly unanimous, for every $c \neq a$, there must exist some voter $i_c$ such that $a >_{\pi_{i_c}} c$. Arbitrarily choose one such voter per $c$ and denote them $i_c$, for all $c \neq a$. Note that it is possible that some such $i_c = i^*$; we will handle this in the proof.

Now, we will construct a pair of instances $\gamma, U$ and $\gamma, U'$ such that $\gamma'$ differs from $\gamma$ only in that $\gamma'_{i^*} > \gamma_{i^*}$, and that three claims hold: *Claim (1):* dist$(f, \gamma', U) >$ dist$(f, \gamma, U)$, *Claim (2):* $\pi \in \Pi_{V(\gamma, U)}$, *Claim (3):* $\pi' \in \Pi_{V(\gamma, U')}$. Together, these claims constitute a violation of instance-wise PS-monotonicity.

**Construction of $\gamma, \gamma'$:** Let $\gamma = \mathbf{0}$ (i.e., all voters have public spirit level 0). Let $\gamma'$ be defined such that $\gamma'_i = \gamma_i = 0$ for all $i \neq i^*$, and let $\gamma'_{i^*} = \epsilon$, where $\epsilon > 0$ is set to some number smaller than $1/2m^2$.

**Construction of $U$:**

- Group 1: For all voters $i \neq i^*$ and $i \notin \{i_c | c \in [m] \setminus \{a\}\}$, let $i$ have 0 utility for all alternatives.

- Group 2: For all voters $i \neq i^*$ and $i \in \{i_c | c \in [m] \setminus \{a\}\}$, let $i$ have utility 1 for $a$ and all alternatives ranked ahead of $a$ in $\pi_i$, and 0 for all other alternatives.

- For $i^*$: starting at the first-ranked alternative in $\pi_{i^*}$, assign utilities starting at 1 and let them descend at intervals of $1/m^2$ until we reach alternative $a$. Then, assign $u_{i^*}(a)$ so that $u_{i^*}(\tilde{a}) - u_{i^*}(a) = \epsilon^2/n$. Now, continuing in order of the $\pi_{i^*}$ after $a$, continue assigning alternatives utilities descending at intervals of $1/m^2$.

**Proof of Claims (1), (2), and (3):**

*Claim (1):* We prove this by proving that $a$ has strictly higher social welfare than any other alternative. Then, the winner changing from $a$ to $b$ from $\pi$ to $\pi'$ must increase the distortion, i.e., dist$(f, \gamma', U) >$ dist$(f, \gamma, U)$.

First, if there is no $c$ such that $i_c = i^*$, then for all $c \neq a$, we have that $\sum_{i \in \text{Group 1}}(u_i(a) - u_i(c)) = 0$, $\sum_{i \in \text{Group 2}}(u_i(a) - u_i(c)) \geq 1$, and $u_{i^*}(a) - u_{i^*}(c) \geq -1/m$. Thus, $\sum_{i \in [n]}(u_i(a) - u_i(c)) > 0$, equivalent to sw$(a, U) >$ sw$(c, U)$.

If there exists $c^*$ such that $i_{c^*} = i^*$, then the previous case holds for all $c \neq c^*$. For $c^*$, we repeat the above analysis: $\sum_{i \in \text{Group 1}}(u_i(a) - u_i(c^*)) = 0$, $\sum_{i \in \text{Group 2}}(u_i(a) - u_i(c^*)) \geq 0$, and $u_{i^*}(a) - u_{i^*}(c^*) \geq 1/m^2$, the final inequality by the fact that $a >_{\pi_{i^*}} c^*$. Thus, again $\sum_{i \in [n]}(u_i(a) - u_i(c^*)) > 0$, equivalent to sw$(a, U) >$ sw$(c^*, U)$.

*Claim (2):* We have assigned voters' utilities in weakly decreasing order according to $\pi_i$ for all $i$, and $\gamma = \mathbf{0}$, meaning that voters' individual utilities fully determine their rankings: thus, $\pi \in \Pi_{V(\gamma, U)}$.

*Claim (3):* The high level proof of this claim is the following: First, for all voters $i \neq i^*$, their $\gamma_i = \gamma'_i$, so their rankings implied by $U\gamma$ and $\gamma, U'$ are the same, as is consistent with $\pi_i = \pi'_i$. For voter $i^*$, the separation between the utilities for all pairs of alternatives other than $\tilde{a}, a$ are too large for an $\epsilon$ increase in public spirit to flip them; however, the separation between the utilities of $\tilde{a}, a$ are small enough for this increase to flip them, realizing the transformation from $\pi_{i^*} \rightarrow \pi'_{i^*}$.

Building on the notation of $\pi \in \Pi_{V(\gamma, U)}$ (meaning that the profile $\pi$ is consistent with the instance $U, \gamma$), we use $\pi_i \in \Pi_{V_i(\gamma, U)}$ to mean a voter $i$'s ranking $\pi_i$ is consistent with the vector of PS-values implied by the $i$th row of the matrix $V(\gamma, U)$.

For all voters $i \neq i^*$, by construction of $\boldsymbol{\pi}, \boldsymbol{\pi}'$ we have that $\pi_i = \pi_i'$. Moreover, $\gamma_i = \gamma_i'$ implies that $\Pi_{V_i(\boldsymbol{\gamma}, U)} = \Pi_{V_i(\boldsymbol{\gamma}, U')}$. By these two equalities, $\pi_i \in \Pi_{V_i(\boldsymbol{\gamma}, U)}$ (as shown in Claim (2)) implies $\pi_i' \in \Pi_{V_i(\boldsymbol{\gamma}, U')}$.

Now, it only remains to show that $\pi_{i^*}' \in \Pi_{V_{i^*}(\boldsymbol{\gamma}, U')}$. First, we observe that for all alternatives $c$,

$$|v_{i^*}(c, \boldsymbol{\gamma}, U) - v_{i^*}(c, \boldsymbol{\gamma}', U)| = |u_{i^*}(c) - (1-\epsilon)u_{i^*}(c) - \epsilon \text{sw}(c, U)/n| = \epsilon|u_{i^*}(c) - \text{sw}(c, U)/n| \leq \epsilon, \quad (11)$$

where the final step holds because all utilities in $U$ are bounded between 0 and 1.

Next, we observe that for all pairs of alternatives $(c, c') \neq (\tilde{a}, a)$, we have that

$$|v_{i^*}(c, \boldsymbol{\gamma}, U) - v_{i^*}(c', \boldsymbol{\gamma}, U)| = |u_{i^*}(c) - u_{i^*}(c')| \geq 1/m^2 > 2\epsilon. \quad (12)$$

Now, fix an arbitrary pair of alternatives $(c, c') \neq (\tilde{a}, a)$ such that $c \succ_{\pi_{i^*}} c'$, and thus $v_{i^*}(c, \boldsymbol{\gamma}, U) \geq v_{i^*}(c', \boldsymbol{\gamma}, U)$. Then, by Equations (11) and (12) we have that $v_{i^*}(c, \boldsymbol{\gamma}', U) \geq v_{i^*}(c', \boldsymbol{\gamma}', U)$:

$$0 < v_{i^*}(c, \boldsymbol{\gamma}, U) - v_{i^*}(c', \boldsymbol{\gamma}, U) - 2\epsilon \qquad\qquad\qquad\qquad\qquad \text{by (12)}$$
$$\leq v_{i^*}(c, \boldsymbol{\gamma}, U) - v_{i^*}(c', \boldsymbol{\gamma}, U) - |v_{i^*}(c, \boldsymbol{\gamma}, U) - v_{i^*}(c, \boldsymbol{\gamma}', U)| - |v_{i^*}(c', \boldsymbol{\gamma}', U) - v_{i^*}(c', \boldsymbol{\gamma}, U)| \quad \text{by (11)}$$
$$\leq v_{i^*}(c, \boldsymbol{\gamma}, U) - v_{i^*}(c', \boldsymbol{\gamma}, U) - (v_{i^*}(c, \boldsymbol{\gamma}, U) - v_{i^*}(c, \boldsymbol{\gamma}', U)) - (v_{i^*}(c', \boldsymbol{\gamma}', U) - v_{i^*}(c', \boldsymbol{\gamma}, U))$$
$$= v_{i^*}(c, \boldsymbol{\gamma}', U) - v_{i^*}(c', \boldsymbol{\gamma}', U)$$

We conclude that for all such pairs $(c, c')$,

$$v_{i^*}(c, \boldsymbol{\gamma}, U) \geq v_{i^*}(c', \boldsymbol{\gamma}, U) \implies v_{i^*}(c, \boldsymbol{\gamma}', U) \geq v_{i^*}(c', \boldsymbol{\gamma}', U). \quad (13)$$

Next, we consider the remaining pair $(a, \tilde{a})$. First, we observe that

$$\text{sw}(a, U) - \text{sw}(\tilde{a}, U) > \epsilon,$$

by the fact that $\sum_{i \in \text{Group 1}}(u_i(a) - u_i(\tilde{a})) = 0$, $\sum_{i \in \text{Group 2}}(u_i(a) - u_i(\tilde{a})) \geq 1$ (note that it cannot be that, given the existence of a $c^* : i_{c^*} = i^*, c^* = \tilde{a}$, because we know that $\tilde{a} \succ_{\pi_{i^*}} a$), and $u_{i^*}(a) - u_{i^*}(\tilde{a}) = -\epsilon^2/n$. Adding up over voters, these inequalities imply that $\text{sw}(a, U) - \text{sw}(\tilde{a}, U) \geq 1 - \epsilon^2/n > \epsilon$.

Then, we show the the inequality

$$v_{i^*}(a, \boldsymbol{\gamma}', U) > v_{i^*}(\tilde{a}, \boldsymbol{\gamma}', U) \quad (14)$$

via the following deduction, where the first inequality uses that $u_{i^*}(a) - u_{i^*}(\tilde{a}) = -\epsilon^2/n < 0$, and the second inequality uses that $\text{sw}(a, U) - \text{sw}(\tilde{a}, U) \geq \epsilon$:

$$v_{i^*}(a, \boldsymbol{\gamma}', U) - v_{i^*}(\tilde{a}, \boldsymbol{\gamma}', U) = (1-\epsilon)(u_{i^*}(a) - u_{i^*}(\tilde{a})) + \epsilon(\text{sw}(a, U)/n - \text{sw}(\tilde{a}, U)/n)$$
$$> -\epsilon^2/n + \epsilon(\text{sw}(a, U) - \text{sw}(\tilde{a}, U))/n$$
$$\geq -\epsilon^2/n + \epsilon \cdot \epsilon/n$$
$$= 0.$$

By Equations (13) and (14), we have that any ranking $\pi$ with the following two properties must be consistent with $\Pi_{V_{i^*}(\boldsymbol{\gamma}, U')}$: First, for all pairs of alternatives $(c, c') \neq (a, \tilde{a})$, $c \succ_{\pi_{i^*}} c' \implies c \succ_\pi c'$, and second, $a \succ_\pi \tilde{a}$. $\pi_{i^*}'$ satisfies these criteria by construction, and thus $\pi_{i^*}' \in \Pi_{V_{i^*}(\boldsymbol{\gamma}, U')}$, as needed, concluding the proof. $\qquad \square$

## B.5 Proof of Lemma 4.15

LEMMA 4.15. If $f$ weakly unanimous and monotonic, then if $f$ is instance-wise PS-monotonic, it must also be swap-invariant.

PROOF. We will prove the contrapositive. Suppose $f$ is not swap-invariant. Then, there exists two profiles $\boldsymbol{\pi}, \boldsymbol{\pi}'$ that differ only in that for some voter $i^*$, $b$ and $c$ are adjacently swapped in their ranking, and $f(\boldsymbol{\pi}) = a$ but $f(\boldsymbol{\pi}') = b$. By the monotonicity of $f$, we know that $c \succ_{\pi_{i^*}} b$: otherwise, going from $\boldsymbol{\pi}' \to \boldsymbol{\pi}$, $b$ would be promoted over $c$ but lose the winning spot, violating monotonicity. Now, we will break into cases depending on the nature of $\boldsymbol{\pi}$, and in either case, show that PS-monotonicity is violated.

CASE 1: $\boldsymbol{\pi}$ contains at least one voter $i$ who ranks $b \succ_{\pi_i} c$.

Now, we will construct $\boldsymbol{\gamma}, U, \boldsymbol{\gamma}'$ such that the following claims hold: *Claim (1):* $\text{sw}(a, U) > \text{sw}(b, U) > \text{sw}(c, U)$; *Claim (2):* $\boldsymbol{\pi} \in \Pi_{V(\boldsymbol{\gamma}, U)}$; and *Claim (3):* $\boldsymbol{\pi}' \in \Pi_{V(\boldsymbol{\gamma}, U')}$. If these claims are true, then by the construction of our example, we have found $\boldsymbol{\gamma} \leq \boldsymbol{\gamma}'$ such that by increasing the public spirit from $\boldsymbol{\gamma}$ to $\boldsymbol{\gamma}'$, we can change the winner from $a$ to $b$, thereby increasing the distortion, a violation of instance-wise PS-monotonicity.

**Construction of $\boldsymbol{\gamma}, \boldsymbol{\gamma}'$.** Let $\boldsymbol{\gamma} = 0$, $\boldsymbol{\gamma}'$ such that $\gamma_i' = \gamma_i = 0$ for all $i \neq i^*$, and $\gamma_{i^*}' = \epsilon$ for some small $\epsilon > 0$ where $\epsilon < 1/(16m)$.

**Construction of $U$.** We set the utilities according to three cases (where latter cases apply only if earlier cases do not hold):

A. If there exists an $i$ who ranks $a \succ_{\pi_i} b \succ_{\pi_i} c$, set $i$'s utilities in weakly decreasing order of $\pi_i$ such that $a$ (and everything before it) gets utility 1, $b$ (and everything after $a$ and before $b$) gets utility 1/2, and $c$ (and everything after) gets utility 0.

   Give all remaining voters besides $i^*$ utility 0 for all alternatives.

B. Else if there exists an $i$ where $b \succ_{\pi_i} a \succ_{\pi_i} c$, set $i$'s utilities in weakly decreasing order of $\pi_i$: give $a$ and everything ranked before it (including $b$) utility 1, and everything after $a$ (including $c$) 0 utility.

   Then, by weak unanimity of $f$, there must be another voter $i'$ where $a \succ_{\pi_{i'}} b$, whose utilities we assign based on two cases:

   B1. If $i' \neq i^*$, set $i'$'s utilities according to $\pi_{i'}$: give all alternatives ranked ahead of $b$ utility 1/2 (this must include $a$ and $c$), and utility 0 to $b$ and all alternatives ranked after.

   B2. If $i' = i^*$, note that $i'$ must have ranking $a \succ_{\pi_{i'}} c \succ_{\pi_{i'}} b$, because $c$ and $b$ must be ranked adjacently. Then, give utility 1 to $a$, 1/2 to $c$, $1/2 - \epsilon^2/n$ to $b$, and set the rest of the alternatives' utilities so they are decreasing at intervals of at least $1/(4m)$.

   Give all other voters except $i^*$ with thus far unset utilities 0 utility for all alternatives.

C. Else, by the falseness of cases A and B and our assumption that there is some $i$ for which $b \succ_{\pi_i} c$, there must exist some voter $i$ who ranks $b \succ_{\pi_i} c \succ_{\pi_i} a$. Set $i$'s utilities in weakly decreasing order of $\pi_i$: Give $b$ and all alternatives before it utility $1/2 + \epsilon^2/n$ (the $+\epsilon^2/n$ is for convenience of arguments later), and all alternatives after it (including $c$ and $a$) utility 0.

Then, by the weak unanimity of $f$, there must exist one voter $i'$ where $a >_{\pi_{i'}} b$ and $a >_{\pi_{i'}} c$, in which case, by the falseness of cases A and B they must have ranking $a >_{\pi_{i'}} c >_{\pi_{i'}} b$.[10] Set $i'$'s utilities in weakly decreasing order of $\pi'_i$, based on two cases:

C1. If $i' \neq i^*$: let $i'$ have utility $1 + \epsilon^2/n$ for $a$ (the $+\epsilon^2/n$ is for convenience of arguments later) and all alternatives ranked before it and utility $0$ for all alternatives ranked after it (including $c$ and $b$).

C2. If $i' = i^*$, set $i'$'s utilities similar to how we did in B2: give utility $3/2$ to $a$, $1/2$ to $c$, $1/2 - \epsilon^2/n$ to $b$, and set the rest of the alternatives' utilities so they are decreasing at intervals of at least $1/(4m)$.

Give all other voters with unset utilities except $i^*$ $0$ utility for all alternatives.

If we have not already set $i^*$'s utilities in cases B or C (we cannot set them in case A), set $i^*$'s utilities in weakly decreasing order of $\pi_{i^*}$: give the alternatives ahead of (and including) $c$ utilities starting at $1/4$ and dropping by additive gaps of $1/(4m)$. Then, set $u_{i^*}(b)$ such that $u_{i^*}(c) - u_{i^*}(b) = \epsilon^2/n$. Then, for alternatives ranked after $b$, continue assigning utilities decreasing by additive gaps of $1/(4m)$.

**Proofs of Claims (1), (2), and (3).**

*Claim (1):* Let $N_A = [n] \setminus \{i\}$, $N_{B1} = [n] \setminus \{i\}$ $N_{B2} = [n]$, $N_{C1} = [n] \setminus \{i\}$, $N_{C2} = [n]$, denote the sets of voters whose utilities are set within cases $A, B$ and $C$, depending on which cases are invoked.

Now, we will show that for any $N \in \{N_A, N_{B1}, N_{B2}, N_{C1}, N_{C2}\}$, we have that

$$\sum_{i \in N} u_i(a) > \sum_{i \in N} u_i(b) > \sum_{i \in N} u_i(c),$$

and moreover, that these inequalities hold by a margin of at least $1/2$.

(Case A): letting $N = N_A$, we have $\sum_{i \in N} u_i(a) = 1$, $\sum_{i \in N} u_i(b) = 1/2$, $\sum_{i \in N} u_i(c) = 0$.

(Case B1): letting $N = N_{B1}$, we have $\sum_{i \in N} u_i(a) = 3/2$, $\sum_{i \in N} u_i(b) = 1$, $\sum_{i \in N} u_i(c) = 1/2$.

(Case B2): letting $N = N_{B2}$, we have $\sum_{i \in N} u_i(a) = 2$, $\sum_{i \in N} u_i(b) = 3/2 - \epsilon^2/n$, $\sum_{i \in N} u_i(c) = 1/2$.

(Case C1): letting $N = N_{C1}$, we have $\sum_{i \in N} u_i(a) = 1 + \epsilon^2/n$, $\sum_{i \in N} u_i(b) = 1/2 + \epsilon^2/n$, $\sum_{i \in N} u_i(c) = 0$.

(Case C2): letting $N = N_C$, we have $\sum_{i \in N} u_i(a) = 3/2$, $\sum_{i \in N} u_i(b) = 1$, $\sum_{i \in N} u_i(c) = 1/2$.

If cases B2 or C2 was the binding case — that is, we set $i^*$ while within the three cases —, then we have already concluded the claim, and $\mathrm{sw}(a, U) - \mathrm{sw}(b, U) \geq 1/2$ and $\mathrm{sw}(b, U) - \mathrm{sw}(c, U) \geq 1/2$. Otherwise, we note that for any pair of alternatives $d, e$, $|u_{i^*}(d) - u_{i^*}(e)| \leq 1/4$; therefore, these social welfare gaps cannot be closed by more than $1/4$, and we conclude that $\mathrm{sw}(a, U) - \mathrm{sw}(b, U) \geq 1/4$ and $\mathrm{sw}(b, U) - \mathrm{sw}(c, U) \geq 1/4$. We will use this lower bound on these gaps later, in Claim (3).

*Claim (2):* We have assigned voters' utilities in weakly decreasing order according to $\pi_i$ for all $i$, and $\gamma = 0$, meaning that voters' individual utilities fully determine their rankings: thus, $\pi \in \Pi_{V(\gamma, U)}$.

*Claim (3):* The proof of this claim follows the same structure as that of Claim (3) in the proof of Lemma 4.13, so we will be slightly more brief here, and invoke parts of that argument when useful.

---

[10]The alternative would be that there would have to exist two voters, the first for whom $c \succ a \succ b$, and the second for whom $b \succ a \succ c$, which is not possible by the falseness of case B.

We again use the notation $\pi_i \in \Pi_{V_i(\boldsymbol{\gamma}, U)}$ to mean a voter $i$'s ranking $\pi_i$ is consistent with the vector of PS-values implied by the $i$th row of the matrix $V(\boldsymbol{\gamma}, U)$.

First, for all voters $i \neq i^*$, by construction of $\boldsymbol{\pi}, \boldsymbol{\pi}'$ we have that $\pi_i = \pi_i'$. Moreover, $\gamma_i = \gamma_i'$ implies that $\Pi_{V_i(\boldsymbol{\gamma}, U)} = \Pi_{V_i(\boldsymbol{\gamma}, U')}$. By these two equalities, $\pi_i \in \Pi_{V_i(\boldsymbol{\gamma}, U)}$ (as shown in Claim (2)) implies $\pi_i' \in \Pi_{V_i(\boldsymbol{\gamma}, U')}$.

Now considering voter $i^*$, we want to show that $\pi_{i^*}' \in \Pi_{V_{i^*}(\boldsymbol{\gamma}, U')}$. To show this, first fix a pair of alternatives $(d, d') \neq (b, c)$. By the same type of reasoning as in Lemma 4.13, we have that $|v_{i^*}(d, \boldsymbol{\gamma}, U) - v_{i^*}(d', \boldsymbol{\gamma}, U)| \geq 1/(4m) > 4\epsilon\}$, and also that $|v_{i^*}(d, \boldsymbol{\gamma}, U) - v_{i^*}(d, \boldsymbol{\gamma}', U)| \leq 2\epsilon$ and $|v_{i^*}(d', \boldsymbol{\gamma}, U) - v_{i^*}(d', \boldsymbol{\gamma}', U)| \leq 2\epsilon$, by the fact that all utilities in $U$ are bounded between 0 and 2. Putting these facts together, we get that for all such pairs $d, d'$,

$$v_{i^*}(d, \boldsymbol{\gamma}, U) \geq v_{i^*}(d', \boldsymbol{\gamma}, U) \implies v_{i^*}(d, \boldsymbol{\gamma}', U) \geq v_{i^*}(d', \boldsymbol{\gamma}', U). \tag{15}$$

Now, finally considering the pair $b, c$, we have the following, using that $\text{sw}(c, U) - \text{sw}(b, U) \geq 1/4$, as shown in the proof of Claim (1):

$$\begin{aligned} v_{i^*}(b, \boldsymbol{\gamma}', U) - v_{i^*}(c, \boldsymbol{\gamma}', U) &= (1 - \epsilon)(u_{i^*}(b) - u_{i^*}(c)) + \epsilon(\text{sw}(b, U)/n - \text{sw}(c, U)/n) \\ &> -\epsilon^2/n + \epsilon(\text{sw}(b, U) - \text{sw}(b, U))/n \\ &\geq -\epsilon^2/n + \epsilon/(4n) \\ &\geq 0. \end{aligned}$$

We conclude that

$$v_{i^*}(b, \boldsymbol{\gamma}', U) - v_{i^*}(c, \boldsymbol{\gamma}', U) > 0. \tag{16}$$

By Equations (15) and (16), we have that any ranking $\pi$ with the following two properties must be consistent with $\Pi_{V_{i^*}(\boldsymbol{\gamma}, U')}$: First, for all pairs of alternatives $(d, d') \neq (b, c)$, $d \succ_{\pi_{i^*}} d' \implies d \succ_\pi d'$, and second, $b \succ_\pi c$. $\pi_{i^*}'$ satisfies these criteria by construction, and thus $\pi_{i^*}' \in \Pi_{V_{i^*}(\boldsymbol{\gamma}, U')}$, as needed, concluding the proof of CASE 1.

**CASE 2:** $\boldsymbol{\pi}$ does not contain a voter $i$ who ranks $b \succ_{\pi_i} c$.

First, observe that $c \succ_{\pi_i} b$ for all $i$ implies that $\boldsymbol{\pi}$ contains at least one voter in which $b \succ_{\pi_i} a$. To see this, first observe that $f(\boldsymbol{\pi}') = b$ implies that $b$ cannot always be ranked behind $a$ in $\boldsymbol{\pi}'$ by weak unanimity; thus there must be a voter $i'$ such that $b \succ_{\pi_{i'}'} a$. Next, since swapping $b$ and $c$ from $\boldsymbol{\pi} \to \boldsymbol{\pi}'$ cannot change the relative ordering of either of these alternatives with $a$, so it must also be the case that $b \succ_{\pi_{i'}} a$ (i.e., there exists such a voter in $\boldsymbol{\pi}$). We let $i'$ be this voter throughout this case.

Now, we will construct $\boldsymbol{\gamma}, U, \boldsymbol{\gamma}'$ such that three claims are true: *Claim (1)*: $\text{sw}(c, U) > \text{sw}(b, U) > \text{sw}(a, U)$, *Claim (2)*: $\boldsymbol{\pi}' \in \Pi_{V(\boldsymbol{\gamma}, U')}$, and *Claim (3)*: $\boldsymbol{\pi} \in \Pi_{V(\boldsymbol{\gamma}, U)}$. If these claims hold, then we will have *decreased* voters' public spirit from $\boldsymbol{\gamma} \to \boldsymbol{\gamma}'$, which realizes the transformation from $\boldsymbol{\pi} \to \boldsymbol{\pi}'$. This transformation changed the winner from $a$ to $b$ — an *increase* in the social welfare and a violation of PS-monotonicity.

**Construction of $\boldsymbol{\gamma}, \boldsymbol{\gamma}'$.** We let $\boldsymbol{\gamma}$ such that $\gamma_i = 0$ for all $i \neq i^*$, and $\gamma_{i^*} = \epsilon$ for some small $0 < \epsilon < 1/m^4$, and let $\boldsymbol{\gamma}' = \boldsymbol{0}$.

**Construction of $U$.** First, for $i^*$, let their utility for the first-ranked alternative in $\pi_{i^*}'$ be $1/m$, and then, in order of $\pi_{i^*}$, assign the alternatives utilities descending at intervals of $1/m^2$ until we reach $b$. Then set $u_{i^*}(b)$ so that $u_{i^*}(b) - u_{i^*}(c) = \epsilon^2/n$. Then, starting after $c$, continue down $\pi_{i^*}'$ assigning alternatives decreasing utilities at intervals of $1/m^2$. For the remaining voters, we break into cases:

- If $i' \neq i^*$, we assign $i'$'s utilities according to $\pi'_{i'}$: let $i'$'s utilities be 1 for $c$ and all alternatives $i'$ ranks ahead of $c$; $1/2$ for $b$ and all alternatives $i'$ ranks between $c$ and $b$, and 0 for all alternatives $i'$ ranks after $b$ (note that this includes $a$, by selection of $i'$). Give all other voters besides $i^*$ and $i'$ 0 utility for all alternatives.

- If $i' = i^*$, then pick another arbitrary voter $i''$ for whom $c >_{\pi'_{i''}} b$. We assign $i''$'s utilities according to their ranking $\pi'_{i''}$: give $c$ and all alternatives ranked ahead of $c$ utility $1/m^3$, give 0 utility to all alternatives they rank after $c$. Give all other voters besides $i^*$ and $i''$ 0 utility for all alternatives.

**Proofs of Claims (1), (2), and (3).**

*Claim (1):* If $i^* \neq i'$, then the only voters with any nonzero utilities are $i^*$ and $i'$; by their utilities, $\mathrm{sw}(c) - \mathrm{sw}(b) = 1/2 - \epsilon^2/n$ and $\mathrm{sw}(b) - \mathrm{sw}(a) \geq 1/2 - 1/m$ (where the $-1/m$ is the maximum possible gap between $u_{i^*}(a)$ and $u_{i^*}(b)$). If $i^* = i'$, then the only voters with any nonzero utilities are $i^*$ and $i''$; by their utilities, $\mathrm{sw}(c) - \mathrm{sw}(b) = 1/m^3 - \epsilon^2/n$, and $\mathrm{sw}(b) - \mathrm{sw}(a) \geq 1/m^2 - 1/m^3$.

*Claim (2):* We have assigned voters' utilities in weakly decreasing order according to $\pi'_i$ for all $i$, and $\boldsymbol{\gamma}' = \mathbf{0}$, meaning that voters' individual utilities fully determine their rankings: thus, $\boldsymbol{\pi}' \in \Pi_{V(\boldsymbol{\gamma}', U)}$.

*Claim (3):* The proof of this claim follows the same structure as that of Claim (3) in Case 1, so we will be slightly more brief here, and invoke parts of that argument when useful. We again use the notation $\pi_i \in \Pi_{V_i(\boldsymbol{\gamma}, U)}$ to mean a voter $i$'s ranking $\pi_i$ is consistent with the vector of PS-values implied by the $i$th row of the matrix $V(\boldsymbol{\gamma}, U)$.

First, for all voters $i \neq i^*$, by construction of $\boldsymbol{\pi}, \boldsymbol{\pi}'$ we have that $\pi_i = \pi'_i$. Moreover, $\gamma_i = \gamma'_i$ implies that $\Pi_{V_i(\boldsymbol{\gamma}, U)} = \Pi_{V_i(\boldsymbol{\gamma}, U')}$. By these two equalities, $\pi'_i \in \Pi_{V_i(\boldsymbol{\gamma}, U')}$ (as shown in Claim (2)) implies $\pi_i \in \Pi_{V_i(\boldsymbol{\gamma}, U)}$.

Now considering voter $i^*$, we want to show that $\pi_{i^*} \in \Pi_{V_{i^*}(\boldsymbol{\gamma}, U)}$. To show this, first fix a pair of alternatives $(d, d') \neq (b, c)$. By the same type of reasoning as in Case 1, we have that $|v_{i^*}(d, \boldsymbol{\gamma}', U) - v_{i^*}(d', \boldsymbol{\gamma}', U)| \geq 1/m^2 > 2\epsilon\}$, and also that $|v_{i^*}(d, \boldsymbol{\gamma}, U) - v_{i^*}(d, \boldsymbol{\gamma}', U)| \leq \epsilon$ and $|v_{i^*}(d', \boldsymbol{\gamma}, U) - v_{i^*}(d', \boldsymbol{\gamma}', U)| \leq \epsilon$, by the fact that all utilities in $U$ are bounded between 0 and 1. Putting these facts together, we get that for all such pairs $d, d'$,

$$v_{i^*}(d, \boldsymbol{\gamma}', U) \geq v_{i^*}(d', \boldsymbol{\gamma}', U) \implies v_{i^*}(d, \boldsymbol{\gamma}, U) \geq v_{i^*}(d', \boldsymbol{\gamma}, U). \tag{17}$$

Now, finally considering the pair $b, c$, we have the following, using that $\mathrm{sw}(c, U) - \mathrm{sw}(b, U) \geq 1/m^4 > \epsilon$, as shown in the proof of Claim (1):

$$\begin{aligned}
v_{i^*}(c, \boldsymbol{\gamma}, U) - v_{i^*}(b, \boldsymbol{\gamma}, U) &= (1 - \epsilon)(u_{i^*}(c) - u_{i^*}(b)) + \epsilon(\mathrm{sw}(c, U)/n - \mathrm{sw}(b, U)/n) \\
&> -\epsilon^2/n + \epsilon(\mathrm{sw}(c, U) - \mathrm{sw}(b, U))/n \\
&\geq -\epsilon^2/n + \epsilon^2/n \\
&= 0.
\end{aligned}$$

We conclude that

$$v_{i^*}(b, \boldsymbol{\gamma}', U) - v_{i^*}(c, \boldsymbol{\gamma}', U) > 0. \tag{18}$$

By Equations (17) and (18), we have that any ranking $\pi$ with the following two properties must be consistent with $\Pi_{V_{i^*}(\boldsymbol{\gamma}, U')}$: First, for all pairs of alternatives $(d, d') \neq (b, c)$, $d >_{\pi'_{i^*}} d' \implies d >_{\pi} d'$, and second, $c >_{\pi} b$. $\pi_{i^*}$ satisfies these criteria by construction, and thus $\pi_{i^*} \in \Pi_{V_{i^*}(\boldsymbol{\gamma}, U)}$, as needed, concluding the proof of Case 2.

$\square$

## B.6 Proof of Lemma 4.17

Lemma 4.17. *If $f$ is monotonic and swap-invariant, then it is Maskin-monotonic.*

Proof. Fix a monotonic and swap-invariant voting rule $f$, and fix a profile $\boldsymbol{\pi}$ such that $f(\boldsymbol{\pi}) = a$. Let $\boldsymbol{\pi}'$ be an arbitrary other profile such that such that $a \succ_{\pi_i'} b$ whenever $a \succ_{\pi_i} b$ for every voter $i$ and for all $b \neq a$. Now, we will show that we can construct $\boldsymbol{\pi}'$ from $\boldsymbol{\pi}$ by promoting $a$ and/or swapping $b$ with alternatives other than $a$. By monotonicity and swap-invariance, this will preserve the winner thus it will hold that $f(\boldsymbol{\pi}') = a$, thereby proving the Maskin monotonicity of $f$.

Fix an $i$, and consider $\pi_i$, from which we must construct $\pi_i'$. First, let $A_1$ be the set of all alternatives ranked ahead of $a$ in $\pi_i$ but behind $a$ in $\pi_i$. Swap the alternatives in $A_1$ with other alternatives ahead of $a$ in $\pi_i$ so that all these alternatives are ranked just ahead of $a$. These swaps didn't change the $f$ winner by the swap invariance of $f$. Then, swap $a$ ahead of all alternatives in $A_1$–this does not change the $f$ winner by the monotonicity of $f$. Finally, swap alternatives other than $a$ to make the relative ordering of all alternatives ahead of and behind $a$, respectively, match their relative ordering in $\pi_i'$; by swap invariance of $f$, this again does not change the $f$ winner. We can do this procedure to the rankings if all $i$, and thereby construct $\boldsymbol{\pi}'$ from $\boldsymbol{\pi}$ while preserving $a$ as the winner. □